# A Survey and Comparison of Post-quantum and Quantum Blockchains

Zebo Yang, *Graduate Student Member, IEEE*, Haneen Alfauri, Behrooz Farkiani, Raj Jain, *Life Fellow, IEEE*, Roberto Di Pietro, *Fellow, IEEE*, and Aiman Erbad, *Senior Member, IEEE*

*Abstract* — Blockchains have gained substantial attention from academia and industry for their ability to facilitate decentralized trust and communications. However, the rapid progress of quantum computing poses a significant threat to the security of existing blockchain technologies. Notably, the emergence of Shor's and Grover's algorithms raises concerns regarding the compromise of the cryptographic systems underlying blockchains. Consequently, it is essential to develop methods that reinforce blockchain technology against quantum attacks. In response to this challenge, two distinct approaches have been proposed. The first approach involves post-quantum blockchains, which aim to utilize classical cryptographic algorithms resilient to quantum attacks. The second approach explores quantum blockchains, which leverage the power of quantum computers and networks to rebuild the foundations of blockchains. This paper aims to provide a comprehensive overview and comparison of post-quantum and quantum blockchains while exploring open questions and remaining challenges in these domains. It offers an in-depth introduction, examines differences in blockchain structure, security, privacy, and other key factors, and concludes by discussing current research trends.

## I. INTRODUCTION

THE emergence of quantum computing threatens the security of current cryptographic systems that underpin blockchain technology. Blockchain technology, which facilitates decentralized trust and communication, has garnered significant attention in both academia and industry. Its recent widespread utilization was seen in Bitcoin [1], a decentralized cryptocurrency platform that operates independently of central banks. The increasing popularity of investing in cryptocurrency highlights the demand for a partially democratic decentralized environment. Over time, blockchain technology has evolved beyond its original role as a cryptocurrency system. It has emerged as a recognized method for establishing decentralized systems, including decentralized clouds and networks [2], [3]. Also, it finds diverse applications, encompassing decentralized applications (dApps) built on smart contracts [4], as well as cost-effective distributed platforms such as Internet of Things (IoT) systems [5].

The rise of quantum computing, however, has threatened the security of the foundational components of blockchains, namely public-key cryptography and hash functions. Shor's algorithm [6] can efficiently factorize numbers, thus compromising the cryptographic foundations of most public-key encryption algorithms. Grover's algorithm [7] enables a quadratic speedup in unstructured search, posing a threat to the hash functions utilized in linking blocks within blockchains. In response to these challenges, the blockchain community has been actively investigating alternative solutions that can withstand attacks from quantum computers. One such solution is post-quantum cryptography [8], which employs encryption methods not based on factorization, thereby mitigating the threat posed by Shor's algorithm. Blockchains that incorporate post-quantum cryptography are referred to as *post-quantum blockchains*. On the other hand, ongoing research is also exploring the utilization of quantum computers and networks to reconstruct the fundamental structure of blockchains. These systems, known as *quantum blockchains*, can be hybrid architectures that combine classical and quantum components or entirely quantum-based systems.

The emergence of quantum computing, coupled with the ongoing development of quantum networks, holds significant potential for integration with classical networks. While such networks' widespread deployment has not yet been achieved, their ongoing development remains closely intertwined with the ongoing debate regarding post-quantum and quantum blockchains. A notable example of this potential integration is quantum key distribution (QKD) [9], where quantum and classical communications are combined to establish a hybrid network for secure key exchange [10]. The discussion of the future blockchain technology is centered around the decision between classical computing with post-quantum cryptography and a quantum internet [11], [12]. While proponents argue that post-quantum cryptography can safeguard classical blockchains, alternative viewpoints assert that the full potential of decentralized networks can only be realized through the

Zebo Yang is with the Department of Computer Science and Engineering, Washington University, St. Louis, MO 63130 USA (e-mail: zebo@wustl.edu).

Haneen Alfauri is with the Department of Electrical and Systems Engineering, Washington University, St. Louis, MO 63130 USA (e-mail: a.haneen@wustl.edu).

Behrooz Farkiani is with the Department of Computer Science and Engineering, Washington University, St. Louis, MO 63130 USA (e-mail: b.farkiani@wustl.edu).

Raj Jain is with the Department of Computer Science and Engineering, Washington University, St. Louis, MO 63130 USA (e-mail: jain@wustl.edu).

Roberto Di Pietro is with the Computer, Electrical and Mathematical Science and Engineering Division, King Abdullah University of Science and Technology, Saudi Arabia, (e-mail: roberto.dipietro@kaust.edu.sa).

Aiman Erbad is with the College of Science and Engineering, Hamad Bin Khalifa University, Doha 5825, Qatar (e-mail: aerbad@hbku.edu.qa).

adoption of quantum blockchains on a quantum internet.

This paper offers an in-depth introduction, comparison, and categorization of the state-of-the-art post-quantum and quantum blockchains. The purpose is to facilitate an understanding of the strengths and weaknesses of each technology, encourage innovation, and guide the development of blockchain systems that can withstand the challenges presented by quantum computing. While there have been several surveys and reviews on post-quantum or quantum blockchains [13], [14], [15], [16], [17], there is a lack of extensive discussions and comparisons, including both post-quantum and quantum blockchains. This paper seeks to fill this gap by presenting a detailed analysis of these two technologies. It offers a more extensive examination of the latest advancements, encompassing both quantum and post-quantum blockchains, along with their possible influence and avenues for exploration toward building secure blockchain systems during the quantum era.

The comparison of the two technologies is crucial for several reasons, yet it is currently lacking. Firstly, the two technologies are both valid solutions to tackle the threat to the cryptographic foundations of classical blockchains. Which one of them is the appropriate path to follow is still undetermined. Both methods should be investigated to determine their strengths and limitations for specific applications and use cases. Comparing them helps to identify which technology is better suited to provide adequate security against quantum attacks. Secondly, the transition from classical to quantum computing will likely be gradual, and both systems will coexist for some time. The comparison can help determine the compatibility and interoperability with the existing computing systems. This can ensure a smoother transition as quantum computing becomes more prevalent. Lastly, evaluating the advantages and limitations of both technologies can provide valuable insights to organizations, businesses, and governments looking to adopt or invest in blockchain technology. This comparison can guide decision-makers in choosing the most appropriate technology for their specific use cases and requirements. Moreover, providing insights into the potential uses and drawbacks of post-quantum and quantum blockchains can assist researchers and developers in creating more robust and secure blockchains for the future.

Table I presents the main differences between existing reviews of quantum-related blockchain technologies and this paper. It is worth noting that the table only indicates the varying emphases of different studies. It does not necessarily indicate superiority or inferiority among them.

With that, the contributions of this paper can be outlined as follows:

- This survey provides a pedagogical introduction to blockchain technology, post-quantum cryptography, and quantum computing to guide beginners.
- This paper surveys the potential and limitations of post-quantum and quantum blockchains by examining the current state-of-the-art in these areas.
- This paper compares the key components of post-quantum and quantum blockchains.
- This paper discusses the research trends and their impacts on current blockchain platforms.

TABLE I
COMPARISON TO RELATED WORK

| Studies | Overview of Blockchain | Overview of Quantum Computing | Post-quantum Cryptography | Post-quantum Blockchains | Quantum Tech for Blockchains | Quantum Blockchains | Categorization of Research | Comparison of the Two Tech | Lessons Learned | Research Directions |
|---|---|---|---|---|---|---|---|---|---|---|
| [13] | ✓ | ✗ | ✓ | ✓ | ✗ | ✗ | ✓ | ✗ | ✓ | ✓ |
| [14] | ✓ | ✓ | ✗ | ✗ | ✓ | ✓ | ✗ | ✗ | ✗ | ✗ |
| [15] | ✓ | ✓ | ✓ | ✓ | ✗ | ✗ | ✓ | ✗ | ✗ | ✗ |
| [16] | ✗ | ✗ | ✓ | ✓ | ✗ | ✗ | ✓ | ✗ | ✗ | ✓ |
| [17] | ✓ | ✓ | ✓ | ✓ | ✓ | △ | △ | ✗ | ✗ | ✗ |
| This Paper | ✓ | ✓ | ✓ | ✓ | ✓ | ✓ | ✓ | ✓ | ✓ | ✓ |

△: *The study covers part of the subject.*

The structure of the rest of the paper is as follows. Section II introduces the foundational technologies, including classical blockchains, quantum attacks, and potential solutions. Section III discusses post-quantum blockchains, while Section IV examines quantum blockchains. Section V compares these two fields and highlights the key differences between them. Section VI discusses the lessons learned and the research trends. Section VII concludes the paper.

## II. PRELIMINARIES

In this section, we provide an overview of classical blockchain technology, quantum computing, and the impacts of quantum computing on current cryptographic systems, i.e., Shor's and Grover's algorithms. We discuss how quantum computing threatens the security of classical blockchains and how post-quantum and quantum blockchains can help thwart the introduced threats.

*A. Classical Blockchains*

Blockchain technology, made famous through its use in Bitcoin, is a decentralized system that facilitates transactions without needing a central authority [18]. It relies on consensus mechanisms and classical cryptographic techniques to ensure the security and immutability of the data housed within its network. In this subsection, we delve into the essential elements of a blockchain system, encompassing how network peers recognize each other for transactions (addresses), the structuring of the decentralized data (data structure), and the protocol for maintaining and agreeing on new data entries (consensus mechanism). In the following subsections, we use Bitcoin as an illustration. While numerous adaptations have been applied to the original blockchain structure over time, most systems retain a shared foundational core.

*A.1. Blockchain addresses*

In a blockchain network, each participant is required to have a unique identifier referred to as a blockchain address. This address is generated through public key cryptography, which creates pairs of public and private keys. Example technologies that employ public key cryptography include RSA [19], DSA [20], and EC-DSA [21]. In a public key scheme, the public key can be disseminated openly while its owner should closely guard its corresponding private key. An individual can use their private key to sign a message or transaction, and this signature can then be authenticated by anyone within the network using the individual's public key. This procedure guarantees the integrity of transactions on the blockchain. It is important to note that a blockchain address is typically derived from a user's public key, often by hashing it to create a more concise and user-friendly form. A user may possess multiple pairs of public and private keys.

In cryptocurrency applications, blockchain nodes or peers represent the individuals engaged in cryptocurrency transactions. To receive coins, a person can instruct that they be sent to their unique blockchain address. Similarly, they can dispatch coins using the same address and validate the transaction using the corresponding private key. The legitimacy of this expenditure can subsequently be confirmed by using their public key. An example of such a transaction is shown in Fig. 1, where User $B$ sends one coin to User $C$ by signing the transaction $T_{BC}$ with their private key $SK_B$, which can be verified by others using User $B$'s public key $PK_B$.

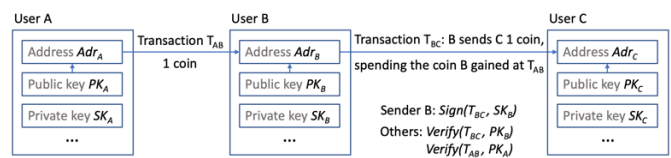

**Fig. 1.** An example of a sequence of blockchain transactions.

It is worth noting that users may have multiple addresses for different transactions to maintain their privacy and anonymity, as using a single address for all transactions could potentially lead to their identification [22]. To protect privacy and prevent identification, users may create a new address for each transaction, which is supported by most blockchain wallet applications [23].

*A.2. Blockchain data*

Within a blockchain network, information (like transactions) is secured in a sequence of blocks that are interconnected using hash pointers. Each block houses multiple transactions and a block header that carries the hash value of the preceding block in the chain. This generates a chain of blocks, depicted in Fig. 2, where the inclusion of any subsequent block bolsters the ones that came before it. If an attempt is made to modify a block in the chain, the successful execution of this would require recalculating all the succeeding blocks, which becomes increasingly challenging when employing proof-of-work consensus algorithms. The block header also encompasses metadata such as a timestamp and a nonce, which play a role in the consensus mechanism. Hash functions, such as SHA-2 and SHA-3 families [24], which convert an input of any length into a fixed-length output, are used to create the hash pointers that point to the previous block [25]. The resulting data, referred to as the hash value, acts as a unique fingerprint for the input data.

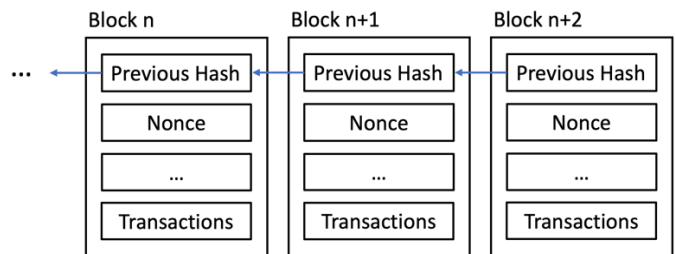

**Fig. 2.** A linked chain of blocks.

Hash functions are generally regarded as non-reversible, illustrating the concept of preimage resistance. Yet, an alternate tactic for attacking this system could involve identifying a hash collision, that is, an input string that produces an identical output. Currently, the sole method for discovering a hash collision involves combing through the entire search space, for which no classical algorithm exists to expedite the process, exemplifying the concept of second preimage resistance.

*A.3. Blockchain consensus*

The consensus mechanism in blockchain technology refers to the process that nodes in a blockchain network use to reach an agreement about the contents of the distributed ledger. This mechanism is essential for preserving the security, integrity, and reliability of the blockchain. In the Bitcoin network, certain nodes called miners strive to earn rewards by securing the privilege to generate blocks. Bitcoin utilizes the proof-of-work consensus algorithm. With this system, miners engage in a race to solve a computationally intensive problem that permits them to add new blocks of verified transactions to the blockchain. This process entails discovering a nonce, which, when combined with the block data and the previous block's hash, results in a value below a predefined limit. The first miner to crack the problem disseminates the solution to the network. If a node verifies the solution as correct, it appends the new block to its version of the blockchain. The hash of the block functions as a pointer reference to the previous block. After a certain number of blocks have been added, it becomes apparent which block has been accepted by the majority of the nodes. This competitive framework fortifies the network's security and encourages miners to participate.

Before a block is finalized, pending transactions are collected into a pool, with miners verifying these transactions' authenticity. If an attacker intends to execute a double-spending attack, they would need to become a miner and complete the proof-of-work faster than the rest of the network to validate both transactions. This scenario is highly improbable, as the network's other nodes collectively possess more computational power than the attacker, enabling them to reach a consensus on the legitimate version of the transactions more rapidly. However, should the attacker gain control over more than 50% of the network's computational power, they might be able to validate both processes. This scenario is referred to as a 51% attack [26].

Discovering a valid hash can include conditions like identifying a nonce that, when combined with the rest of the block and hashed, yields a hash starting with a specific number of zeros [1]. Although this task can be computationally intensive and time-consuming, its verification is straightforward and involves running a single hash function. Once miners discover a valid hash, they forge a new block with it and distribute it to all other nodes in the network. The other miners accept the block only if the hash is valid and all the transactions in the block are verified. Their acceptance is signified by creating the next block in the chain using the hash of the approved block. Occasionally, this may result in forks where different nodes validate block sets. In such scenarios, the block followed by a predetermined number of blocks (for example, six blocks in Bitcoin's case) is acknowledged as the final block, and any shorter forks from that block are discarded. For instance, in Fig. 3, the orange block is recognized as the final block, while the two light blue forks are disregarded.

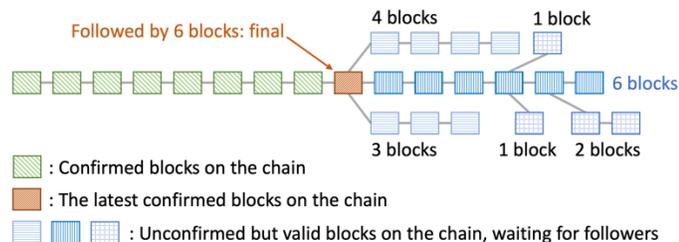

**Fig. 3.** An example of finalizing a new block.

The proof-of-work consensus mechanism offers robust security against classical computers, yet it is relatively energy-intensive. Here, it is also worth mentioning the Byzantine generals' problem, which led to one of the earliest consensus protocols and is commonly referenced and utilized in quantum blockchain solutions. A Byzantine agreement protocol addresses the Byzantine generals' problem [27]. It is often used as a metaphor to represent the challenges of reaching a consensus in a decentralized and potentially distrustful network, as it embodies the task of coordinating efforts among multiple possibly untrustworthy parties. By equating these challenges to the Byzantine generals' problem, researchers can conceptualize the hurdles of distributed consensus, paving the way to devise solutions applicable to a wide array of decentralized systems, including blockchain technologies.

In the Byzantine generals' problem, a group of generals commanding separate divisions of the Byzantine army must decide collectively whether to attack or retreat. They can only communicate via messengers, and some generals may be traitorous, sending false messages or trying to confuse others. The generals need an algorithm to satisfy: 1) All loyal (honest) generals agree on the same action plan. 2) A small fraction of traitors cannot lead loyal generals to a bad plan.

The Byzantine agreement protocol [28] is designed to tackle the Byzantine generals' problem. It is a protocol among $n$ entities in which one entity (sender) holds an input value $X$ (picked from a finite domain), and all other entities eventually decide on an output value (picked from the same finite domain). The protocol is said to achieve the Byzantine agreement (or consensus) if it guarantees that: 1) all honest receiver entities decide on the same output value Y; and 2) if the sender is honest then $Y = X$ [27], [28]. It is necessary to point out that the sender himself may be dishonest. An example of Byzantine Agreement protocol is presented in [27] and referred to as the Broadcast protocol [25] or classical Byzantine agreement protocol in literature. In addition, the proof-of-work consensus algorithm, which is used in Bitcoin, solves the Byzantine problem probabilistically [29].

Additionally, we concisely describe the Broadcast protocol, frequently referred to in the quantum blockchain literature. Assume $n$ nodes are connected by pairwise authenticated channels. Of these, $m$ nodes may be dishonest ($m < n/3$). Assume each node $i$ wants to make other nodes its decision or action $X_i$. The protocols execute in a sequence of communication rounds. In the first round, each node $i$ sends its $X_i$ to all other nodes. In the subsequent rounds, all nodes send all the information they received in all previous rounds from other nodes to all other nodes. It is proved that this protocol achieves the Byzantine agreement in no more than $m + 1$ rounds. However, its communication overhead is considerably high.

Beyond proof-of-work, Byzantine, and Broadcast protocol, there exists a diverse range of other consensus methodologies, including proof-of-stake [30], delegated proof-of-stake [31], [32], proof-of-capacity, and proof-of-authority. Unlike proof-of-work, which is based on solving complex mathematical problems, proof-of-stake chooses the block producer based on their stake (e.g., the number of tokens they hold and are willing to stake for the chance to validate a block). Delegated proof-of-stake is a variation of proof-of-stake where token holders vote for the delegates to create new blocks. It is often faster than proof-of-stake and proof-of-work because it relies on a smaller set of trusted validators rather than the entire network. Proof-of-capacity is a mechanism that allows miners to mine blocks and verify transactions based on the amount of storage space they have. The more storage capacity a miner allocates to the network, the higher their chances of being selected to mine a new block. Proof-of-authority is typically used in permissioned or private blockchains where trust is distributed among a limited subset, and the identity of validators is known and can be proven.

All these consensus mechanisms serve the same purpose of validating transactions and maintaining the integrity of the blockchain. The choice of consensus mechanism depends on the specific needs of the blockchain in question, including its required speed, security level, and decentralization. For a more comprehensive exploration of different consensus mechanisms, detailed information can be found in [31], [33], [34].

*A.4. Smart contracts*

A smart contract in a blockchain is a self-executing contract with the terms of the agreement between parties directly written into lines of code. The code and the agreements contained therein exist across a blockchain network. Smart contracts eliminate the need for a trusted third party or arbitrator to facilitate the execution of conditional services (i.e., the fulfillment of contract terms under specific conditions) [4], [35]. When the conditions specified in the contract are met, the actions specified in the contract are automatically triggered. In this way, blockchain smart contracts can be considered an "if-then" statement in a programming language. Smart contracts can be used for various purposes, from simple transactions like sending cryptocurrency from one person to another to more complex operations such as running dApps on a blockchain.

*B. Quantum Computing*

Quantum computing harnesses the fundamental principles of quantum mechanics to perform computations. In contrast to classical computers that rely on classical physics, quantum computers operate based on the principles of quantum mechanics. Central to quantum computing are qubits, which differ from the classical bits used in conventional computers. Qubits possess two distinguishable states and are represented by quantum particles, like the spin of an electron (spin up and spin down) or the polarization of a photon (horizontal and vertical). Notably, qubits exhibit two fundamental properties of quantum mechanics: superposition and entanglement.

In contrast to classical computers, where a bit can only be in the states of 0 or 1, a qubit can exist in the state of 0, 1, or a combination of both with assigned probabilities. This phenomenon is known as superposition. While a qubit remains in a superposition state, its specific state is uncertain. Superposition lets us perform computations on multiple states simultaneously, offering an exponential advantage over classical computing. A qubit can maintain its superposition state until it undergoes measurement. Upon measurement, it collapses into a definite 0 or 1 state, leading to the loss of the quantum information encoded in the probabilities of the superposition [36]. In particular, when a qubit has an equal chance of being measured as either 0 or 1, it is described as being in a *uniform* superposition state. This concept can be extended to multiple qubits. For example, a 2-qubit state can exhibit an equal probability of being measured in any of the four potential states (00, 01, 10, 11) in a uniform superposition.

Additionally, quantum entanglement refers to a phenomenon in which the states of two particles become interdependent, even when they are physically separated. This means that the state of one particle can instantly influence the state of the other particle, regardless of the distance between them. Thus, when a measurement is performed on one particle, the other reacts correspondingly and instantaneously, exemplifying the entanglement between the two particles [37]. This is the "spooky action at a distance" popularized by Einstein and formally introduced in their EPR paper [38], which has now been proven by repeated experiments and for which a Nobel prize in physics was awarded in 2022 [39].

Due to entanglement and superposition, quantum computers can perform operations on all $2^n$ possible states of $n$ qubits as the input simultaneously. Nevertheless, qubits are extremely unstable and require isolated areas with very low temperatures. This is one of the biggest impediments to commercializing quantum computers [36].

The quantum state of a qubit can be written as a linear combination of basis states $|0\rangle = \begin{bmatrix}1\\0\end{bmatrix}$ and $|1\rangle = \begin{bmatrix}0\\1\end{bmatrix}$ as $|\psi\rangle = \begin{bmatrix}\alpha\\\beta\end{bmatrix} = \alpha\begin{bmatrix}1\\0\end{bmatrix} + \beta\begin{bmatrix}0\\1\end{bmatrix} = \alpha|0\rangle + \beta|1\rangle$ in Dirac notation [40], in which $\alpha$ and $\beta$ are complex numbers. After measurement, $|0\rangle$ and $|1\rangle$ always convert into classical bits 0 and 1, respectively. For quantum state $|\psi\rangle$, the probability of measuring classical bit 0 is $|\alpha|^2$ and classical bit 1 is $|\beta|^2$ and $|\alpha|^2 + |\beta|^2 = 1$. The quantum state of a qubit can also be represented by a Bloch sphere, as shown in Fig. 4 (sphere generated by QuTiP, a toolbox in Python) [41]. The quantum state is formulated as $|\psi\rangle = cos\frac{\theta}{2}|0\rangle + e^{i\varphi}sin\frac{\theta}{2}|1\rangle$ [42].

Quantum gates are basic quantum circuits operating on qubits. They are building blocks of quantum algorithms and can be represented by matrices. As examples of quantum gates, one can name NOT, AND, OR, CNOT, and Hadamard gates. For a deep understanding of quantum gates and other aspects of quantum computing, readers can refer to [42] and [43].

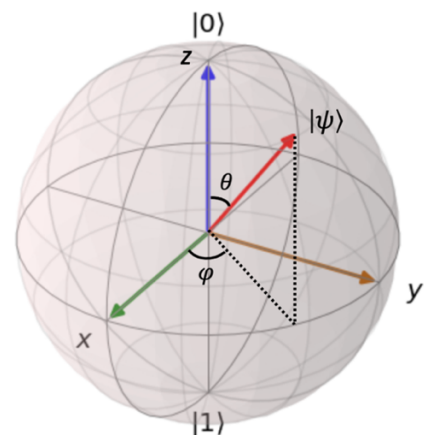

**Fig. 4.** State $|\psi\rangle$ on a Bloch Sphere.

*C. Impacts of Quantum Computing*

Quantum computing offers an exponential advantage over classical computing, enabling more efficient solutions to computationally challenging problems. While this presents significant benefits, it also introduces potential risks when in the hands of adversaries.

Numerous blockchains, including Bitcoin, employ the elliptic curve digital signature algorithm (ECDSA) to generate public-private key pairs and SHA-256 for generating hash values. These hash values serve various purposes, such as linking blocks via hash pointers, generating unique transaction identifiers, and creating account addresses. If a technology were to emerge, such as Shor's algorithm, capable of efficiently retrieving a private key from a public key (thus reversing the one-way function), it would jeopardize the security of blockchains and the internet as a whole. Likewise, if a technology like Grover's algorithm could efficiently invert hash functions and overcome the hash difficulty, the proof-of-work consensus used in many blockchains would become vulnerable. This would allow an attacker to easily rewrite the entire blockchain and substitute it with a counterfeit chain. Quantum computing possesses the potential to achieve such capabilities, hence posing a significant concern.

Fig. 5 illustrates potential attacks on a blockchain using Shor's algorithm and Grover's algorithms. While users engage in transactions and rely on the blockchain as a ledger to trace transaction history, a quantum computer-powered attacker can compromise all blockchain accounts using Shor's algorithm. As indicated by the orange text in Fig. 5 (bottom left), the attacker analyzes the public blockchain transactions and obtains the public key of the targeted user, such as User A's $p(35)$. Utilizing Shor's algorithm, the attacker then deduces the private key, exemplified by deriving A's $s(5,7)$ by factoring 35. Subsequently, the attacker employs the acquired private key to spend A's cryptocurrency.

On the other hand, as shown in Fig. 5 with blue text (bottom right), to modify a transaction, an attacker uses Grover's algorithm to search for a valid nonce that generates a hash meeting the required difficulty level. The attacker then reconstructs all subsequent blocks after the altered one by finding a valid nonce for each block. The attacker continues creating this chain of counterfeit blocks until it becomes the longest chain. Other miners will then adhere to the longest chain. It is important to note that current advancements in quantum computing are not yet powerful enough to attack blockchains effectively. However, the execution of the outlined attacks in Figure 5 is likely to become feasible in the near future.

At this point, we have discussed the potential threats Shor's and Grover's algorithms pose to blockchains. In the subsequent paragraphs, we briefly explain how these algorithms function. Specifically, we discuss how Shor's algorithm addresses the widely-used cryptographic primitive, i.e., factorization, and how Grover's algorithm can effectively search for a hash function that meets specific criteria.

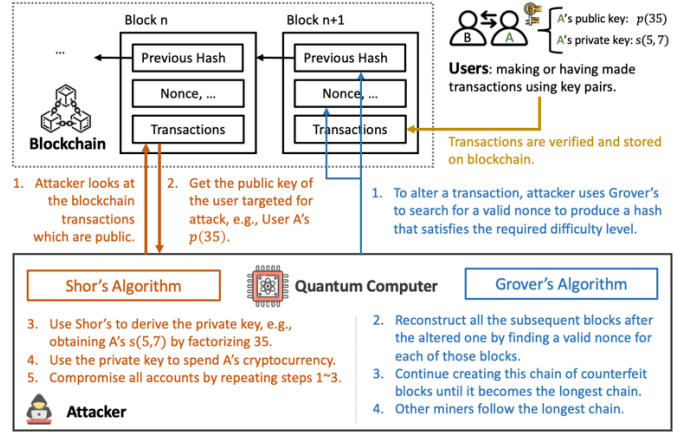

**Fig. 5.** Potential quantum attacks on a blockchain.

*C.1. Shor's algorithm*

Shor's algorithm can efficiently solve the factorization of integers and the discrete logarithm problem, which are fundamental components in designing digital signature schemes such as RSA, DSA, and EC-DSA. RSA relies on factoring integers, while DSA and EC-DSA employ discrete logarithm problems. In a blockchain, nodes sign their transactions using their private keys. Other nodes in the network have access to the corresponding public keys and can verify the authenticity of the transaction's originator. However, if a malicious node can generate the private key from the public key, it can impersonate the legitimate originator, posing a significant security risk to the blockchain network [26].

Here, we briefly describe Shor's algorithm. Assume we need to find factors of integer $N$. We use function $x^p \bmod N$, in which $x$ is a co-prime to $N$, i.e., $\gcd(x,N) = 1$. Also, $mod$ represents modulus in modular arithmetic. Therefore, if $x^p \bmod N = b$, then $x^p - b = Nk$ for an integer $k$. For example, $3^2 \bmod 91 = 9$ and $3^8 \bmod 91 = 9$. The period $r$ is defined as the smallest positive integer $r$ that satisfies $x^p \bmod N = x^{p+r} \bmod N$. Shor's algorithm uses the period $r$ to calculate factors of $N$ as described below.

1. Find a co-prime to $N$ on a classical computer;
2. Use two quantum registers: One for $p$ and another one for storing the result of $x^p \bmod N$;
3. Apply quantum Fourier transform and find period $r$ in parallel;
4. If $r$ is odd or $x^{\frac{r}{2}} \bmod N = -1$, then go back to step 1;
5. Calculate $\gcd(x^{\frac{r}{2}} + 1, N)$ and $\gcd(x^{\frac{r}{2}} - 1, N)$ on a classical computer. These are the factors of $N$.

For example, for $x = 3$ and $N = 91$, $r = 6$ and we have $\gcd\left(3^{\frac{6}{2}} + 1, 91\right) = 7$ and $\gcd\left(3^{\frac{6}{2}} - 1, 91\right) = 13$. We can see that 7 and 13 are both factors of 91.

Shor proved that his algorithm runs in polynomial time. It is shown that a quantum computer operating at 10 MHz can break RSA2048 in only 10 minutes. However, this number is 65+ million years for a classical computer operating at 1 THz. Among current popular blockchain implementations, Bitcoin,

Ethereum [44], Zcash, Litecoin, and Monero are all vulnerable to Shor's algorithm [45].

*C.2. Grover's algorithm*

In classical computing, the complexity of finding a solution (that is unique or absent) in any search space of cardinality $N$ is currently of order $O(N)$. For instance, consider an unsorted array of random integers of size $N$, with the task to check if the integer $x$ is in the array. That would take $O(N)$ steps. However, Grover's algorithm can find a solution in $O(\sqrt{N})$ on a quantum computer. Practical implementation of Grover's algorithm could potentially have significant security implications. For example, it can help find hash collisions and mining. In a blockchain, hashes are used for linking a block to its parent, block validation, and for generating user addresses. For instance, miners must find a hash value that satisfies specific conditions in the proof-of-work consensus algorithm. To date, this problem can only be solved through brute-force searching in the space of hashes. However, Grover's algorithm can speed up the process; thus, quantum miners can find the hash faster than others. In this way, they would always earn the reward, hence monopolizing the network. In addition, quantum nodes can rewrite the past block; they can find a hash collision and change the contents of a block. Then, they move forward and change all blocks, resulting in a blockchain compromise [25], [26].

Grover's algorithm is based on the diffusion operator, a quantum operator that acts on a multi-qubit state and spreads the probability amplitude across all possible measurement results. The amplitude of a quantum state indicates is the square root of the probability that the state will be measured as that particular outcome, e.g., $\alpha$ in $|\psi\rangle = \alpha|0\rangle + \beta|1\rangle$ is an amplitude and $|\alpha|^2$ is the probability being measured as 0. In Grover's algorithm, the diffusion operator is used to amplify the amplitude of (probability of finding) the desired measurement outcomes (i.e., the value we are searching for) among all outcomes. The steps of Grover's algorithm can be briefly described as follows [37]:

1. Initialize all qubits in a uniform superposition.
2. Apply an oracle gate to flip the amplitude of the target;
3. Apply Grover's diffusion operator;
4. Perform steps 2 and 3 at about $\sqrt{N}$ times, where $N$ is the size of the search space;
5. At this point, the amplitude of the value we are searching for is almost one, and we can measure it.

It is worth noting that while the implementation of Grover's algorithm is still being researched, its practicality for large-scale issues is yet to be determined. Researchers have been able to execute small-scale Grover's algorithm using methods such as trapped ions, photonics, and superconducting qubits. However, these small-scale demonstrations are still far from achieving the high number of qubits required to run Grover's algorithm for larger problems. In particular, Bitcoin, Ethereum, Zcash, Monero, and Litecoin are all known to be vulnerable to Shor's algorithm but not so much to Grover's algorithm [45]. Monero uses RandomX as its new proof-of-work consensus algorithm to combat quantum supremacy in mining, which makes it safe from Grover's algorithm [45].

*D. Solutions to Quantum Threats*

In recent years, researchers and practitioners have shown considerable interest in the potential threats posed by Shor's and Grover's algorithms to blockchains [13], [14], [15], [16], [46], [47]. As mentioned, two distinct approaches have gained momentum in addressing the long-term security and resilience of blockchain technology. The first approach centers around substituting the susceptible components of blockchains, such as cryptographic algorithms and hash functions, with robust classical post-quantum algorithms that can withstand quantum attacks [13], [15], [16]. The second approach revolves around leveraging the capabilities of quantum computing itself by integrating quantum computing algorithms, concepts, and tools into the architecture and deployment of blockchains. This approach explores the potential benefits that quantum computing can offer to enhance the efficiency and functionality of blockchains [14], [48], [49].

Post-quantum blockchains (i.e., the first approach) are fundamentally built upon post-quantum cryptography. Post-quantum cryptography replaces the vulnerable parts of blockchains, namely factorization-based public key schemes (for encryption and signature), with counterparts immune to attacks from quantum computers. It is also known as quantum-resistant or quantum-proof cryptography. They can be divided into the following categories: code-based cryptosystems, multivariate-based cryptosystems, lattice-based cryptosystems, and supersingular elliptic curve isogeny cryptosystems. However, these algorithms generally cannot provide small key and hash sizes and low computational complexity. Researchers are still investigating possible ways to tackle these issues. In Section III, we provide an overview of post-quantum cryptography and its related technologies. Detailed information of post-quantum cryptosystems can be found in [8], [13], [16].

Quantum blockchains (i.e., the second approach) utilize various quantum concepts and techniques, including QKD, quantum signature, quantum teleportation, and quantum bit commitment (QBC). They construct quantum systems that are fully or partially resistant to quantum algorithms like Shor's and Grover's. These blockchains are built on top of a quantum network infrastructure, which some researchers believe to be a crucial part of the future internet. In general, a quantum network can be built by using quantum teleportation, which allows the transmission of quantum information (in qubits) between two parties. However, the teleported qubit is destroyed on the sender's side after transmission.

While a global quantum internet is still a way off, recent breakthroughs in quantum networks have shown promising results [11]. The existence of a quantum network has been commonly assumed as a way of reasoning when introducing quantum applications. In Section IV, we briefly describe these concepts and provide a review of the studies that used them to design their proposed blockchain frameworks.

TABLE II
POST-QUANTUM CRYPTOGRAPHIC PRIMITIVES

| Type | Description |
|---|---|
| Hash-based | Hash-based schemes (HBS) utilize hash functions as the underlying one-way functions. Many HBS schemes are based on the Merkle signature scheme, which involves using hash trees (also known as Merkle trees) [50]. HBS schemes are notable for their simplicity but often require large key and signature sizes. |
| Lattice-based | The lattice-based scheme is one of the most popular post-quantum cryptographies known for its efficiency (relatively fast operation speed). It constructs signatures based on the difficulty of solving lattice problems [51], [52]. A lattice is a discrete mathematical structure composed of points and a set of mathematical operations that can be applied to those points. |
| Code-based | Code-based schemes are based on one-way functions that utilize error-correcting codes (ECC) [53]. ECC involves adding errors to a word or verifying a parity check matrix of the word. A parity check is often used to confirm the accuracy of data during communication through the use of a parity bit. It is based on the difficulty of decoding in a random linear code [54], [55]. |
| Multivariate | Multivariate schemes are asymmetric (e.g., public key) cryptosystems based on multivariate polynomials. They have short signatures, making them relatively fast. They rely on the hardness of solving multivariate (nonlinear) equations or polynomials (usually quadratic) over a finite field [56]. |
| Isogeny-based | Isogeny-based schemes are also known as supersingular elliptic curve isogeny schemes. An isogeny is a specific mathematical function that connects two algebraic objects, like elliptic curves, known as the domain and codomain. When it comes to an isogeny-based scheme, the security of the system is dependent on the difficulty of solving some mathematical issues that it is generally hard to find an isogeny of a specific form (depending on the protocol in question) between two elliptic curves [57], [58]. |

## III. POST-QUANTUM BLOCKCHAINS

In this section, we review the state-of-the-art approaches in post-quantum blockchains and discuss the advantages and disadvantages of applying different types of post-quantum cryptography algorithms to blockchains.

*A. Post-quantum Cryptography*

Post-quantum cryptography aims to create classical cryptographic systems that can withstand attacks from quantum computers. This involves exploring alternative solutions that do not rely on the current computational hardness of factorization, which is the foundation of many existing cryptographic systems. The field of quantum computing and its impact on cryptography are continually evolving, and the long-term resilience of current cryptographic techniques, including hash functions, against the immense computing power of fully-developed quantum computers remains uncertain. Moreover, post-quantum cryptography does not typically involve the redesign of hash functions, as they are relatively robust to quantum threats in their present state [4], [9]. Increasing the key size is a straightforward way to address Grover's search algorithm [11]. This subsection provides a brief overview of the various types of post-quantum cryptography, summarized in Table II.

Generally, post-quantum cryptography is considered more secure against quantum threats than factorization-based methods. This is due to the complexity of their mathematical models and the lack of efficient quantum algorithms to solve them. Lattice-based cryptography, for instance, is founded on the mathematical structure of lattices, which are grids formed by regularly spaced points in multiple dimensions. Its security is derived from the computational difficulty of specific lattice problems, such as the shortest vector problem (SVP) [51] and the learning with errors (LWE) problem [59], which become exponentially harder to solve with an increase in lattice dimensions. Similarly, code-based cryptography relies on the mathematical properties of error-correcting codes used to store and transmit information with the ability to recover data reliably in the presence of noise or errors. The security of code-based cryptography depends on the computational difficulty of some code-related problems, such as the decoding problem [53]. These problems remain challenging even for quantum computers. However, it is important to note that research in post-quantum cryptography is ongoing. As quantum computing technology advances, the security of these systems will continue to be explored and validated.

There have been numerous proposals for post-quantum cryptography using the cryptographic primitives mentioned in Table II. In particular, the National Institute of Standards and Technology (NIST) has been working to standardize post-quantum public-key cryptography [60]. It has recently selected the first group of winners following a six-year competition [61]. The four selected algorithms are Kyber [62] for general encryption and Dilithium [63], FALCON [64], and SPHINCS+ [65] for digital signatures. Kyber has relatively small encryption keys, making it more efficient and suitable for client-server communications, such as when users access websites. Dilithium and FALCON have high efficiency, with NIST recommending Dilithium as the primary algorithm and FALCON as a choice for applications that require smaller signatures (than Dilithium). Although SPHINCS+ has a larger key size and is slower than the other two for the same security level, it was selected mainly because it is based on a different cryptographic primitive (hash-based) than the other three selections (lattice-based) [61]. Table III reviews the example technologies in different types of post-quantum cryptographic primitives.

TABLE III
EXAMPLE TECHNIQUES OF POST-QUANTUM CRYPTOGRAPHY

| Type | Name | Description | Ref |
|---|---|---|---|
| Hash-based | XMSS | XMSS (extended Merkle signature scheme) is a one-time signature scheme based on a hash tree structure (similar to a Merkle tree). It is efficient and suitable for applications that require long-term security, such as the certification of documents. | [66] |
| | Lamport | Lamport signatures are based on a "hash chain," where messages are hashed multiple times to create a series of hash values used to generate the signature. It is known for its simplicity (easy to implement). | [67] |
| | SDS | SDS (smart digital signatures) is similar to Lamport but uses the chain's final hash value to generate the signature. | [68] |
| | W-OTS+ | W-OTS+ is a variant of the Winternitz-style one-time signature scheme (W-OTS), which promises to reduce the key sizes more than other W-OTSs. W-OTSs generate several function chains over random inputs. The random inputs are used as the secret key. The final outputs of the chains are used as the public key. | [69] |
| | SPHINCS+ | SPHINCS+ is an upgraded version of SPHINCS [70], a tree-based stateless signature scheme. A stateless scheme (c.f., stateful scheme [70]) does not require generating a new key for every message. It signs messages by creating a tree of height $h$ and using the $2^h$ leaves as private keys and the root as the corresponding public key. | [65] |
| Lattice-based | NTRU | NTRU (n-th degree truncated polynomial ring units) consists of an encryption scheme (based on the SVP in a lattice) and a signature process (based on the closest vector problem in a lattice). It is known for its speed and efficiency. | [71] |
| | Ring-LWE | Ring-LWE is based on the hardness of "learning with errors (LWE)" problems that are as hard as the worst-case lattice problems [72], [59], [73]. Ring-LWE was developed to address the inefficiency of LWE applications by considering them in a broad class of rings and identifying ideal lattices within these rings. It has also been widely used for homomorphic encryption [74], [75], and key exchange protocols [76], [77]. | [78] |
| | GLP | GLP (authors' surname initials: Güneysu-Lyubashevsky-Pöppelmann) scheme is based on the Ring-LWE problem. They are known for having the smallest public key and signature sizes among the post-quantum signatures. | [79] |
| | Kyber | Kyber is a key exchange scheme using the LWE lattice problem as its basic trapdoor function. | [62] |
| | Dilithium | The security of Dilithium is based on the hardness of lattice problems over module lattices. It uses the "Fiat-Shamir with Aborts" [80] technique for compactness and security. It is known for its small key sizes and signature sizes. | [63] |
| | FALCON | FALCON (fast Fourier lattice-based compact signatures over NTRU) is based on the NTRU lattices, with a trapdoor function called "fast Fourier sampling." Its security relies on the hardness of the short integer solution problem (SIS) [51] over NTRU lattices. It is known for its efficiency and scalability. | [64] |
| Code-based | McEliece | McEliece is a public key cryptosystem that uses randomization for encryption. It is based on the difficulty of decoding a general linear code. | [81] |
| | Niederreiter | Niederreiter is a version of the McEliece code scheme that is faster in terms of computational time, being able to run about ten times faster than the encryption process of McEliece. Despite this, both of these schemes provide the same level of security. | [82] |
| | UOWHF | UOWHF (universal one-way hash function) is a type that is particularly resistant to hash collision attacks. | [83] |
| Multi-variate | HFE | HFE (hidden fields equation) is a public key cryptosystem based on polynomials over finite fields of different sizes. Its basic version was experimentally proven to be broken in expected polynomial time [84], [85], [86]. Still, all its variations (e.g., the minus variant and the vinegar variant [87]) were also proven to be strengthened against all known attacks. | [88] |
| | UOV | The UOV (unbalanced oil and vinegar) scheme is a signature scheme based on a minimal quadratic equation system [89]. Despite its simplicity and resistance to quantum attacks, the large key sizes make it impractical for many applications. | [90] |
| Isogeny-based | SIDH | The SIDH (supersingular isogeny Diffie-Hellman) key exchange is based on the mathematics of supersingular elliptic curves. It creates a key exchange similar to the Diffie-Hellman key exchange [91], which can be utilized as a simple and quantum-resistant replacement for the Diffie-Hellman key exchange techniques. | [57] |
| | CSIDH | CSIDH is a non-interactive and commutative SIDH scheme that utilizes supersingular elliptic curves defined over a large prime field rather than regular elliptic curves. This makes it faster in practice. | [58] |

*B. Studies of Post-quantum Blockchains*

In this subsection, we provide an overview and classification of the current state-of-the-art post-quantum blockchains, categorizing them into three distinct groups based on their intended objectives. These categories include 1) signature approaches, addressing quantum threats by the use of post-quantum public key schemes for their signature processes; 2) privacy-enhancing approaches, addressing quantum threats to the security of private transactions via techniques such as zero-knowledge proofs (ZKPs); 3) decentralized applications: developing dApps for post-quantum blockchains by using smart contracts.

*B.1. Signature approaches*

Numerous proposals have emerged for post-quantum blockchains that employ post-quantum public key cryptography in their signature processes. This subsection presents a range of notable signature approaches commonly utilized in post-quantum blockchains.

*B.1.1. Industrial approaches*

The blockchain and cryptography sectors have started to evaluate the possible implications of quantum computing on their systems, recognizing the need to address potential quantum threats [92], [93], [94], [95]. For example, Hedera [96], a so-called "third generation blockchain" that claims to be a high-throughput, high-speed, and low-fee platform, has expressed interest in FALCON [64] due to its balance between security level and key size. They have been waiting to complete the post-quantum standardization to implement their post-quantum blockchains. Also, IOTA Tangle [97], a blockchain-based data exchange platform, claims to be resistant to quantum attacks such as the nonce search for the proof-of-work consensus [13], [98], [99]. It uses one-time hash-based signatures (i.e., Winternitz signatures [100]) rather than code-based (i.e., ECC) and is expected to make use of ternary hardware and a new hash function called CURL-P [101]. Even though IOTA's approach has been controversial, it has sparked interesting discussions on post-quantum blockchains.

Abelian is another post-quantum blockchain platform that attempts to use lattice-based cryptography, which consists of multiple cryptographic building blocks [102]. These cryptographic primitives are based on the SIS problem [51] and the LWE problem [59]. They offer strong security against quantum attacks but have relatively large key sizes. The authors consider the trade-off between security level and key size and find that the variants of SIS and LWE, called Module-SIS and Module-LWE [103], provide both a high level of security and relatively small key sizes. They also propose an alternative key exchange protocol between the sender and receiver to generate a one-time key for signature based on Kyber. The authors also envision using homomorphic lattice-based commitment schemes for privacy with linkable signatures and verifiable encryption.

In addition, some cryptocurrencies have been developed to replace quantum-vulnerable cryptocurrencies like Bitcoin. For example, a quantum-resistant ledger (QRL) [46] uses a hash-based signature scheme, specifically XMSS, to design a cryptocurrency platform where users can mine and spend their assets. It proposes using SHA-256 with a 16-bit word size to build up the Merkle tree in XMSS, which is believed to offer 196-bit security and remain resistant to quantum computers until 2164 [104]. QRL currently has over 700 active nodes worldwide. It claims to be compatible with almost any cryptographic solution through variable key heights and signature space, allowing updates to the blockchain address format to support different hash functions or signature schemes.

Moreover, Bitcoin post-quantum (BPQ) [47] employs a post-quantum signature scheme using a trial version of Bitcoin's main blockchain. BPQ proposes to use a variant of the W-OTS+ [69], [100]. It generates a new one-time key (a set of 256-bit randomly-generated numbers by parametric W-OTS+) for each message because it is insecure to use the same private key to sign multiple messages. It also uses XMSS to combine a set of W-OTS+ public keys into one public key, making multiple private keys correspond to one public key, a common scheme used in hash-based Merkle trees. It randomly generates a seed value to build the extended Merkle tree in XMSS of height $h$, from which $2^h$ private keys and their corresponding public keys ($2^h$ leaves) are generated. With such a public key scheme, it hard-forks from the Bitcoin network. It keeps Bitcoin's historical transactions and consensus rule (using ECDSA signatures) before the block height of 555,000. The new signature scheme based on XMSS was applied for later blocks. To switch their Bitcoin accounts to BPQ, the users of Bitcoin need to create new keys for XMSS and make a transaction from their old addresses to their new ones. The authors stated that about one year after the launch of the main BPQ network, support for the ECDSA signatures would be completely disabled. Any coins that have not been transferred to the new addresses would be burned, which prevents the lost ECDSA keys from compromising BPQ.

Applying post-quantum cryptography in blockchains also raises the possibility of using post-quantum blockchains in financial technology [105]. For example, Corda [106], a blockchain-based financial platform for regulated markets, published a new post-quantum signature algorithm based on

SPHINCS for blockchains [107], [70]. Corda offers users a variety of cryptographic options, i.e., different signature key types, including RSA, ECDSA, EdDSA (Edwards-curve digital signature algorithm) [108], and SPHINCS. EdDSA is the default scheme, while SPHINCS serves as a post-quantum option. It is designed to secure long-term contracts, such as mortgage and pension contracts. It is particularly well-suited for situations where assets must remain locked or inactive for many years and remain secure even when powerful quantum computers become widely available. However, SPHINCS is not practical for many applications due to its relatively slow signing process and large signature size (41 KB). This is why it is offered as an option rather than the default signing scheme in Corda. It is useful in cases where signature keys are used once or only a few times, as is common with one-time addresses in public blockchains, and when there is an immutable data structure that can serve as a global cache for keys. An example of this is in permissioned blockchains [105], [109], where blockchain nodes repeatedly sign with the same key.

*B.1.2. Academic approaches*

Thus far, this section has examined integrating post-quantum techniques into commercial blockchains. Now, let us explore the academic literature on post-quantum blockchain methods. Academic research predominantly focuses on lattice-based and hash-based signature schemes, which offer robust mathematical foundations and well-established formalizations.

An example of a proposed solution in literature for post-quantum blockchains is a lattice-based signature scheme with a compact key size [110]. This scheme was implemented using the bonsai tree algorithm [111]. The approach emphasizes generating a new address for each transaction, employing a distinct public key as a countermeasure against statistical attacks. However, this requires blockchain users to store many public and private key pairs, which can make their wallets bulky. The authors proposed a lightweight wallet design for this problem, as it only requires users to store a single root key (root of the bonsai tree) for all transactions. The root key is the root lattice basis used to generate multiple pairs of sub-public and sub-private key pairs which are used to sign and verify transactions. The wallet only needs to keep the root key and specific parameters of the bonsai tree algorithm, reducing the storage space required [110].

Moreover, a lattice-based double-signature scheme that promises to satisfy the correctness and unforgeability under the SIS assumption with a shorter signature and key size was proposed [112]. The double-signature scheme was then utilized to construct a cryptocurrency system. The system operates similarly to other blockchain approaches: only its signature scheme is replaced by the double-signature scheme. It involves typical blockchain processes such as generating a public and private key pair, using the public key to create an account address, signing transactions, and having miners check for double-spending. The system has smaller key and signature sizes than other lattice-based signature algorithms because it uses basic linear operations such as modular multiplication and addition [112].

In particular, it is expected that signatures based on LWE and Ring-LWE can effectively counter quantum computers executing Shor's algorithm [113]. Thus, its variant TESLA# [114] was proposed to replace Bitcoin's SHA-256 and the Koblitz curve secp256k1 [1], [113] in the ECDSA signature. TESLA#, which uses BLAKE2 [115] and SHA-3 [116], is secure through a provable security reduction to Ring-LWE [114].

Having discussed the lattice-based approaches in the previous paragraphs, let us now focus on exploring hash-based signature schemes for blockchains. These hash-based methods primarily focus on improving efficiency by reducing the sizes of both keys and signatures. For example, SDS was proposed to decrease the time required for creating a hash tree and the sizes of keys and signatures [68]. SDS is a condensed hash-based signature method designed to address the typical problems of large key and signature sizes in hash-based signature schemes. The authors achieve this by building a key compression tree instead of directly using XMSS. The authors also proposed a model incorporating SDS within a distributed ledger system.

Furthermore, introducing a signature scheme named BPQS (blockchained post-quantum signatures) [107] deserves attention. BPQS is built upon the XMSS framework and has demonstrated superior efficiency compared to other existing hash-based algorithms, particularly when a key is utilized repeatedly for a moderate number of signatures. The scheme also includes a fallback mechanism, which makes it possible to use a practically unlimited number of signatures if needed. The main objectives of this scheme are to decrease the costs of key generation, signing, and verification and the size of signatures.

To further shorten the signature size, a new hash-based signature called WOTS-S was proposed [117], a variant of the W-OTS scheme. The signature sizes of WOTS-S are 59% and 24% shorter than the IOTA and QRL schemes, respectively. Thus, WOTS-S is more computationally efficient than other variants of WOTS that use randomizations and bitmasks to reduce signature sizes, which can be costly. The authors claimed that this is because WOTS-S uses collision-resistant hash functions and does not require using bitmasks. They plan to integrate WOTS-S into blockchains beyond just cryptocurrencies.

While lattice-based and hash-based signatures are prevalent in post-quantum blockchain literature [110], [112], [113], [118], alternative post-quantum techniques have also been proposed. One such example is the utilization of the Rainbow algorithm [119], which is based on the multivariate public key scheme (specifically, the UOV scheme). The Rainbow algorithm is known for its efficient signature generation. In a specific case, it was applied to the Ethereum platform, and its feasibility was evaluated by creating a private blockchain. The authors further compared the signature efficiency of the Rainbow algorithm and ECDSA.

*B.2. Privacy-enhancing approaches*

Including private transactions in blockchains using ZKPs has been a subject of extensive discussion [120], [121]. ZKPs allow one party (prover) to demonstrate to another party (verifier) that they know something without revealing the information. It can be used to guarantee the integrity of confidential data for efficient computations, replacing the need for human auditors and eliminating the potential for corruption while reducing costs. It is worth noting that certain types of ZKPs may also be vulnerable to attacks from quantum computers.

In general, there are two types of techniques for ZKPs: interactive and non-interactive. Interactive ZKPs require live communication between the verifier and the prover. The following is an example of an interactive ZKP process using the Schnorr protocol [122]:

1. A prover wants to prove the knowledge of the discrete logarithm $x$ of some group $h = g^x \bmod p$ where $g$ and $p$ are public. $h$ is known to both parties.
2. The prover picks a random value $1 \leq k \leq p$ and sends $u = g^k \bmod p$ to the verifier.
3. The verifier picks another value, $1 \leq c \leq p$, and sends it to the prover.
4. The prover sends $s = (xc + k) \bmod p$ to the verifier.
5. The verifier checks if $g^s \equiv h^c u \bmod p$. If yes, it shows that the prover knows $x$.

However, interactive ZKPs require the participation of a verifier who is online and willing to engage in the process. Thus, non-interactive ZKPs were developed to implement zero-knowledge proofing without needing interaction between the prover and verifier (using a hash function) [123]. Take the above example, the non-interactive version is to pick $c = H(g^k)$ by using a hash function $H$ (instead of getting $c$ from the verifier). There are two popular schemes of non-interactive ZKPs: zero-knowledge scalable transparent arguments of knowledges (zk-STARKs) [124] and zero-knowledge succinct non-interactive argument of knowledge (zk-SNARKs) [125]. Both schemes can prove the validity of computations without revealing the underlying data. However, they have different properties and use cases. There is a debate between STARKs and SNARKs, with one of the main points of contention being their efficiency and post-quantum properties. STARKs, based on hash-based systems, are resistant to quantum attacks, while SNARKs, based on elliptic curves, are not. Moreover, zk-STARKs typically produce larger proofs than zk-SNARKs, which can result in higher verification overhead. However, in certain situations, such as when proving large datasets, zk-STARKs may be more cost-effective than zk-SNARKs as they do not require a trusted setup and instead rely on public randomness. They have a simple verification process, providing transparency and scalability. Ethereum developer communities support SNARKs and STARKs, but the Ethereum Foundation [44] has shown particularly strong support for STARKs by providing a $12 million grant to StarkWare for this emerging post-quantum technology [126]. StarkWare's zk-STARKs use an off-chain prover not exposed on the blockchain to protect users' privacy. Off-chain provers enable private transactions on blockchains by performing transactions off-chain and verifying them on-chain (on-chain verifiers).

The Zerocash protocol [127] uses zk-SNARKs to maintain the integrity of a decentralized registry that hides the commitments of unspent funds, which provides privacy and confidentiality. It is possible that zk-STARKs could be used to achieve the same level of efficiency and confidentiality as zk-SNARKs while preserving the post-quantum property [124]. In addition, zero-knowledge systems have the potential to solve blockchain scalability issues by reducing the verification time of a blockchain process [128], [129]. Specifically, a prover can generate proof of the validity of the newly added transactions in quasi-linear time. This proof can be stored on the blockchain instead of requiring all nodes to blindly keep copies of all transactions. Many post-quantum ZKP applications for blockchains are currently being researched and investigated [130]. For example, there have been discussions about the potential for using ZKPs to create blockchain-based identity management systems. Researchers seek ways to minimize the exposure of sensitive identification data [131], [132].

Moreover, a ZKP system for post-quantum blockchains was proposed [133] [134], using the ring-confidential transaction (RingCT) protocols [135] based on the lattice problems. Instead of disclosing the exchanged amount of cryptocurrency, RingCT includes a mathematical proof that the transaction is balanced, meaning that the recipient did not receive more money than the sender spent. Only the transaction recipient can reveal the exchanged value and spend the transaction that has been recorded on the blockchain. For instance, if a user wants to spend money from $M$ of their accounts while hiding among $N$ other users, they will gather $M \times N$ accounts (including their own) to implement the transaction and then prove that the total value of the input accounts is equal to the total value of the output accounts. All these applications require a robust post-quantum ZKP scheme. In addition, Abelian, mentioned in the last subsection, also proposed to use lattice-based ZKPs as their privacy-preserving building blocks [102], [136].

*B.3. Post-quantum decentralized applications*

Determining the optimal post-quantum cryptography for blockchains is an ongoing endeavor. However, the consensus among experts is that replacing the existing public key schemes of blockchains with post-quantum alternatives would render them resilient against quantum computer attacks. Guided by this assumption, researchers have been actively exploring the potential applications of post-quantum blockchains. This subsection delves into various approaches for post-quantum decentralized applications, highlighting the advancements made in this field.

Like classical blockchain-based applications, post-quantum decentralized applications are renowned for their immutability, transparency, and decentralization properties. To illustrate, a blockchain-based online voting protocol [137] was introduced, utilizing Niederreiter's code-based system [82] to safeguard against quantum attacks. Electronic voting (e-voting) has gained popularity as an efficient, transparent, and accessible method for conducting elections. The proposed protocol leverages blockchain technology's inherent features to enhance the voting process's security and integrity. The protocol aims to provide a convenient and low-cost democratic decision-making system that is transparent and tamper-proof. The approach can audit the correctness of the voting operations and resist quantum attacks using Niederreiter's cryptosystem. The authors evaluated and found their approach suitable for small-scale elections. There are four roles in this voting system: a regulator as the key generation center, a voting initiator, voters and candidates that the voting initiator sets. The candidates are also nodes in the blockchain. The regulator would generate necessary public-private key pairs for all potential voters to start a vote. Then, the initiator announces the voting content and candidate list and collects and publishes the voter list by publishing their public key addresses. Voters can then register and vote by sending transactions on the blockchain. After that, the verifiers interested in the election results would view and count the voting content in the transactions. The regulator can verify the voter signatures if anyone wants to audit the voting. Niederreiter's scheme works at the key generation and verification processes. An additional example involves a certificate-less ring signature scheme based on the NTRU lattice, designed for e-voting systems with post-quantum resistance [138]. In e-voting systems, it is essential to guarantee voters' security, privacy, and anonymity to preserve the voting process's integrity. This scheme removes the certificate requirement, simplifying the process and decreasing the management burden typically linked to traditional public key infrastructure (PKI) systems.

Furthermore, a proposal was made to establish a framework for sharing sensitive industrial data in public distributed networks by combining the Inter-Planetary File System (IPFS) with Ethereum. This framework aims to create a post-quantum decentralized file system, facilitating secure and efficient sharing of confidential industrial data within the network [139]. They implemented a key exchange protocol with SIDH. With their framework, users can encrypt their data and choose with whom they want to share it. They implemented the framework with Diffie-Hellman key exchange, Elliptical Curve Diffie-Hellman key exchange, and SIDH. They found that SIDH is the most appropriate implementation even though it requires an off-chain (not on the blockchain) computation for encryption. They cannot do encryption on a smart contract because the source codes of smart contracts are transparent and thus could leak the private variables used to encrypt data. They showed the feasibility of SIDH in a blockchain-based decentralized application to provide a quantum-resistant property.

Efforts have also been made to develop a post-quantum blockchain infrastructure encompassing authentication and access control mechanisms. One notable approach is proposing a quantum-resistant transaction authentication scheme [140]. This scheme incorporates a lattice-based bonsai tree signature and adheres to a standard transaction model, effectively mitigating the risks of quantum attacks. They proposed constructing a lightweight cryptocurrency wallet by a transaction authentication model that extends the lattice space to multiple lattice spaces accompanied by the corresponding key in bonsai trees. This approach shares a technique similar to [110] but specializes in signature authentication for blockchain wallets. On the other hand, a lattice-based (specifically ring-LWE) GLP signature [79] was modified and used for a blockchain-based PKI called QChain [141]. PKI is an infrastructure for managing user identities using public key schemes and a certificate authority. PKIs heavily depend on the security of the underlying public key schemes and thus need post-quantum alternatives badly. The authors compared their approach to a centralized PKI system and found that they perform similarly to the centralized one while providing decentralized post-quantum benefits.

The convergence of blockchain and the internet of things (IoT) has been a subject of study, including its application in the context of post-quantum IoT [142], [143]. Post-quantum blockchains have been implemented in the Social Internet of Things (SIoTs), which amalgamate social networks with IoT functionalities [144]. The IoT domain has been extensively explored, encompassing a diverse array of technologies, as evidenced by numerous research studies [145], [146]. Blockchain is a promising technology that could revolutionize IoT by ensuring data integrity and privacy. In [144], a multivariate scheme called ring signature was proposed and used to design a blockchain-based SIoT platform. A main feature of the ring signature is to provide private transactions by signing a transaction with the user's private key and mixing the signature up with a group of other users' public keys. In this way, the public can only verify that someone in the group has signed the transaction but does not know exactly who. The public keys of the users in the group are connected as a ring and are used as one ring public key for the sender. The sender's private key is then used to sign the transaction. When someone (e.g., a miner) wants to verify the transaction, they verify it with the ring public key. Thus, the sender's identity is fused into this group of people.

IoT nowadays encompasses billions of interconnected devices that collect and share data. The diverse nature of these devices and their inherent resource constraints present unique challenges in ensuring security and privacy. To address these challenges while pursuing post-quantum properties, a modified key exchange protocol has been proposed [147]. This protocol enables the reuse of keys, reducing the overhead and resource consumption typically associated with traditional key exchange methods. By reusing keys, the proposed protocol minimizes computational and communication costs while maintaining a high level of security for IoT devices. The authors demonstrate the resilience of the protocol against various attacks, such as replay, man-in-the-middle, and impersonation attacks.

Additionally, blockchain technology is well-suited for securely connecting and managing smart devices within smart city infrastructure, offering convenient and secure features like autonomous authentication and decentralized operations. It is no surprise that post-quantum properties of blockchain-based smart cities are also being researched [148], [149]. A post-quantum proof-of-work consensus algorithm was proposed, and a lightweight transaction proposal was developed upon it [148]. Their consensus replaces the hashing problem based on SHA-256 in Bitcoin proof-of-work with a mining problem based on the hardness of solving multivariate quadratic equations (c.f., multivariate in Table III). Miners execute the Gröbner basis solution algorithm [150] to try to solve the random quadratic multivariate equations generated based on a seed number and compete for the right to generate a block. The authors proposed to embed an identity-based signature scheme and the IPFS to construct lightweight transactions and showed their efficient performance.

There have been more approaches focusing on different aspects of the future form of blockchains in the quantum era. For example, a parametric hash function for blockchains based on the problem of solving a set of polynomial equations in integers was proposed [151]. The authors claim that collision with their hash function is impossible. In [152], the authors analyzed the integration of IoT with cloud and blockchains in the quantum era. In [153], the authors describe the implementation of a post-quantum dApp that uses Ethereum smart contracts, the IPFS, and lattice-based key exchange to provide end-to-end encryption for digital data sharing or storage, such as for maintaining digital certificates and cryptographic keys. Instead, they use classical technologies for distributed systems, such as multi-party computation (MPC), to accomplish such post-quantum properties. In addition, the authors of [154] explore information security in the post-quantum era for 5G networks, including the potential impact of quantum machine learning on classical networks. With the emergence of blockchain and machine learning, it is crucial to examine how these technologies may be influenced by quantum technologies and vice versa [155], [156], [157]. Moreover, in [158] and [159], the authors use the concurrent preprocessing model in multi-party computation to achieve an asynchronous quantum-resistant blockchain consensus instead of using post-quantum cryptography.

*B.4. Summary of post-quantum blockchains*

Lattice-based post-quantum cryptography is important in redesigning blockchains for the upcoming quantum computing era. This is because it is efficient and relatively simple to implement. In addition, hash-based signatures have also been adopted due to their simplicity (e.g., constructing hash trees) and higher security level. However, there are also investigations (even though not many) on code-based, multi-variate, and Isogeny-based cryptography. This is because the trade-offs between security level and key size are subject to change for different applications. Also, approaches using post-quantum cryptography on privacy enhancement have been proposed and brought about a considerable amount of research work on post-quantum ZKPs. Even though post-quantum cryptography is not yet widely applied, researchers have begun to consider how and what decentralized applications need a quantum-resistant property and propose post-quantum blockchain schemes for IoT and other application scenarios. Table IV summarizes the studies introduced in this section.

TABLE IV
SUMMARY OF STUDIES IN POST-QUANTUM BLOCKCHAINS

| Types | Ref | Features | Description | Implementation |
|---|---|---|---|---|
| Lattice-based signature approaches (Industry) | [96] | • Efficient<br>• Low-fee<br>• Compatible with other signature schemes | Proposed by a blockchain company called Hedera proposes to use the lattice-based FALCON signature to balance between security level and key size. It is not yet finalized and may follow the post-quantum cryptography standardization for their post-quantum blockchains. | Hedera shows significant interest but has not yet implemented any of these signature algorithms despite the completion of the standard. |
| | [102] | • Strong security<br>• Relatively large key sizes in general but smaller than similar approaches<br>• With a key exchange protocol | A blockchain company called Abelian uses a lattice-based signature based on the SIS problem and the LWE problem to develop a blockchain. It proposes an alternative key exchange protocol to generate a one-time key for the signature process based on Kyber. It also proposes a post-quantum ZKP scheme for homomorphic encryption. | Abelian has launched its post-quantum blockchain, which currently has an average block mining time of 266 seconds and a total transaction count of about 150,000. |
| Hash-based signature approaches (Industry) | [98] | • To replace both the signature scheme and hash function<br>• Legacy post-quantum proposal | Proposed by a blockchain company called IOTA, it optimizes the proof-of-work consensus for quantum-resistant properties by using the W-OTS signature and CURL-P hash function. Addressing post-quantum concerns became a long-term goal for IOTA. | IOTA has replaced the W-OTS signature in its protocol with the more standard ed25519, resulting in a lighter-weight protocol [160]. |
| | [46] | • Compatible with almost any cryptographic | Proposed by the QRL Foundation, it uses an XMSS (hash-based) signature scheme. It also uses SHA-256 with a 16-bit word size to build up the Merkle tree in XMSS. QRL currently has over 700 active nodes | QRL's blockchain features an average block mining time of 58 seconds, with about 2,500,000 blocks in total. Each |
| | | solution<br>• Flexible on address format | worldwide and is believed to be resistant to quantum attacks until 2164. | block typically contains one or two transactions. |
| | [47] | • Hard-forked from Bitcoin<br>• Keep old transactions and add new post-quantum transactions | The authors employ an XMSS (hash-based) signature based on a variant of the W-OTS+ to construct a blockchain scheme called Bitcoin Post-Quantum. It keeps Bitcoin's historical transactions and consensus rule (using ECDSA signatures) before the block height of 555,000. | Bitcoin Post-Quantum ceased operation at the time this paper was being revised. |
| | [106] | • Offers a variety of cryptographic options for users<br>• Good for long-term smart contracts<br>• Relatively slow signing process | A FinTech company called Corda proposed a new post-quantum signature algorithm based on SPHINCS. It is suitable for long-term contracts, such as mortgage and pension contracts. SPHINCS is not practical due to its low efficiency but is optional for users. It is suitable in permissioned blockchains, where blockchain nodes repeatedly sign with the same key. | There has been no ongoing update since Corda introduced the post-quantum signature algorithm designed explicitly for blockchains in 2018 [161]. |
| Lattice-based signature approaches (Academia) | [110] | • Small key size<br>• Lightweight wallet<br>• To counter statistical attacks: new address for each transaction | The authors propose a lattice-based signature scheme with the bonsai tree algorithm with a small key size. They propose a lightweight wallet design that stores a single root key (the root of the bonsai tree) for all transactions. The root key generates multiple pairs of sub-public and sub-private keys. | The research paper focuses on the proposed scheme's theoretical aspects and security analysis rather than providing implementations or experiments. |
| | [112] | • Correct and unforgeable to the SIS assumption<br>• Small signature and key size | The authors propose a lattice-based double-signature scheme with a relatively short signature and key size. It uses basic linear operations such as modular multiplication and addition to reduce the key size. | The paper emphasizes the conceptual analysis of the proposed system rather than offering implementations or experimental results. |
| | [113] | • To replace both the signature scheme and hash function<br>• Relatively efficient (compared to LWE) | The authors propose a lattice-based signature scheme based on LWE and Ring-LWE. They propose to replace Bitcoin's SHA-256 in ECDSA with TESLA#, which is based on BLAKE2 and SHA-3. | The paper presents a security analysis of Bitcoin in the context of post-quantum cryptography, with no implementations. |
| Hash-based signature approaches (Academia) | [68] | • Efficient<br>• Relatively small key sizes (compared to hash-based approaches) | The authors use hash-based SDS to reduce the hash tree-creating time and the sizes of keys and signatures. The relatively small key size is achieved by building a key compression tree instead of directly using XMSS. | The authors developed a testbed on a Windows device using Python, and their results indicate that SDS outperforms two instances of XMSS in terms of efficiency. |
| | [107] | • Efficient<br>• Support an unlimited number of signatures | The authors present a hash-based signature scheme for blockchains called BPQS based on XMSS. It aims to decrease the time of key generation, signing, and verification. BPQS outshines existing hash-based algorithms when a key is reused for a reasonable number of signatures. An open-source implementation of the scheme is provided, along with benchmarking results. | The authors implemented a prototype of BPQS, and the results demonstrate that it compares favorably to both classical and post-quantum signatures. BPQS outperforms XMSS regarding the signature size. |
| Multivariate signature approaches (Academia) | [119] | • Efficient<br>• Suitable for a private blockchain | The authors use the Rainbow algorithm based on the multivariate scheme (specifically, the UOV scheme), known for its high signature efficiency. They show that replacing the ECDSA with the Rainbow signature algorithm does not interfere with the normal functioning of its original features, such as creating a new account or sending transactions. | The authors carried out experiments using Ethereum as a basis. Results show replacing the ECDSA with the Rainbow signature algorithm does not affect blockchain processes and performance. |
| Privacy-enhancing | [127] | • Support private transactions | The authors propose a ZKP protocol called the Zerocash [127], based on zk-SNARKs. It keeps the transaction | The authors built Zerocash, in which transactions are less |

| approaches | | • Provide confidentiality<br>• Efficient in the verification process | integrity while hiding the commitments of unspent funds, which provides privacy and confidentiality. | than one kB and require less than 6 ms for verification, making them comparable to plain Bitcoin in terms of efficiency. |
|---|---|---|---|---|
| Decentralized applications | [137] | • Post-quantum decentralized application | The authors use Niederreiter's code-based cryptosystem for a blockchain-based online voting protocol. Niederreiter's scheme is for key generation and verification. | The experiment shows that their scheme is suitable for small-scale elections and provides benefits when the number of voters is limited. |
| | [139] | • File system<br>• A key exchange protocol | The authors propose a framework for sharing sensitive industrial data based on the Inter-Planetary File System (IPFS) and Ethereum and implementing a key exchange protocol with SIDH (Isogeny-based). | The paper focuses on the theoretical aspects of the proposed scheme rather than implementations. |
| | [140] | • A lightweight cryptocurrency wallet | The authors present a transaction authentication scheme based on a lattice-based bonsai tree signature and a standard transaction authentication model used to design a lightweight wallet. | The paper's primary focus is on the theoretical aspects instead of implementations. |
| | [141] | • Post-quantum PKI | The authors propose a post-quantum blockchain scheme called QChain. They propose to use a variant of the lattice-based (specifically ring-LWE) GLP signature. It is to build a PKI with quantum-resistant properties. | The research paper focuses on the theoretical aspects instead of implementations. |
| | [144] | • SIoT<br>• Support private transactions | The authors aim to create a post-quantum SIoT platform based on the proposed ring signature using post-quantum blockchains. It mixes up the user's private key and a group of other users' public keys to provide privacy. | The research paper focuses on the theoretical aspects rather than experiments. |
| | [148] | • Smart cities<br>• Post-quantum consensus | The authors propose to use post-quantum blockchains for smart cities, which provide connection and authentication for smart devices. It includes a post-quantum proof-of-work consensus algorithm and a lightweight transaction proposal. | The authors performed simulations to demonstrate the stability of their proof-of-work algorithms. |

## IV. QUANTUM BLOCKCHAINS

This section describes the quantum computing concepts and tools to secure blockchain frameworks against quantum computing threats. We first briefly review the related concepts and then study the related literature.

### A. Quantum Cryptography and Protocols

The role of quantum cryptography in quantum blockchains is to strengthen security and privacy within blockchain systems by leveraging the unique properties of quantum mechanics. Key quantum cryptographic techniques employed in quantum blockchains include QKD, quantum signatures, quantum teleportation, and quantum bit commitment. In particular, QKD is the fundamental layer for key exchange and authentication in quantum blockchains. It establishes a secure channel for transmitting data within the blockchain network, ensuring confidentiality and integrity. Quantum signatures, on the other hand, rely on quantum mechanics for signing and verifying messages. They offer advanced security measures against forgery and tampering, surpassing classical digital signatures.

Quantum teleportation, although still in its early stages, holds promise for facilitating future communication methods in the quantum internet, which in turn supports the development of fully quantum blockchains. These blockchains would operate entirely based on quantum processes. However, both fully quantum blockchains and the quantum internet are still in the nascent stages of exploration. Additionally, quantum bit commitment enables nodes in a quantum blockchain to ensure the immutability of committed transactions. Once a transaction is finalized and committed, it becomes tamper-proof, preserving the overall integrity of the blockchain.

This subsection provides a review of quantum algorithms and protocols that are used in the design and development of quantum blockchains. There are multiple implementations for each concept. Here, we describe the most-cited and well-known implementations.

### A.1. Quantum key distribution

A QKD protocol uses the properties of quantum mechanics to share a secret key between a sender and a receiver. It is based on quantum mechanics principles, specifically quantum states' properties such as measurement and entanglement.

TABLE V
PHOTON POLARIZATION EXAMPLE IN BB84

| Basis | 0 | 1 |
|---|---|---|
| Rectilinear (R) | ↕ | ↔ |
| Diagonal (D) | ↗ | ↘ |

TABLE VI
AN EXAMPLE OF THE BB84 PROTOCOL

| Alice random bits | 0 | 1 | 0 | 1 | 1 | 1 | 0 | 0 | 1 |
|---|---|---|---|---|---|---|---|---|---|
| Alice basis | R | D | D | R | D | R | D | R | R |
| Alice photons | ↕ | ↘ | ↗ | ↔ | ↘ | ↔ | ↗ | ↕ | ↔ |
| Bob random basis | R | D | R | D | R | D | R | D | R |
| Bits as received by Bob | 0 | 1 |  |  |  |  |  |  | 1 |
| Alice sends its basis to Bob over the classical channel. ||||||||||
| Shared Secret | 0 | 1 |  |  |  |  |  |  | 1 |

QKD protocols can be grouped into two main categories: discrete and continuous. Discrete-variable protocols, such as BB84 [162], E91 [163], B92 [164], SARG04 [165], and DPS [166], use discrete states like photon polarization or spin to transmit keys. Meanwhile, continuous-variable protocols use continuous variables like light beam amplitude or phase for key transmission. One example of a continuous-variable QKD protocol is the Gaussian-modulated coherent-state protocol. In this protocol, information is encoded using continuous variables by applying Gaussian modulation to coherent states of light. Coherent states represent quantum states of the electromagnetic field that are highly similar to classical light waves. Gaussian modulation introduces random Gaussian fluctuations to the amplitude and phase of these coherent states. For a more in-depth understanding of such protocols, readers can consult [167] and [168].

Here, we focus on one of the most widely recognized QKD protocols, the BB84. BB84 uses photon polarization to transfer data. Assume Alice wants to share a secret key with Bob. They are connected by a quantum channel that allows the transmission of quantum states and a classical channel. The classical channel is assumed to be prone to passive eavesdropping but not to the injection or alternation of messages (that is, the adversary model considers a passive adversary only). If the classical channel is subject to active eavesdropping, they can still use this method if they share a small secret key beforehand, which is used to generate authentication tags. BB84 uses the no-cloning theorem that states measuring a quantum state disturbs it, and as a consequence, information will be lost. The state of a single polarized photon can be described by a combination of two-unit vectors $(1,0)$ and $(0,1)$, which is denoted as the rectilinear basis. Moreover, it also can be described as a combination of $(0.777, 0.777)$ and $(0.777, -0.777)$ vectors representing a 45-degree and 135-degree photon, respectively. This second basis is named the diagonal basis. An instance of such a protocol is reported in Table V.

The BB84 protocol works as follows (where Alice wishes to establish a secure key with Bob):
1. Alice generates a random string. An example of this is shown in Table VI.
2. Alice randomly selects a basis for each bit and sends a photon according to the polarization of that base over the quantum channel. The second row of Table VI represents the photons sent by Alice.
3. For each of the received photons, Bob uses a random basis and measures the photon. If the basis is the same one that was selected by Alice, Bob will measure the same value as Alice. Otherwise, the measurement will return a random answer. After the measurement, the photon is polarized, and its initial state will be lost.
4. Alice and Bob communicate the correct basis and measured basis over the public channel.
5. They discard the photons measured in the incorrect bases and use the remaining as the shared secret.

A key advantage of QKD is its ability to detect unauthorized attempts to access the key. If someone tries to intercept the key, the system can recognize this and alert the authorized parties. Such detection is possible since the parties should get the same reading if they use the same basis. If this is not the case and the number of bit mismatches exceeds a given threshold (that should also account for plain errors due to the transmission media), the communication is considered compromised, the run is aborted, and started again. This makes it highly challenging for anyone to compromise a key while escaping detection. The security of QKD is based on the principles of quantum physics rather than computational assumptions, making it highly secure. However, it requires trust among nodes and an authenticated classical channel. The latter can be achieved using a secret key between parties before the first QKD session [25], [169]. It is important to note that QKD is considered to be the most practical application of quantum technology, with numerous implementations worldwide using either fiber or free-space communication channels, such as satellite-based QKD [170], [171]. Free-space QKD is gaining momentum due to the availability of high transmission windows in the atmosphere, such as the 650 nm to 670 nm wavelength window with a small diffraction spread, which facilitates the detection of photons [170]. However, both quantum and classical wireless communications are susceptible to environmental factors [172], [173]. Regardless of the communication channel used, QKD is gaining attention as the first practical application of quantum technology. Its immediate applicability is significant in the development of quantum blockchains.

*A.2. Quantum signature*
Digital signatures serve as a means to ensure the authenticity and integrity of digital communications and transactions by verifying the sender's identity and confirming the message's integrity. As the era of quantum computing approaches, the development of quantum signature schemes is underway to provide secure digital signatures that can withstand the computational power of quantum computers.

Digital signatures in the classical world are based on the principles of asymmetric cryptography. Formally, we can describe classical signatures as follows. Suppose $f(x)$ is a one-way function with the property that calculating $f(x)$ is easy, but computing $x$ given $f(x)$ is computationally difficult. Suppose Alice sends the set $\{f, (0, f(k_0)), (1, f(k_1))\}$ to Bob. $k_0$ and $k_1$ are private keys. Then, for message B, if $B = 0$, Alice sends B along with $k_0$ to Bob. Otherwise, she sends B along with $k_1$. Bob receives B and $k_B$ and it computes $f(k_B)$ and compares it with the values Alice already announced. Then he can verify whether the sender was Alice or not [67], [174]. In this way, we sign every bit of the message.

The quantum signature scheme employs a process where a public key is generated from a private bit string or a sequence of qubits. In this context, we adopt a widely recognized quantum signature scheme [174] and define a one-way quantum function as $k \mapsto |f_k\rangle$ in which $k$ is a bit string and $|f_k\rangle$ is a quantum state. In comparison to classical signature, there are two limitations in the scheme proposed in [174]; first, we cannot exactly compare quantum states to see whether they are equal or not. We use the swap test, which uses a combination of quantum gates. If the states are the same, the test always passes. However, if they differ, the test may fail to conclude the difference with a nonzero probability. For the other difference, let us assume that each bit string $k$ is of length $L$ and that to each one of them a quantum state $|f_k\rangle$ of length $n$ qubit is assigned. These states are nearly orthogonal and $L \gg n$. According to Holevo's theorem [175], if we measure $n$-qubits, it gives at most $n$ classical bits of information. Therefore, if there are $T$ copies of $|f_k\rangle$, it gives at most $Tn$ bits of information about $k$, and if $(L - Tn) \gg 1$, the chance of finding $k$ remains very small. However, this theorem also limits the number of copies we can distribute, which contrasts with classical signatures.

Considering the above specifications and definitions, the quantum signature protocol works as follows. Alice wants to transfer an $L$ bit message $B$ and chooses at most $M$ pairs ($2M$ keys) of private keys $\{k_0^i, k_1^i\}$ and corresponding quantum public keys $\{|f_{k_0^i}\rangle, |f_{k_1^i}\rangle\}$. $k_0^i$ and $k_1^i$ are randomly and independently used to sign bits 0 and 1 for each $i$, respectively. Assume all receivers have Alice's public keys. One way to transfer them is to use quantum teleportation or a trusted third party [176]. Then Alice sends $\{B, k_b^1, k_b^2, \dots, k_b^M\}$ to all receivers over an insecure classical channel. This way, all receivers can calculate half of the Alice public keys. Then, they check these public keys and verify $k_b^i \mapsto |f_{k_b^i}\rangle$ through swap test. Suppose $f$ is the number of incorrect keys. Bob considers the message valid and transferrable if $f < c_1 M$. If $f > c_2 M$ the message is invalid, and if $c_1 M < f < c_2 M$, the message is valid but not transferrable, which means it is possible another receiver may categorize it as invalid. $c_1$ and $c_2$ are two thresholds that entities are aware of. Finally, all used and unused keys are discarded.

*A.3. Quantum teleportation*

Quantum teleportation leverages quantum entanglement to facilitate the transfer of quantum information to a distant location without physically moving the particles or using quantum channels. It is crucial to understand that this process should not be confused with the instant transmission, which might be implied due to the nature of entanglement. While the state change happens instantaneously, quantum teleportation does not provide instantaneous transmission. In practice, quantum teleportation cannot occur faster than light speed because the sender needs to transmit the outcome of their measurement to the receiver via a classical communication channel. This measurement result supplies details about the necessary operations that must be carried out to rebuild the quantum state at the destination.

Before delving into quantum teleportation, it is essential to understand the primary form of entanglement, specifically Bell states, upon which the quantum teleportation process relies. A Bell state is a unique type of quantum state consisting of two entangled qubits. There are four distinct Bell states, each representing a different entanglement combination between the two qubits: $|\psi\rangle = \frac{1}{\sqrt{2}}(|00\rangle \pm |11\rangle)$ and $|\psi\rangle = \frac{1}{\sqrt{2}}(|01\rangle \pm |10\rangle)$. As mentioned in Subsection II.B, the notations $|00\rangle$, $|01\rangle$, $|10\rangle$, and $|11\rangle$ denote the possible two-qubit basis states, while the coefficients ($\frac{1}{\sqrt{2}}$) ensure the probability amplitudes are normalized. Bell state measurement (BSM) is a crucial process in quantum information processing that identifies which of the four Bell states a given two-qubit system belongs to. This procedure requires a joint measurement on both qubits, projecting the system onto one of the four Bell states. BSM is a vital component of numerous quantum communication protocols, such as quantum teleportation, QKD, and quantum networks.

We provide a brief overview of the quantum teleportation process [177], [178]. Alice wants to teleport the unknown state of qubit $q$ to Bob. In addition, suppose Alice and Bob share an entangled pair of qubits, each one possesses half of it. Alice makes a joint BSM on the qubit she wants to send and half of the entangled qubit she possesses. This process will output one of the four possibilities $\{00, 01, 10, 11\}$ with equal probabilities. Alice transfers the result to Bob over a classical channel. After receiving the result, Bob can apply a series of quantum gates to the qubit he possesses, and the state of his qubit will become the exact copy of the initial (unknown) state of the qubit $q$. It is worth mentioning that the state of Alice's qubit $q$ is destroyed because of the measurement and no-cloning theorem. Therefore, there is just one copy of $q$, and that one is in possession of Bob.

*A.4. Bit commitment*

The goal of a bit commitment protocol is to allow a sender, Alice, to commit to a specific bit value (0 or 1) so that the receiver (Bob) cannot learn the value of the committed bit until the sender chooses to reveal it. Once the value is revealed, the sender cannot change the committed bit value. This allows for creating protocols that require the commitment of a value without revealing it.

Many bit commitment protocols have been introduced, and one of them is presented in this subsection [179], [180], [181].

Suppose Alice and Bob share a pseudo-random generator function $R$ that maps from $n$ bits to $3n$ bits. Alice wants to commit a bit $x$ to Bob.

1. Bob sends a random $3n$-bit string $B$ to Alice.
2. Alice selects a random $n$-bit string $A$ and feeds it into the function $R$. The output will be $R_A$
3. Commit phase: If $x = 1$, Alice sends $R_A$ to Bob; otherwise, she sends $R_A \oplus B$, where $\oplus$ denotes the XOR function.
4. Reveal Phase: Alice sends $A$ to Bob. Bob can easily calculate $R_A$, and since he knows $B$, he can understand whether $x$ was one or zero by comparing the XOR result.

Bob cannot know the value of $x$ before the revealing phase. Thus, we define a bit commitment protocol as concealing if the receiver cannot recover the value of the committed bit before the revealing phase. In addition, we define the protocol as binding if the sender cannot change the value of the committed bit after the commitment phase. If Alice can change the bit after the commitment phase or Bob can understand the value of the committed bit before the revealing phase, we say they are cheating. A bit commitment protocol is unconditionally secure if cheating can be detected with a probability arbitrarily close to one. However, in both classical and quantum worlds, unconditionally secure bit commitment is impossible [181]. Instead, researchers have proposed cheat-sensitive quantum bit commitment protocols that detect cheating with a probability greater than zero (instead of close to one) [182].

*A.5. Quantum blockchain overview*

In this subsection, we focus on acquainting the readers with two distinct categories of quantum blockchains: hybrid and fully. This serves to provide readers with a preliminary comprehension of quantum blockchains. A more in-depth discussion on the state-of-the-art approaches in quantum blockchains will follow in Subsection IV.B.

*A.5.1. Hybrid quantum blockchain overview*

Hybrid quantum blockchains typically employ Quantum Key Distribution (QKD) as a fundamental layer to generate secret keys for blockchain users [25]. Unlike the existing blockchain structure, which utilizes asymmetric cryptography, hybrid quantum blockchains are mostly rooted in symmetric cryptography. QKD would facilitate the exchange of keys. In this context, the symmetric key blockchain transforms into a private or permissioned blockchain, allowing selected participants to join the network.

Typically, these hybrid quantum blockchains operate in two stages: the deployment phase and the block production phase. The deployment phase is achieved through the QKD network, as shown in Fig. 6(a). Following this, every node authenticates with the other nodes in the network, and they can jointly maintain the private or permissioned blockchain. During the block production phase, new transactions are executed and authenticated via a consensus algorithm (for example, the original BFT state-machine replication [27]) and a keyed hash function (e.g., the Toeplitz hashing [183]) using the secret keys produced in the deployment phase.

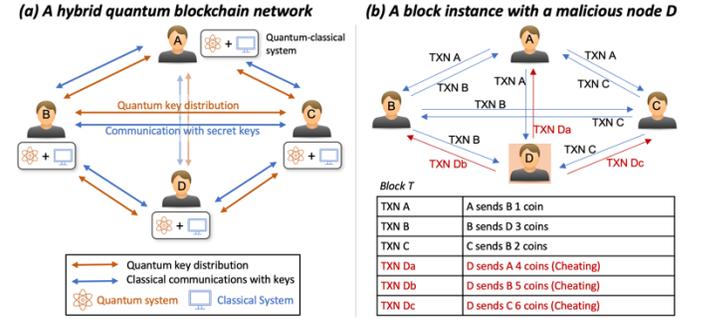

**Fig. 6.** A hybrid quantum blockchain network and consensus.

The process of authenticating new valid transactions between two nodes, Alice and Bob, could proceed as follows: Alice forwards her version of new transactions to Bob, accompanied by a hash value computed using the keyed hash function and the secret key. Upon receiving these transactions, Bob calculates a hash value of the new transactions using the shared key. Bob then compares these two hash values. If they match, Bob accepts the new transactions.

Despite being a private blockchain network, not every node within the blockchain network is trustworthy. For instance, consider node D in Fig. 6(b). Initially, researchers contemplated adopting the original BFT state-machine replication [25], [27]. Assume a network of $n$ nodes that are pairwise connected and have access to each other's secret key. Let $T$ represent a set of transactions set to be incorporated into a new block, which is also labeled $T$. Each node is given a copy of the set $T$ for verification before producing a new block. Let $T_i$ be the set received by the $i$th node. Following the broadcast of all the intended transactions within a specific period, the nodes start to send their blocks $T_i$ to each other. In the subsequent communications, the nodes exchange all the received $T_i$ with other nodes until all the online nodes have communicated with one another. It has been proven that $n$ network nodes can reach a consensus on $T_i$ in no more than $m + 1$ communications where $m$ is less than $n/3$. This suggests that consistency can be reliably attained if the number of dishonest nodes in the network is less than $n/3$ [27]. A downside of this consensus algorithm is that it becomes exponentially data-heavy if many dishonest or non-operational nodes are present. Therefore, more research on an efficient consensus algorithm is necessary.

*A.5.2. Fully quantum blockchain overview*

Now, we discuss a typical example of a fully quantum blockchain that utilizes temporal entanglement [49]. Note that this is a significantly simplified rendition of the classical blockchain. A single quantum state makes up one block. To form a chain of quantum states, multi-qubit entangled states are progressively appended to an expanding chain in a time-ordered manner, thanks to entanglement in time. This chain of quantum states can be used as the foundation of a blockchain. For simplicity, we use a 2-bit string $b_1 b_2$ to represent the data in a block. In this context, it is prepared into a temporal Bell state:

$$|\psi_{b_1 b_2}\rangle = \frac{|0\rangle |b_2\rangle + (-1)^{b_1} |1\rangle |\overline{b_2}\rangle}{\sqrt{2}}$$

where $|\bar{b}\rangle$ denotes the negation of $|b\rangle$. Its subsequent state, i.e., its following block, would be formulated to be temporally entangled with this state.

Since a time delay of $T$ is imposed upon the creation of each state, the creation time can be used as the block timestamp. Assume the creation time of the first state is $t = 0$, and for the subsequent state (next block), it would be $t = T$. The first and second blocks can be merged to form a four-photon GHZ state:

$$|\psi_{b_1 b_2}\rangle^{0,T} = \frac{|0\rangle^0 |b_2\rangle^T + (-1)^{b_1} |1\rangle^0 |\overline{b_2}\rangle^T}{\sqrt{2}}$$

Where 0 and $T$ are the timestamps of the first and second blocks, respectively. The blocks can recursively evolve into a chain of entangled states. At time point $t \in \{0, T, 2T, \ldots, nT\}$, the chain of quantum states can be expressed as:

$$|\psi_{b_1 b_2 \ldots b_{2n}}\rangle^{0,T,T,2T,2T,3T\cdots,(n-1)T,(n-1)T,nT} = \frac{1}{\sqrt{2}}(|0\rangle^0 |\psi_{b_2}\rangle^T |\psi_{b_3}\rangle^T |\psi_{b_4}\rangle^{2T} \cdots |\psi_{b_{2n-1}}\rangle^{(n-1)T} |\psi_{b_{2n}}\rangle^{nT} + (-1)^{b_1}|1\rangle^0 |\overline{\psi_{b_2}}\rangle^T |\overline{\psi_{b_3}}\rangle^T |\overline{\psi_{b_4}}\rangle^{2T} \cdots |\overline{\psi_{b_{2n-1}}}\rangle^{(n-1)T} |\overline{\psi_{b_{2n}}}\rangle^{nT}).$$

where $b_i \in \{0,1\}$. Each pair of $|\psi_{b_{2i-1}}\rangle^{(i-1)T} |\psi_{b_{2i}}\rangle^{iT}$ where $i \in \{2, 3, \ldots, 2n\}$ constitutes a block, as shown in Fig. 7(a). The 2-bit data in a block is represented by $b_{2i-1} b_{2i}$. The superscripts $iT$ act as the timestamps of the blocks.

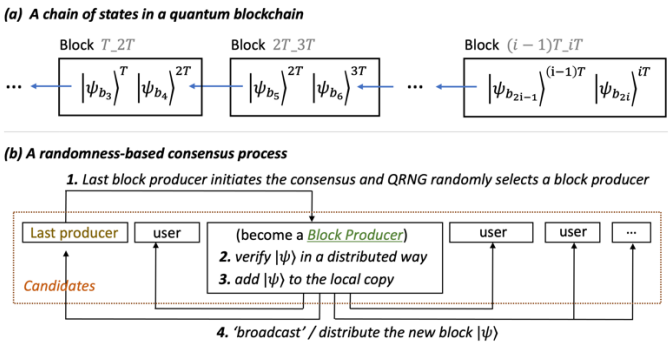

**Fig. 7.** A fully quantum blockchain and consensus.

As depicted in Fig. 7(b), a fully quantum blockchain could employ a quantum random number generator (QRNG) [184] and distributed verification tests (e.g., θ-protocol [185]) to achieve a consensus protocol. QRNG ensures randomness in selecting a block producer. The true randomness of QRNG takes the place of the equitable election offered by proof-of-work or proof-of-stake. The chosen block producer then performs a distributed verification test to validate the new state (i.e., the new block). Once the verification tests are completed, the new block is incorporated into the blockchain. Then, the updated chain is disseminated to all network nodes.

## B. Studies of Quantum Blockchains

In this subsection, we provide a survey of research efforts that explore the integration of quantum computing into blockchains. While the field of quantum blockchain studies is still relatively limited compared to post-quantum blockchains, we can categorize the existing literature into two main groups. Firstly, we will examine papers that primarily delve into the design aspects of quantum blockchains. Following that, we will summarize studies that center around the practical applications of quantum blockchains.

### B.1. Designing quantum blockchains

Designing a quantum blockchain involves utilizing two approaches: combining quantum and classical systems (hybrid quantum-classical blockchains) and implementing purely quantum systems (fully quantum blockchains).

While fully quantum blockchains offer unparalleled security against all known cyber-attacks, their practical implementation is currently unfeasible due to the limited availability of quantum computers. As a more pragmatic approach, integrating classical systems with quantum techniques provides adequate security against quantum attacks, albeit at a lower level of security. Consequently, there are more hybrid solutions than fully quantum implementations in the quantum blockchains field. Most of the approaches for fully quantum blockchains remain in the realm of theory, lacking practical implementation or experimental validation. In the following sections, we will explore QKD-based approaches and other forms of hybrid quantum blockchains before delving into fully quantum blockchains.

#### B.1.1. QKD-based hybrid quantum blockchains

JPMorgan has recently disclosed its research on a QKD-based blockchain network [48], a hybrid quantum-classical approach. While the underlying blockchain platform remains classical, the key exchange processes are facilitated by quantum channels. In their proof-of-concept infrastructure, JPMorgan employs a multiplexed QKD system capable of supporting data rates of up to 800 Gbps. This high-speed capability is crucial for mission-critical applications, enabling optical-layer encryption and open application programming interfaces (APIs) to be utilized in real-world scenarios.

A similar QKD-based blockchain system was proposed in 2018 using a two-layer network [25], as discussed in Subsection IV.A.5. The first layer is a QKD network, which shares secret keys. The second layer transfers messages with authentication tags produced by the Toeplitz hash function [183]. The secret keys generated in the first layer are used in the Toeplitz hash function to generate hashtags. Transactions are created by those nodes who want to transfer their money to other nodes. Each transaction contains the following information: information about the sender and receiver, timestamp, amount to be transferred, and a list of reference transactions to show the sender has enough money. The sender sends the transaction to all other nodes, and each of them verifies that the transaction is correct. To create a block of transactions, the authors utilized the Broadcast protocol [27] to achieve a Byzantine agreement. At certain points of time, for example, every ten minutes, all network nodes (and not just some miner nodes) execute the Broadcast protocol to achieve consensus about the correct version of transactions (to eliminate double spending cases) and admissibility of the transactions. Then, every node sorts admissible transactions based on their timestamp creates a block of them, and adds it to its local copy of the ledger. This way, all nodes will create the same block, and there will be no forks in the chain. The authors mentioned the overhead of transferred messages for reaching a consensus. However, they also mentioned that these messages are carried over the classical channel, not the quantum channel. Since the Broadcast protocol is unconditionally secure, an attacker must hack at least one-third of the nodes to affect the consensus process if the number of dishonest nodes is less than one-third. Also, the Toeliptz hash function is prone to Grover's algorithm. The authors used their urban fiber QKD network and experimentally tested their blockchain.

Implementing a QKD network requires a dedicated quantum channel, and due to its relatively low key-generation rate and the limitation of generating only one pair of keys at a time, it may not be well-suited for large-scale networks or systems. To address this issue, some researchers propose utilizing the concept of a quantum key pool (QKP) [186] to enable secure and efficient key management. The QKP is a collection of pre-generated keys shared among users to ensure secure communication. Subsequently, the authors of [187] introduce QKP as part of the proposed security solution. By leveraging the inherent security advantages of QKD, the authors ensure that the keys used for encryption and query processes are securely distributed among the participating entities within the smart grid system. This further strengthens the privacy-preserving and security aspects of the proposed solution, making it more resistant to potential attacks and eavesdropping.

In addition, a logic-based blockchain based on the QKD and QSDC was proposed [188]. They utilized the idea of [25] and enhanced it by replacing the classical Byzantine agreement protocol with their honest-success quantum Byzantine protocol (QHBA). The authors also introduced multiple enhancements to the previous studies. The authors assumed that each pair of nodes in the blockchain, half of which must be honest, is connected by an authenticated quantum channel and a not-necessarily authenticated classical communication channel. Each pair of nodes can generate secrets by using the quantum channel according to the circumstances described by QKD. The authors proposed a coin named qulogicoins and used the following format for protected transactions:

$$T_x = (x; y; j; \alpha; \phi; \beta; \psi)$$

In the format mentioned above, $x$ represents the hash value of the transaction that is obtained by applying Toeliptz Hashing [25], which uses a private key obtained through the quantum channel. $T_y$ is the transaction from which the coin was redeemed, and $j$ is the receiver of the transaction. $\alpha$ is the classical certificate, and $\phi$ is the quantum certificate, which can be transferred only over the quantum channel. The qulogicoins that are transferred as a part of the transaction can only be used when both $\alpha$ and $\phi$ hold. Finally, $\beta$ is classical data, and $\psi$ is quantum data. If we remove the $\alpha; \phi; \beta; \psi$ part of the transaction, a plain transaction will be yielded. Each transaction is sent to all miners, and each miner validates each transaction. Then, all miners perform the QHBA protocol, and if at least half of them agree that it is admissible, the transaction is added to the ledger of every node. QHBA differs from the classic Byzantine protocol in the following aspects: If the sender is honest, all other entities reach the same output value equal to the sender input value, similar to the classic Byzantine protocol. However, if the sender is dishonest, all other entities abort the protocol, or all honest entities decide on the same output value Y. The authors claimed that their QHBA is more efficient than classical Byzantine protocol in the presence of many dishonest nodes. The authors proposed three phases for their QHBA protocol, in which they use quantum secure direct communication to distribute correlated lists of numbers. The authors proposed a quantum secure direct communication protocol based on Shamir's three-pass protocol that allows messages to be deterministically and securely sent over a quantum channel. Finally, the authors showed the application of their blockchain in the design of quantum bit commitment.

Moreover, QKD was used to design a quantum-secured permissioned blockchain [189]. The authors assumed a secret key is shared between each pair of nodes in the network, which is distributed through a QKD protocol. In addition, they designed the Toeplitz Group Signature Scheme (TSG), which is based on the Toeplitz hash function [182], and used it for signing messages. They designed and utilized a quantum vote-based consensus algorithm named QSYAC based on another consensus algorithm named YAC, which is used in the Hyperledger Iroha Blockchain framework. The authors divided network nodes into clients and peers. Clients generate transactions, sign them with TSG, and send it to all peer nodes. Peers participate in the voting and consensus process. QSYAC is executed in three phases, and in each phase, one of the peers acts as the proposing node, and the rest act as voting nodes. Authors assumed among $n$ peers, at least ¾ of them are honest. QSYAG works as follows: 1) The proposing phase: The proposing node verifies the transaction signature, generates a proposal block that contains a list of transactions, signs it with the TSG, and sends it to the voting nodes. 2) The voting phase: Every voting node calculates a verified proposal block which contains a sub-list of the transaction from the proposals it received. It also contains the hash of all transactions in that block. However, the authors did not describe the hash function for generating it. Then, the voting peer generates a vote on the block and sends it to all peers. A vote is the hash value of the block and the TSG of the hash value. 3) The decision phase: When a peer node receives at least $3n/4$ votes for a block, it sends an accepting message to all peers. Each peer that receives the accepting message adds the block to its ledger and broadcasts the accepting message to all other peers, and the round ends. If the peer does not receive at least $3n/4$ votes, it broadcasts the rejecting message. Finally, if no accepting message is broadcasted, the round ends and the block is rejected.

The authors claimed that their QSYAC has the communication complexity of $O(n^2)$, which is lighter than the Broadcast protocol [27]. In the rest of their papers, the authors also proposed a simple scripting language for implementing smart contracts, similar to the Bitcoin scripting language. By utilizing the scripting language, the authors also implemented a lottery protocol. However, their lottery does not benefit from QBC, and the authors did not show that it has the properties of an ideal lottery protocol described in [190].

*B.1.2. Other hybrid quantum blockchains*

Having discussed QKD-based hybrid blockchains, let us now examine the utilization of quantum encryption and signature schemes in constructing quantum blockchains. For example, a hybrid blockchain paradigm based on asymmetric quantum encryption and a variant of the delegated proof-of-stake [31], [32] consensus algorithm was proposed [191]. The proof-of-stake variant is called the delegated proof-of-stake with node's behavior and Borda count (DPoSB) [192]. In DPoSB, the behavior of a node in creating blocks is used as a metric for selecting block producers, with those that exhibit poor behavior receiving low rankings. This is combined with the Borda count voting method, in which voters rank candidates in order of preference to ensure robust and fair selections that align with the preferences of user nodes. The candidate nodes' malicious behaviors are assigned a weight factor and a threshold that indicates the maximum number of times a behavior can be tolerated. For instance, failing to hash transactions is met with a harsher penalty than failing to verify blocks, allowing a higher number of occurrences. The amount of stake a node holds, evaluating the node's behaviors, and the Borda count are all considered to select a block producer group. In addition, a quantum digital signature scheme builds on quantum state computational distinguishability and fully flipped permutation (QSCD$_{ff}$) problem [193] to construct a one-way quantum function, ensuring the security of transactions. The authors justified their selection of QSCD$_{ff}$ by comparing the security level of QSCD$_{ff}$ with nonorthogonal-encoding-based and entanglement-based signature schemes.

Moreover, securing blockchain transactions with a quantum signature instead of a classical one was explored [176]. Here, we briefly describe the process of signing transactions in this approach. Assume that Alice wants to send a transaction to miners. Alice uses quantum signing to sign every bit of the transaction. First, she creates a pair of private keys, generates corresponding public keys, and teleports all public keys to all miners. Since the teleportation of quantum information needs an interactive process between the two parties, there are currently no effective broadcasting methods in quantum networks. Here, they assume all miners are known to the blockchain users, and new miners need to start the process by getting all public keys from the users. This seems impractical in a public blockchain with millions of users but is suitable for a permissioned or private blockchain where users are only admitted by request. Back to the process, Alice uses her private keys, signs the transaction, and transmits the transaction and related public keys to miners. Miners do a swap test to verify the transaction. If the number of incorrect keys is small, the miner accepts the transaction and adds it to the blockchain. Otherwise, it discards the transaction and informs Alice. Since Alice needs to generate and distribute a set of keys for every transaction, the computation overhead would be considerable. In addition, using a classical channel for transferring the transaction would result in the difference between the time miners receive the transaction and the corresponding private keys of Alice. The authors did not delve into the scenario where a miner acts maliciously and attempts to pass off a message as coming from Alice. It would be interesting to investigate how the receiving node would detect and handle such a situation and how to design proper transaction formats to address this concern. Finally, securing the transmission of the result of the BSM as a part of teleportation was not considered.

*B.1.3. Fully quantum blockchains*

In this subsection, we explore studies that focus on the design of purely quantum systems for quantum blockchains. These approaches provide valuable insights into applying quantum mechanics to construct a blockchain based on a chain of quantum states, similar to classical transaction chains. However, it should be noted that fully quantum blockchains have a limited capacity to store data compared to classical blockchains. Nevertheless, their potential for unconditional security makes them an attractive prospect for the future of blockchain technology, thus warranting ongoing research efforts in this area.

Since fully quantum blockchains are still in the early phase, there are not many variants on the design of fully quantum blockchains. Here, we introduce several primitive quantum blockchain designs. The first type discussed in Subsection IV.A.5 is a fully quantum blockchain design using entanglement in time [49]. It designs a quantum blockchain system by exploiting the concept of entanglement in time. The authors develop a new protocol that uses a series of entangled qubits arranged in a temporal sequence to create a blockchain-like structure. This temporal entanglement ensures that the quantum states of the qubits are correlated across time, forming an interconnected chain of information. The key advantage of this approach is that it provides inherent security and tamper resistance, as any attempt to alter the information within the quantum blockchain would cause the entangled qubits to collapse, rendering the tampering evident. This feature makes the proposed quantum blockchain intrinsically secure against various attacks, including those from potential quantum adversaries.

Alternatively, there is ongoing research work on building upon these fully quantum blockchain designs. In particular, we have developed a quantum blockchain-based infrastructure for access control and authentication in our previous work [194], which theoretically analyzes the use of fully quantum blockchain technologies for decentralized identity authentication. The approach utilizes a fully quantum blockchain structure based on entanglement in time (i.e., a chain of quantum states). We proposed a quantum blockchain identity framework (QBIF) with three roles: identity owners, identity issuers, and identity verifiers. In QBIF, attestations are kept as quantum states and chained together by entanglement, called quantum identity attestations (QIAs). Identity owners who want authorized actions from verifiers must go through authentication and attestation processes if they have not done that with the issuers. The network of QBIF is composed of multiple quantum nodes connected via a quantum network and a classical network dealing with off-chain communications in a peer-to-peer fashion. The proposed consensus protocol uses a QRNG and a set of distributed verification tests called θ-protocol [185]. The QRNG provides true randomness in selecting a block producer that creates and verifies a block (GHZ states) in the quantum blockchain. The selected block producer executes the verification tests, θ-protocol, to validate the newly added block. The protocol is valid even in the presence of untrusted network nodes [185]. Once the verification tests are done, the new block is added to the blockchain and broadcast to all network nodes. Though the proposed infrastructure currently has various limitations and challenges, it presents a unique and likely decentralized perspective of quantum applications.

In addition, quantum teleportation and quantum signatures have been utilized to design qbitcoin, a novel cash system [195]. While certain quantum processes, like quantum teleportation, still rely on classical communications, these approaches are considered fully quantum since they are designed based on a chain of quantum states rather than classical data. In qbitcoin, each coin is defined as a pair $c_i = (r_i, |\psi_i\rangle)$ in which $i$ is a serial number, $r_i$ is a transaction record as a string of bits, and $|\psi_i\rangle$ is a quantum state corresponding to the transaction. We also have $r_i \neq r_j \Leftrightarrow |\psi_i\rangle \neq |\psi_j\rangle \forall i,j$. A transaction is done by transferring the coin to another entity. The quantum part of a coin is transported by quantum teleportation. A transaction is accepted if it has the same serial number with respect to $r$ and $|\psi\rangle$. Because of the teleportation, the owner will not possess the coin after teleporting, and this would prevent the double-spending problem. In addition, the identity of the sender also needs to be verified. The authors utilize quantum signatures to tackle this issue. The sender needs to distribute its public keys to randomly chosen receivers in the first place. Then, the sender sends $r$ along with the public keys used to sign the transaction to those random receivers. If one of the receivers verifies the signature, it records it and tells others. The verifier also receives some rewards. The receiver can check the serial number and verify the transaction. However, the detailed solutions also suffer from a few drawbacks: the authors did not mention the relation, or mapping, between $r_i$ and $|\psi_i\rangle$. In addition, the authors did not mention how the sender can distribute its public keys to receivers. Using serial numbers and pairing between a coin's classic and quantum parts is vague because no one can check the quantum state. No one can either know whether they possess a valid coin because they cannot check whether the quantum state matches the classic part of the coin. If the authors defined a mapping function, it would have been possible through swapping.

Furthermore, some approaches specifically focus on redesigning certain blockchain components. For instance, a blockchain framework was proposed that replaces the traditional hash function with a quantum hash function [196]. It utilizes the inspired quantum walk to chain the blocks. They also use the hash function in the encryption of the messages. A transaction comprises sender and receiver IDs, timestamps, and cipher data. Every blockchain node exchanges its public key parameters with each other over a secure quantum channel. Then, the sender of a transaction sends it to the receiver. The receiver verifies it, and then every node adds it to its current copy of the blockchain. It is not clear why the nodes do not execute a consensus algorithm.

Quantum-weighted hypergraph states have also been proposed as an alternative to implementing the hash pointers in a blockchain [197]. For simplicity, they represent one block with one qubit. Then, a weighted hypergraph of $n$ qubits forms the chain of $n$ blocks. Consensus is achieved based on the phases carried by the hyperedges of the quantum-weighted hypergraph states, which encode data in the hypergraph state. Creating a block creates a conditional relative phase. The network nodes need to agree on the preset conditions on the relative phase when verifying new blocks.

The other blockchain components that quantum systems can replace are the consensus mechanism and ZKP for privacy. Quantum technology has the potential to fundamentally transform the way consensus is established in a blockchain. Quantum computers, for instance, can generate true randomness, surpassing the capabilities of classical computers. This randomness plays a crucial role in consensus protocols for selecting block producers or validators. Additionally, quantum systems can offer an advanced variant known as quantum ZKP, which provides more secure and private methods to verify transaction correctness without revealing additional details.

For instance, a new mining and consensus approach for quantum blockchains was proposed in [198]. They proposed an architecture consisting of quantum servers, clients equipped with quantum modems, and optical fibers connecting clients' modems to quantum servers. Quantum servers are equipped with a modular laser source, a source of photonic entanglement, a qubit measuring device, and a quantum random number generator. Quantum servers generate entanglement and perform a consensus algorithm based on the Byzantine fault-tolerant protocol. However, the authors solely focused on mining and did not improve the vulnerable parts of blockchains, i.e., the hash function and signatures. Following this line of innovation, a quantum blockchain consensus mechanism based on quantum measurements and quantum ZKP was proposed [199]. The authors explained how to use the true randomness of a quantum measurement as a secret value for miners to compete to find, but the mechanism for using quantum ZKP to verify the equality of the miner's value and the secret value is not clearly explained.

*B.2. Building quantum blockchain applications*

While quantum blockchains may currently seem far-reaching due to the limited capacity and stability of quantum systems, QKD-based hybrid quantum blockchains persist as a viable method for safeguarding blockchains against quantum attacks. As mentioned, QKD facilities have been built worldwide, making the deployment of QKD-based hybrid quantum blockchains foreseeable. Consequently, there has been a surge of interest in researching and developing blockchain applications based on quantum blockchains. This subsection presents the latest advancements in creating quantum blockchain applications.

An instance of research involved studying a voting application that combines blockchain technology with QKD [200]. In this voting platform, users must register using their identification and other relevant information before participating in regular voting. A block is generated and appended to the blockchain upon casting a vote. The block contains essential details, such as the voter's ID, their vote, their digital signature, a timestamp, and the hash (SHA256) of the previous block. Notably, the block's hash is encrypted in a database using QKD technology. It is important to note that while the authors' proposed blockchain framework shares similarities with conventional blockchain frameworks, it remains equally susceptible to quantum attacks. The utilization of QKD in this application is limited to securing the storage of block hashes within the database and is not integrated into the chain's design.

A different voting protocol leveraging quantum blockchains was proposed [182]. According to the authors, a voting protocol should meet the following conditions: anonymity (only the voter knows what they voted for); binding (no one can change their ballot after voting); non-reusability (each voter can only vote once); verifiability (each voter can verify whether their ballot is considered or not); eligibility (only eligible voters can vote); fairness (no one should gain information about the tally of ballots before the tallying phase); and self-tallying (those interested in the results should be able to tally them themselves). The authors proposed utilizing QKD, QHBA [187], and QBC in the blockchain to adhere to these conditions. They developed a straightforward voting protocol wherein voters could choose between agreement or disagreement. To illustrate their implementation, let us consider an example with three voters denoted as $V_i \in \{0,1\} \; \forall i \in I$ where $I$ represents the set of voters and $|I| = n$. Suppose the original votes are $V_1 = 1$, $V_2 = 1$, and $V_3 = 0$. There are two phases in the approach: ballot commitment and ballot tallying. In the ballot commitment phase, each voter $i$ generates row $i$ ($r_i$) of $n \times n$ matrix $V$ in a way that $\sum_{j \in I} r_{i,j} = 0 \; (mod \; n + 1)$. Therefore, suppose we have the following matrix $\begin{bmatrix} 2 & 1 & 1 \\ 1 & 1 & 2 \\ 2 & 0 & 2 \end{bmatrix}$. Each node only knows its row, and they are not aware of other rows of the matrix. Then, each voter $i$ communicates the corresponding value $r_{i,j}$ to all $j \in I \setminus i$ via secure quantum communication, which utilizes QKD. Therefore, for example, voter 2 sends "2" to voter 3. After that, every voter $i$ knows the value of column $i$. Then, each node $i$ computes its masked ballot as $\hat{V}_i = V_i + \sum_{j \in I} r_{j,i} \; (mod \; n + 1)$. Therefore, voter 1 calculates $1 + 2 + 1 + 2 \; (mod \; 4) = 2$. Voter 2 calculates $1 + 1 + 1 + 0 \; (mod \; 4) = 3$, and Voter 3 calculates $0 + 1 + 2 + 2 \; (mod \; 4) = 1$. After that, all voters commit their masked ballots to miners using a QBC protocol.

In the ballot tallying phase, the revealing phase of QBC is executed, and each voter reveals their masked ballot to all miners. Then, miners use a quantum honest-success Byzantine agreement protocol and would have been reach a consensus about the marked ballots. After that, the result of the voting can be calculated by $\sum_{i \in I} \hat{V}_i \; (mod \; n + 1)$. Therefore, the result will be $2 + 3 + 1 \; (mod \; n + 1) = 2$, which is equal to $\sum_{i \in I} V_i$ because $\sum_{j \in I} r_{i,j} = 0 \; (mod \; n + 1)$. The authors claimed that their voting protocol satisfies all described properties of an ideal voting protocol. However, the authors did not mention how agents can determine the number of all voters. In addition, how the system can determine whether someone did not vote remained unanswered.

In addition to voting protocols, proposals for lottery and auction protocols were put forward [190], building upon a redesigned quantum blockchain. In the lottery protocol, QKD and QBC are utilized. The lottery protocol operates as follows: 1) Each player purchases a ticket and commits it to all miners. 2) Each player reveals their ticket to the miners using a QBC protocol. 3) All miners work collectively to agree on the purchased tickets. 4) The miners perform an XOR operation on the tickets to determine the winning ticket. 5) The prize for each player is determined by calculating the Hamming distance between their ticket and the winning ticket.

The authors asserted that their protocol ensures an equal chance of winning the lottery for every ticket, making it unpredictable. Additionally, the protocol safeguards against ticket forgery or alteration, especially after the winning ticket is revealed. The transparency of the winning ticket and associated awards allows for public verification. However, while unconditional security was claimed, it is worth noting that both the QBC and consensus mechanisms are susceptible to cheating nodes, albeit within certain limitations. The authors did not provide an analysis regarding this aspect.

The authors further introduced a sealed-bid auction in their paper, aiming to achieve bid privacy, ensuring no bids are known until the revealing phase and posterior privacy, where only the seller knows the losers and their corresponding bids. Additionally, the proposal ensures that bids cannot be altered after the commitment. The auction involves three entities: a seller, a group of buyers, and a set of miners. The auction process unfolds as follows: 1) Each buyer commits their bids to the seller and all miners. 2) Each buyer reveals their bid to the seller. 3) The seller calculates the highest bid and the winner. 4) The seller sends all miners the information about the winning bid and the winning buyer. 5) The seller sends all miners a permutated list of the remaining bids. 6) Each miner compares the winning bid's value to the other offers. If the calculation of the winning bid is correct, it sends the winning and the rest of the bids to all buyers. Otherwise, it recognizes the seller as a cheater. 7) All buyers verify that the winning bid is calculated correctly and their bid is also listed among the received bids. The buyer sends a valid message to the miner if the verification passes successfully. Otherwise, it reveals its bid to the miner. Then, the miner recognizes the seller as a cheater. 8) If a miner does not verify the seller as a cheater, it will output all bids and the winner. 9) All miners launch a consensus process to agree on the winner. Then, the result will be added to the blockchain. However, similar to their lottery process, the authors did not analyze their protocol's sensitivity to the cheating nodes.

Further research has been conducted on quantum blockchains, exploring aspects beyond those previously proposed. For instance, in [201], the authors present a blockchain-based application for the Internet of Vehicles (IoV), assuming the utilization of quantum communication devices on mobile vehicles. They aim to integrate quantum cryptography, including quantum state signature and encryption [202], with classical blockchains. In [203], a smart contract for blockchain was developed by integrating light-weighted quantum blind signature [204], [205], and QKD into classical blind signature [206] to make blockchain smart contracts resistant to quantum threats. Besides, quantum blind signatures have also been employed to improve security and privacy for blockchain-based applications, such as healthcare [207]. This approach aims to maintain the confidentiality of users' data while acquiring digital signatures from trusted authorities. The authors showcase the feasibility and effectiveness of the proposed hybrid scheme using simulations and case studies across a range of healthcare scenarios, including remote patient monitoring and telemedicine.

In [208], a hybrid classical-quantum payment system was developed by integrating a classical blockchain with smart contracts and quantum lightning technology, enhancing public-key quantum money presented in [209]. In [198], a quantum-based mining protocol that aims to solve the energy waste problem of proof-of-work by using quantum entanglement as a resource to secure new blocks on a blockchain was proposed. The protocol is integrated with classical blockchains for energy-efficient block production. The authors propose to use quantum optical devices in the blockchain architecture and call it proof-of-entanglement. In [210], the authors propose a scheme called "quantum Bitcoin" to construct a distributed quantum currency using quantum money [211], which cannot be counterfeited due to the non-cloning theorem of quantum mechanics. A classical blockchain is used as a timestamped dictionary that matches the quantum money's serial numbers to users' public keys. The classical blockchain does not store transactions but descriptors of newly minted quantum money. Unlike quantum money, which assumes that a trusted bank mints the currency, the miners of this system can be untrusted. It appends the pair of the serial number of a newly minted quantum money and the user's public key to the classical blockchain, which should fail if the quantum money has already been in the blockchain.

*B.3. Summary of quantum blockchains*

We observed that QKD plays an inseparable role in the design of quantum blockchains. This is because it is the most practical quantum solution we currently have. It enables nodes to generate a shared secret key, which can be used to encrypt other messages. In addition, quantum signatures, hash functions, and new consensus algorithms enhance the security of the proposed blockchain frameworks. However, few approaches propose replacing classical hash functions in the reviewed studies, and the researchers still consider using vulnerable hash functions. This is because attackers still need to achieve more computing power than the rest of the network to replace valid blocks, which is impossible within the next few years with the current quantum technology. In addition, such threats can be mitigated by doubling the current hash sizes.

Moreover, there are fully quantum blockchain approaches that tend to chain quantum states using graph states or entanglements, which promise a high level of security but are impractical in the current stage of quantum computing development. Finally, blockchain-based applications using quantum blockchains have also been experimented with, and some interesting protocols adapting quantum properties have been developed. Table VII summarizes the related studies in quantum blockchains.

TABLE VII
SUMMARY OF STUDIES IN QUANTUM BLOCKCHAINS

| Type | Ref. | Features | Description | Implementations |
|---|---|---|---|---|
| Hybrid quantum blockchains (QKD-based) | [25] | • QKD<br>• Experimentally implemented and tested | The approach uses QKD for key exchange and the Toeplitz hash function for authentication. It uses the Broadcast protocol to achieve a Byzantine agreement. | The authors conducted experimental tests of their protocol using a three-party urban fiber network QKD, and they propose that their protocol could form the basis of scalable blockchain platforms. |
| | [48] | • Multiplexed QKD<br>• A proof-of-concept network infrastructure | The approach uses optical-layer encryption and open APIs and proposes a proof-of-concept infrastructure. The blockchain is classical, but its key exchange processes are done by quantum channels. | The article describes a hybrid quantum-classical approach proposed by JPMorgan. There is no mention of any specific implementation of the protocol. |
| | [188] | • QKD<br>• Quantum secure communication<br>• QHBA | The approach enhances the approach in [25] by replacing the Byzantine protocol with a quantum version, QHBA. It is assumed that at least half of the nodes in the blockchain are honest. Each pair of nodes can share keys by QKD. | The authors assert their approach can be implemented with existing technology, as their blockchain does not use multiparticle entanglements. They have provided a framework to support this claim. |
| | [189] | • QKD<br>• Quantum vote-based consensus<br>• A lottery protocol | This uses a signature scheme based on the Toeplitz hash function. It also proposes a vote-based consensus algorithm. However, it did not analyze the security aspect. | The authors developed a conceptual implementation for the quantum-resistant lottery protocol by presenting a toy script language. |
| Hybrid quantum blockchains (quantum-signature-based) | [176] | • Quantum signature<br>• Quantum teleportation<br>• A set of keys per transaction | The approach uses a quantum signature scheme but creates considerable computation overhead since a set of keys needs to be distributed for each transaction. It did not securely transfer the result of the BSM and did not answer how geographically distributed miners can detect whether the sender is genuine. | The study proposes a transaction mechanism that utilizes quantum digital signatures transported via teleportation to enhance the security of key pairs. Although this research holds promise for providing secure blockchains in the future quantum era, it does not offer any implementations or simulations. |
| | [191] | • Quantum signature scheme<br>• DPoSB Consensus algorithms | The approach is based on the asymmetric quantum encryption $QSCD_{ff}$ and a variant of the delegated proof-of-stake called DPoSB. Node behaviors in creating blocks are rated and used for electing the next block producer. | The analysis shows that integrating their blockchain with modern blockchains can improve security. It also shows that the blockchain offers quantum information-theoretical security. |
| Fully quantum blockchains | [194] | • Entanglement-based blockchains<br>• Identity management<br>• QRND consensus algorithm | The approach presents a quantum-blockchain-based infrastructure for access control and authentication. It uses a chain of quantum states based on entanglement in time to design a quantum blockchain with an application on identity management. It also proposes a consensus algorithm based on QRND and quantum state verification tests. | The paper presents a conceptual design for a quantum blockchain identity framework (QBIF). Although the proposed design is not ready for immediate deployment, it outlines a clear path towards creating a decentralized quantum application to achieve secure pseudonymization, ensuring privacy and preventing forgery. |
| | [195] | • Quantum signature<br>• Quantum teleportation | The approach uses quantum teleportation and quantum signature to design a cryptocurrency system. However, mapping between a coin's classic and quantum parts is not described. The owner of a coin cannot tell whether they have a valid coin. The approach for distributing public keys is not specified. | The authors successfully implemented a decentralized online quantum cash system called qBitcoin, which demonstrated faster transactions and solved privacy concerns compared to Bitcoin. |
| | [196] | • A quantum hash function based on the quantum-inspired quantum walk | The work proposes to use quantum walk and a quantum hash function to chain the blocks in a blockchain and encrypt messages, but it does not present any consensus protocol. It also uses the hash function for communication encryption. | The authors introduced a quantum-inspired blockchain framework and evaluated its effectiveness by conducting randomness tests, sensitivity analysis, and security analysis. They demonstrated its potential |

| | [197] | • A chain of quantum states<br>• Relative phases for consensus | The work proposes to use the quantum-weighted hypergraph states for chaining. Consensus is achieved based on the phases carried by the hyperedges of the quantum-weighted hypergraph states. | The study proposed a protocol for building a quantum blockchain and conducted a proof-of-concept by implementing it on IBM's 5-qubit computer. The effectiveness and security of the protocol were discussed. |
|---|---|---|---|---|
| | | | | application in the context of blockchain IoT-based Smart Cities. |
| | [198] | • A new mining and consensus | The approach presents a blockchain architecture of quantum systems serving as servers and clients. Still, it did not improve the vulnerable parts of current frameworks, hash function, and digital signature. | The authors implemented their interactive mining protocol using a Sagnac interferometer and demonstrated its real-world feasibility. The implementation shows the robustness and energy efficiency of the protocol. |
| | [199] | • Randomness-based consensus<br>• Quantum ZKP | The work proposes a consensus protocol based on quantum ZKP and the randomness of quantum measurement. However, details on using quantum ZKP to verify the random value are not provided. | The paper presents a security analysis of the proposed consensus protocol. However, it does not provide implementations or simulations. |
| Quantum blockchain applications | [200] | • QKD<br>• A voting application | The work proposes a blockchain-based voting application that relies on QKD. However, the use of QKD is limited to storing block hashes in a database. The blockchain framework still uses vulnerable cryptographic components. | The study implements a decentralized blockchain-based e-voting system to address the issue of vote forgery during electronic voting. The implementation demonstrates the realistic and reliable application for e-voting systems. |
| | [182] | • QKD<br>• QBC<br>• QHBA | The work proposes a set of requirements for a voting protocol and shows that it satisfies all of them. However, it did not mention how agents can determine the number of voters. | The paper proposes a simple voting protocol based on Quantum Blockchain. However, it does not include any implementation or simulation. |
| | [190] | • QKD<br>• QBC<br>• A consensus algorithm | The approach presents two protocols for blockchain-based lottery and auction using QKD. It claimed unconditional security, but QBC and consensus algorithms are prone to cheating and dishonest nodes. | The paper introduces a lottery and an auction protocol that rely on quantum bit commitment and quantum blockchain and includes a formal analysis. However, no implementations or simulations were provided. |

## V. COMPARISON

This section examines the pros and cons of post-quantum and quantum blockchains. Then, it compares them in terms of their hash functions, signature schemes, consensus algorithms, security level, and privacy. In this section, quantum blockchains are separated into hybrid and fully quantum blockchains to make a better comparison. Hybrid quantum blockchains are blockchain approaches that utilize both classical and quantum systems. Fully quantum blockchains only use or assume quantum systems and networks.

### A. Retrospective of post-quantum blockchains

Post-quantum blockchains incorporate cryptographic algorithms into classical blockchain technology to safeguard against potential quantum attacks. These systems maintain compatibility with current blockchain infrastructure and are relatively easier to implement and adopt, preserving existing performance levels. However, they do not fully harness the potential of quantum technologies, and their security relies on unproven cryptographic algorithms since post-quantum cryptography's effectiveness remains theoretical.

At first glance, post-quantum blockchains appear to be the most practical and feasible solution for addressing quantum threats. Replacing the widely used cryptography algorithms with post-quantum alternatives may be the easiest to implement. However, most post-quantum blockchains do not address the potential for hash function brute force attacks by quantum computers. While these attacks may not be as significant as cryptography vulnerabilities, it is possible that quantum computers will be so powerful in the future that even hash functions with large key sizes can be broken. In that case, using post-quantum hash functions or applying hybrid-quantum or fully quantum blockchains may be better solutions. However, fully quantum blockchains may still be decades away, considering the small-scale instability of current quantum systems.

### B. Retrospective of hybrid quantum blockchains

Hybrid quantum blockchains blend classical blockchain technology with aspects of quantum cryptography, offering enhanced security and performance. These systems employ a combination of classical and quantum algorithms, enabling the gradual integration of quantum technologies for more manageable adoption. Although hybrid quantum blockchains provide a higher level of security compared to purely classical blockchains, their implementation can be complex due to the integration of both classical and quantum technologies, and they may introduce new attack vectors and vulnerabilities.

Hybrid quantum blockchains primarily rely on QKD, making them more practical than fully quantum blockchains but less efficient than post-quantum blockchains. Typically, one of the first QKD-based blockchain approaches, the quantum-secured blockchains [25], [189], utilize QKD to establish a foundation for a classical blockchain's key exchange network. They differ from conventional blockchain approaches, which use asymmetric cryptography instead of symmetric cryptography. Once all nodes possess symmetric keys, they communicate using those keys. The consensus is a mostly Byzantine agreement, which is basic but effective, though involving a relatively high message overhead [34]. It is believed that optimized consensus algorithms can be developed for better performance, and there have been multiple follow-ups on such consensus algorithms [194], [212], [213]. On the one hand, the challenges of a hybrid-quantum blockchain depend on the specific role of the quantum components in a blockchain. For example, in the approaches mentioned above, QKD addresses the vulnerability of current cryptography, but it brings up consensus challenges.

Moreover, instead of using QKD, hybrid quantum blockchains may concentrate on substituting a specific aspect of blockchain technology with its quantum equivalent. For example, if a technique prioritizes a quantum-based consensus algorithm, it becomes less feasible and less reliable (owing to quantum noise [214]), as evidenced in the protocol for energy-efficient block production [198].

A hybrid-quantum blockchain can be a transition from a classical blockchain to a fully quantum blockchain. As is expected, the most plausible form of the future network infrastructure is a hybrid classical-quantum infrastructure [42], [215], where hybrid-quantum blockchains may be the most suitable to fit in.

*C. Retrospective of fully quantum blockchains*

Fully quantum blockchains are designed from scratch to leverage quantum computing and quantum communication technologies, aiming to maximize the benefits of quantum technologies while delivering exceptional security and performance. These systems are optimized for quantum technologies, offering potential performance improvements and novel applications not possible with classical blockchains. However, their adoption and implementation require significant advances in quantum computing and communication infrastructure, along with high complexity and cost. Additionally, they exhibit limited compatibility with existing blockchain systems.

Another significant constraint of fully quantum blockchains is the lack of practical approaches, as current research is purely theoretical. Implementations are barely available due to the lack of accessible and powerful quantum computers. Even quantum simulations for blockchains are underdeveloped. Researchers thus mainly propose theoretical designs and potential applications under certain assumptions. Additionally, current quantum blockchain approaches are in a very early stage and only focus on mimicking the classical blockchain structure by "chaining" quantum states in a desired order and a distributed manner. This theoretically provides blockchain properties such as decentralization, immutability, transparency, and security but typically assumes one qubit as one quantum state (one block), which is a good starting point but still far from being competitive with current classical blockchains. The number of qubits in a quantum computer is rapidly increasing [216], but still not comparable to classical computers with terabytes of memory, let alone the number of qubits needed to integrate into a blockchain block. This field has many open issues, such as developing foundational components like quantum broadcast networks [217]. Until these foundational components are fully developed, quantum blockchains will remain confined to laboratories. However, the continuous breakthroughs in quantum blockchains are highly inspiring, and they are believed to gradually realize the vision of an unconditionally secure decentralized network.

*D. Comparisons of the technologies*

Each of these technologies offers unique advantages and limitations. Post-quantum blockchains provide a more accessible solution compatible with existing infrastructure, while hybrid quantum blockchains offer a balance between classical and quantum technologies. Fully quantum blockchains, on the other hand, have the potential to deliver the most significant benefits, but their implementation and widespread adoption require substantial advancements in quantum computing and communication technologies. Organizations, businesses, and governments must carefully consider their needs and resources when choosing the most appropriate technology for their specific use cases. The comparison of the three types of blockchains is summarized in Table VIII. Please note that this table content is meant to provide a general comparison and may not cover all aspects or nuances of the respective technologies.

TABLE VIII
COMPARISON OF THE TECHNOLOGIES

| Aspect | Post-quantum blockchains | Hybrid quantum blockchains | Fully Quantum Blockchains |
|---|---|---|---|
| Compatibility | Highly compatible with existing blockchain infrastructures, allowing for seamless integration with minimal changes to current systems. | Moderate compatibility due to the combination of classical and quantum technologies, requiring adjustments to existing infrastructures. | Limited compatibility with existing blockchain systems, necessitating significant modifications and new infrastructures. |
| Security | Theoretically secure against quantum attacks due to the use of post-quantum cryptography. Long-term security is uncertain as these algorithms remain susceptible to future attacks. | Enhanced security levels compared to classical blockchains, leveraging quantum technologies like QKD for secure communication. However, new attack vectors and vulnerabilities may emerge. | Offers exceptional security due to inherent quantum properties, such as the no-cloning theorem and entanglement, making unauthorized duplication or tampering difficult. However, practical security is yet to be demonstrated. |
| Expected Performance | Maintains the current performance levels of classical blockchains, as post-quantum cryptography does not introduce significant overhead. However, there are trade-offs on the key size, signature size, and key generation efficiency. | Potential performance improvements due to the integration of quantum technologies, such as more efficient cryptographic algorithms. However, these benefits may be offset by additional complexity and potential bottlenecks in hybrid systems. | Optimized for quantum technologies, promising significant improvements and novel applications, such as faster transactions and secure computation. However, this potential is currently theoretical and depends on the deployment of quantum systems. |
| Scalability | Easier to adopt and scale, as they rely on existing infrastructure and technology, reducing the need for new hardware or significant changes to current systems. | Moderate adoption and scalability challenges due to the gradual integration of quantum technologies into classical systems, requiring phased updates and a longer transition period. | Difficult to adopt and scale, as they require significant advancements in quantum infrastructure, as well as the development of new standards and protocols for wide-scale deployment. |
| Cost | Relatively lower cost, as they leverage existing blockchain infrastructure and technology, with the primary investment being the integration of post-quantum cryptographic algorithms. | Higher cost than classical blockchains due to the integration of quantum technologies, such as QKD, but lower than fully quantum blockchains, as they still utilize some classical infrastructure. | Substantial cost associated with the deployment of quantum computing and communication infrastructure, as well as the research and development of new protocols and standards to support fully quantum systems. |

Even though there may be significant variations in implementing different blockchains within the same type (such as the many variants of post-quantum blockchains), they generally share similar characteristics specific to that type of blockchain. In Table IX, we compare post-quantum and quantum blockchains in terms of the typical blockchain components and several other factors. Note that while there is currently no explicit research on the privacy aspect of fully quantum blockchains, it is expected that privacy can be achieved using blind quantum computation [218].

TABLE IX
COMPARING THE BLOCKCHAIN COMPONENTS OF THE TECHNOLOGIES

| Type | Identifier | Hash | Signature | Consensus | Privacy | Examples |
|---|---|---|---|---|---|---|
| Post-quantum blockchains | Public-key-based addresses | Post-quantum hash functions | Post-quantum signatures | Any existing classical consensus | ZKPs | [96], [46], [47], [119], [127], [137]. |
| Hybrid blockchains | QKD-based symmetric keys | Classical hash functions | Replaced by symmetric keys | Byzantine agreement | Not mentioned | [25], [48], [188], [189], [191]. |
| Fully quantum blockchains | Identifier by quantum states or entangled qubits, e.g., [219] | Replaced by entanglement | Quantum signatures | QRNG and state verification | Blind computation [218] | [49], [197], [194], [200]. |

## VI. LESSONS LEARNED AND RESEARCH TRENDS

In this section, we summarize the lessons we learned during the review of post-quantum and quantum blockchains and present the research trends we concluded from the industrial products and literature.

### A. Research Challenges and Lessons Learned

In this subsection, we delve into the research challenges in the field and lessons learned through the exploration of post-quantum and quantum blockchains. By identifying the obstacles encountered and the valuable insights gained during the research process, we aim to provide a better understanding of the current state of these technologies, as well as to illuminate potential avenues for future investigation. The discussion of these challenges and lessons learned serves not only to guide researchers and practitioners in their endeavors but also to foster the ongoing development and refinement of these innovative blockchain technologies.

#### A.1. Critical challenges in these fields

As the main objective of these research areas is to prepare for the eventual integration of quantum computers into the current internet, which will occur when they become advanced enough, the domains of post-quantum and quantum blockchains are still in their early stages of development. Through the comparisons presented in the previous section, we have identified several critical research challenges in these areas, which are listed below:

1. The key challenge of post-quantum blockchains is to balance trade-offs among post-quantum cryptographic approaches within the blockchain context. For instance, popular approaches for post-quantum blockchains involve using hash-based or lattice-based cryptographies. While hash-based signature schemes provide greater security, they are less practical. On the other hand, lattice-based schemes offer higher efficiency but possess comparatively lower resistance to quantum attacks. Other post-quantum cryptographic methods also exhibit varying trade-offs, such as those between efficiency and key sizes, as elaborated in Section III.
2. One of the primary challenges in designing hybrid quantum blockchains is to ensure seamless integration with existing classical blockchain infrastructures and networks. This requires developing methods for incorporating quantum technologies (e.g., QKD) without disrupting the operation of current systems. Additionally, achieving a balance between the benefits of quantum technologies and the maintenance of system efficiency and performance is essential, as integrating quantum components may introduce complexities and overhead that could impact overall system performance.
3. One of the significant challenges in fully quantum blockchains is the lack of available implementations. The current state of quantum computing technology is still in its infancy, with only a few small-scale to medium-scale quantum computers available, which are not yet capable of running complex quantum algorithms or supporting quantum blockchains.
4. Raising understanding and awareness of post-quantum and quantum blockchain technologies is a key challenge in driving their adoption and fostering ongoing development. This entails creating educational resources and platforms that cater to various audiences, including researchers, practitioners, and stakeholders from different industries. Collaboration between academia, industry, and policy-makers is essential for developing curricula, workshops, and conferences focusing on post-quantum or quantum blockchains and their implications.

#### A.2. Lessons learned

We have summarized below the lessons we learned through our review and survey of these fields:

1. The extent to which quantum computers will impact the current blockchains is still uncertain. In [220], the authors suggest proof-of-work may be relatively resistant to significant quantum computational acceleration for the next decade (as of 2017). They argue that specialized mining devices based on application-specific integrated circuits (ASIC) are already highly efficient and comparable in speed to current estimates of quantum computer clock speeds. Therefore, if the current mining devices cannot compromise the current blockchain, it is unlikely that near-future quantum computers will be able to.
2. Instead of using a newly-designed post-quantum hash function, we can mitigate the threat of quantum-based hash function attacks simply by increasing the key size of the current hash functions, but this will come at the cost of increased computation and storage.
3. Although post-quantum blockchains currently represent the most practical approach, the long-term resistance of post-quantum cryptography being used against quantum attacks remains uncertain. This is due to the constant evolution and advancements in the field of quantum computing. New developments may challenge previously established concepts [197].
4. Despite the potential threat that quantum computing poses to traditional blockchains, it also has the potential to bring breakthroughs and advancements to the field. Quantum supremacy could potentially lead to the development of an efficient and secure decentralized internet. As discussed, current blockchains are facing limitations in terms of efficiency. Bitcoin's transaction speed is limited to 7 transactions per second (TPS), and a block's final confirmation can take up to an hour (even though the time required to mine a block is approximately 10 minutes, a block must wait for an additional six subsequent blocks before it is considered finalized, as discussed in Subsection II.A.3). This is a significant limitation compared to the high-traffic demands of modern online systems. While new blockchain paradigms promise to increase throughput to millions of TPS, there are currently no practical implementations of these concepts. In the future, quantum computing may provide the necessary breakthroughs in throughput and security to bring about a decentralized internet.

5. In some cases, researchers have proposed a post-quantum signature scheme and applied it directly to the current blockchain paradigm without considering any additional techniques or determining the most appropriate use cases. They have demonstrated the security of their signature schemes but have not considered how they can be best integrated into blockchains.

*B. Research directions*

In this subsection, we summarize the trends in research for post-quantum and quantum blockchains based on the review and comparison of the industrial products and literature surveyed in this paper.

*B.1. Ideal Post-quantum Cryptography*

The ideal solution for post-quantum cryptography and post-quantum blockchains would be the development of a post-quantum cryptographic algorithm that is both efficient and highly resistant to quantum attacks. However, current post-quantum cryptography faces many challenges, such as large key and signature sizes [13] and interactive processes [122]. For example, hash-based signature schemes may offer higher security against quantum threats, but they are less practical than lattice-based signature schemes. Lattice-based schemes offer fast performance and small signatures and keys, but their security levels are uncertain and may be subject to change as research progresses. Moreover, unlike lattice-based and hash-based signatures, multi-variate-quadratic signatures may offer smaller signature sizes and faster performance, but their long-term security is unclear. Code-based signatures can also provide short signatures, but to be secure against quantum computers, they may eventually need to use large key sizes.

Moreover, the isogeny-based and multi-variate-based schemes have relatively large key and signature sizes. In short, existing post-quantum cryptography may not be sufficient for blockchains due to limitations such as large key and signature sizes [221], [222]. Post-quantum blockchains are not yet widely adopted due to the lack of an ideal solution. If a suitable post-quantum cryptographic solution is found, it would be the best candidate for post-quantum blockchains and could be widely applied.

*B.2. Post-quantum blockchain paradigms*

Research in post-quantum blockchains is focused on integrating secure and efficient post-quantum signature schemes into existing blockchain platforms [13], [15], [16], [102]. This includes investigating how the different operations required by these schemes, such as key generation, message signing, and signature verification, may affect the overall functioning of the blockchain [13]. For example, if a signature scheme requires an interactive process between the two communicating parties, a new blockchain process will need to be devised [122]. Nonetheless, we never want an interactive process in a blockchain since it can negatively impact performance. Additionally, efforts are being made to develop new post-quantum consensus mechanisms [148]. Researchers are exploring combining traditional cryptographic methods and quantum-resistant techniques to balance security and performance in post-quantum blockchain solutions [46], [106], [113]. The goal is to utilize the strengths of both types of cryptography to develop a secure blockchain against potential attacks from quantum computers. Research is also looking into alternative proof-of-work mechanisms resistant to quantum attacks and more energy efficient.

*B.3. Post-quantum blockchain applications*

While post-quantum blockchains operate similarly to current blockchains, there are differences in developing dApps based on post-quantum smart contracts [137], [139], [140], [141]. Research is needed to explore how to transition current smart-contract-based dApps to post-quantum versions [106]. Additionally, blockchain application infrastructure, such as access control and authentication, must be developed to provide foundational functions for dApps [140]. This can be achieved by incorporating data, such as ZKPs for identities, into blockchain storage [127]. However, this incorporation depends heavily on the cryptographic security provided by post-quantum versions of hash functions and public key schemes. Thus, understanding how these changes will affect blockchain-based applications and how they should be adapted to post-quantum blockchains remains an important area of research. For example, it is important to investigate the potential of post-quantum blockchains in various sectors such as finance, healthcare, supply chain management, and IoT, where enhanced security against quantum threats is essential. In addition, it is crucial to contribute to developing standards and regulations for post-quantum blockchain technologies, which will be essential for fostering adoption, ensuring interoperability, and maintaining the security and stability of these systems.

*B.4. Hybrid quantum blockchains*

Currently, most hybrid quantum blockchains use QKD as the foundation [25], [48], [188], [189], [191]. These systems layer classical blockchains on top of a QKD network, which can lead to issues such as low scalability and low efficiency. QKD only enables the exchange of symmetric keys, which means applying QKD to a blockchain scheme would require a redesign of the blockchain to be based on symmetric keys [25]. This, however, comes with the disadvantage of increased overhead for exchanging symmetric keys for every pair of nodes, making it less efficient and less scalable. Thus, it is both crucial and sought-after to design a hybrid blockchain structure that optimally leverages the capabilities of QKD, particularly within the context of permissioned or private blockchains.

Conversely, delving into quantum cryptographic methods beyond QKD, such as quantum state encryption or quantum public key protocols, is a promising frontier for innovation within hybrid quantum blockchains. It's also crucial to devise strategies for seamlessly incorporating quantum-resistant cryptographic algorithms and quantum communication tech into established classical blockchain systems without triggering disruptions or compatibility concerns. An enticing prospect is the creation of hybrid quantum blockchains that can interface with classical, post-quantum, and entirely quantum blockchains alike, promoting cross-chain applications and strengthening the overall resilience and adaptability of the blockchain ecosystem.

*B.5. Quantum blockchain paradigms*

Current quantum blockchain paradigms, highly simplified versions of blockchains, are still in the early stages of development [14], [194]. They assume a limited number of qubits in a block and use randomness and verification tests for consensus, which is more of a synchronization protocol for an immutable chain of quantum states [49], [194]. Implementing transactions with quantum states, identifying blockchain users, and associating assets to user accounts are still unclear. These current quantum blockchain approaches have limited functionalities and scalability with high complexity. Fully quantum blockchain technology necessitates quantum computers capable of executing specific tasks at a significantly faster pace and with greater efficiency than classical computers, essentially demonstrating substantial quantum advantage [223]. Therefore, fostering the development of quantum computing also contributes to the progress and potential breakthroughs in quantum blockchains.

While this may be due to the current limitations of quantum computing technology, it is important to conduct a theoretical analysis to explore the possibilities of a pure quantum-based blockchain system [194]. Other methods beyond blockchain can be considered for creating decentralized quantum networks. Blockchain may not always be the ultimate solution for decentralization, and other solutions may be more suitable with quantum computing [42]. Furthermore, the lack of powerful quantum hardware for experimentation makes the development of useful quantum blockchain simulators an important area of research [14].

*B.6. Post-Quantum and quantum supporting infrastructure*

Given the different stages of development of post-quantum and quantum solutions, as well as the different requirements and investments that such solutions can demand, it can be said that adopting these solutions will not be uniform. Hence, while the current internet allows different types of solutions to co-exist, the future of a post-quantum and quantum-supportive internet will depend on its capability to host a heterogeneity of solutions: some post-quantum, some purely quantum, and some still legacy. In the worst scenario, these different technological architectures developing in their silos may result in constituting different internets on their own, partitioning the current (quasi) homogeneous logical vision of the internet. To avoid this, it would be rewarding to start discussing the requirements, limitations, and needs for such a hybrid internet infrastructure that would support, in the long run, the transition to a full post-quantum or quantum world.

## VII. CONCLUSION

Blockchains have become increasingly popular in recent years due to their ability to support decentralized trust and knowledge. However, the rapid advancement of quantum computing poses a significant threat to the security of current blockchain technologies. In this paper, we summarize the core concepts and algorithms of blockchains and quantum computing. We explore how the advancement of quantum computers could potentially compromise the security of current blockchains, specifically through the effects of Shor's and Grover's algorithms. We review existing research on addressing these security threats, which can be divided into two categories: using post-quantum cryptography (i.e., post-quantum blockchains) and using quantum technologies to assist in developing blockchains (i.e., quantum blockchains, including hybrid and fully quantum), and provide a comprehensive overview and comparison of these different types of blockchains and assess the pros and cons of various approaches. We then discussed the open questions and remaining challenges and highlighted the research trends in these two fields. Overall, it appears that the development of quantum-resistant blockchains is necessary to ensure the security and integrity of decentralized systems in the face of the growing threat from quantum computing. It is anticipated that quantum computing will eventually be advanced enough to solve practical problems. Thus, investing efforts in addressing the challenges and open issues related to combining quantum computing and blockchains is important. Given the extent of scope and richness in details, we believe this work is posed to become a blueprint for development and research in post-quantum and quantum blockchains.

ACKNOWLEDGMENT

This work has been supported under the grant ID NPRP11S-0109-180242, funded by the Qatar National Research Fund (a member of The Qatar Foundation). The statements made herein are solely the responsibility of the authors.

REFERENCES

[1] S. Nakamoto, "Bitcoin: A Peer-to-Peer Electronic Cash System," *Bitcoin*. https://bitcoin.org/bitcoin.pdf (accessed Jan. 29, 2023).
[2] K. Gai, J. Guo, L. Zhu, and S. Yu, "Blockchain Meets Cloud Computing: A Survey," *IEEE Commun. Surv. Tutor.*, vol. 22, no. 3, pp. 2009–2030, 2020, doi: 10.1109/COMST.2020.2989392.
[3] K. Yue *et al.*, "A Survey of Decentralizing Applications via Blockchain: The 5G and Beyond Perspective," *IEEE Commun. Surv. Tutor.*, vol. 23, no. 4, pp. 2191–2217, 2021, doi: 10.1109/COMST.2021.3115797.
[4] B. K. Mohanta, S. S. Panda, and D. Jena, "An Overview of Smart Contract and Use Cases in Blockchain Technology," in *2018 9th International Conference on Computing, Communication and Networking Technologies (ICCCNT)*, Jul. 2018, pp. 1–4. doi: 10.1109/ICCCNT.2018.8494045.
[5] A. Reyna, C. Martín, J. Chen, E. Soler, and M. Díaz, "On blockchain and its integration with IoT. Challenges and opportunities," *Future Gener. Comput. Syst.*, vol. 88, pp. 173–190, Nov. 2018, doi: 10.1016/j.future.2018.05.046.
[6] P. W. Shor, "Algorithms for quantum computation: discrete logarithms and factoring," in *Proceedings 35th Annual Symposium on Foundations*

*of Computer Science*, Nov. 1994, pp. 124–134. doi: 10.1109/SFCS.1994.365700.

[7] L. K. Grover, "A fast quantum mechanical algorithm for database search," in *Proceedings of the twenty-eighth annual ACM symposium on Theory of computing - STOC '96*, Philadelphia, Pennsylvania, United States: ACM Press, 1996, pp. 212–219. doi: 10.1145/237814.237866.

[8] D. J. Bernstein and T. Lange, "Post-quantum cryptography," *Nature*, vol. 549, no. 7671, Art. no. 7671, Sep. 2017, doi: 10.1038/nature23461.

[9] R. Alléaume *et al.*, "Using quantum key distribution for cryptographic purposes: A survey," *Theor. Comput. Sci.*, vol. 560, pp. 62–81, Dec. 2014, doi: 10.1016/j.tcs.2014.09.018.

[10] Y. Cao, Y. Zhao, Q. Wang, J. Zhang, S. X. Ng, and L. Hanzo, "The Evolution of Quantum Key Distribution Networks: On the Road to the Qinternet," *IEEE Commun. Surv. Tutor.*, vol. 24, no. 2, pp. 839–894, 2022, doi: 10.1109/COMST.2022.3144219.

[11] C. Simon, "Towards a global quantum network," *Nat. Photonics*, vol. 11, no. 11, Art. no. 11, Nov. 2017, doi: 10.1038/s41566-017-0032-0.

[12] A. Singh, K. Dev, H. Siljak, H. D. Joshi, and M. Magarini, "Quantum Internet—Applications, Functionalities, Enabling Technologies, Challenges, and Research Directions," *IEEE Commun. Surv. Tutor.*, vol. 23, no. 4, pp. 2218–2247, 2021, doi: 10.1109/COMST.2021.3109944.

[13] T. M. Fernández-Caramès and P. Fraga-Lamas, "Towards Post-Quantum Blockchain: A Review on Blockchain Cryptography Resistant to Quantum Computing Attacks," *IEEE Access*, vol. 8, pp. 21091–21116, 2020, doi: 10.1109/ACCESS.2020.2968985.

[14] M. Edwards, A. Mashatan, and S. Ghose, "A review of quantum and hybrid quantum/classical blockchain protocols," *Quantum Inf. Process.*, vol. 19, no. 6, p. 184, May 2020, doi: 10.1007/s11128-020-02672-y.

[15] A.-T. Ciulei, M.-C. Crețu, and E. Simion, "Preparation for Post-Quantum era: a survey about blockchain schemes from a post-quantum perspective." 2022. Accessed: Apr. 27, 2023. [Online]. Available: https://eprint.iacr.org/2022/026

[16] M. Buser *et al.*, "A Survey on Exotic Signatures for Post-Quantum Blockchain: Challenges & Research Directions," *ACM Comput. Surv.*, Dec. 2022, doi: 10.1145/3572771.

[17] R. K. Dhanaraj, V. Rajasekar, S. H. Islam, B. Balusamy, and C.-H. Hsu, Eds., *Quantum Blockchain: An Emerging Cryptographic Paradigm*, 1st edition. Wiley-Scrivener, 2022.

[18] B. Bhushan, P. Sinha, K. M. Sagayam, and A. J, "Untangling blockchain technology: A survey on state of the art, security threats, privacy services, applications and future research directions," *Comput. Electr. Eng.*, vol. 90, p. 106897, Mar. 2021, doi: 10.1016/j.compeleceng.2020.106897.

[19] R. L. Rivest, A. Shamir, and L. Adleman, "A method for obtaining digital signatures and public-key cryptosystems," *Commun. ACM*, vol. 21, no. 2, pp. 120–126, Feb. 1978, doi: 10.1145/359340.359342.

[20] Information Technology Laboratory, "Digital Signature Standard (DSS)," National Institute of Standards and Technology, NIST FIPS 186-4, Jul. 2013. doi: 10.6028/NIST.FIPS.186-4.

[21] D. Johnson, A. Menezes, and S. Vanstone, "The Elliptic Curve Digital Signature Algorithm (ECDSA)," *Int. J. Inf. Secur.*, vol. 1, no. 1, pp. 36–63, Aug. 2001, doi: 10.1007/s102070100002.

[22] Q. ShenTu and J. Yu, "Research on Anonymization and De-anonymization in the Bitcoin System." arXiv, Oct. 27, 2015. doi: 10.48550/arXiv.1510.07782.

[23] Q. Feng, D. He, S. Zeadally, M. K. Khan, and N. Kumar, "A survey on privacy protection in blockchain system," *J. Netw. Comput. Appl.*, vol. 126, pp. 45–58, Jan. 2019, doi: 10.1016/j.jnca.2018.10.020.

[24] Q. H. Dang, "Secure Hash Standard," National Institute of Standards and Technology, NIST FIPS 180-4, Jul. 2015. doi: 10.6028/NIST.FIPS.180-4.

[25] E. O. Kiktenko *et al.*, "Quantum-secured blockchain," *Quantum Sci. Technol.*, vol. 3, no. 3, p. 035004, May 2018, doi: 10.1088/2058-9565/aabc6b.

[26] V. Gheorghiu, S. Gorbunov, M. Mosca, and B. Munson, "Quantum-Proofing the Blockchain," *University of Waterloo*, 2017. https://evolutionq.com/quantum-safe-publications/mosca_quantum-proofing-the-blockchain_blockchain-research-institute.pdf (accessed Jan. 29, 2023).

[27] L. Lamport, R. Shostak, and M. Pease, "The Byzantine generals problem," in *Concurrency: the Works of Leslie Lamport*, New York, NY, USA: Association for Computing Machinery, 2019, pp. 203–226. Accessed: Jan. 29, 2023. [Online]. Available: https://doi.org/10.1145/3335772.3335936

[28] M. Fitzi, N. Gisin, and U. Maurer, "Quantum Solution to the Byzantine Agreement Problem," *Phys. Rev. Lett.*, vol. 87, no. 21, p. 217901, Nov. 2001, doi: 10.1103/PhysRevLett.87.217901.

[29] "Proof of Work and how it solves the Byzantine Generals Problem | The Radix Blog | Radix DLT," *Radix*. https://www.radixdlt.com/post/what-is-proof-of-work (accessed Jan. 29, 2023).

[30] "Proof-of-stake (PoS)," *ethereum.org*. https://ethereum.org/en/developers/docs/consensus-mechanisms/pos/ (accessed Jan. 29, 2023).

[31] C. T. Nguyen, D. T. Hoang, D. N. Nguyen, D. Niyato, H. T. Nguyen, and E. Dutkiewicz, "Proof-of-Stake Consensus Mechanisms for Future Blockchain Networks: Fundamentals, Applications and Opportunities," *IEEE Access*, vol. 7, pp. 85727–85745, 2019, doi: 10.1109/ACCESS.2019.2925010.

[32] "What Is Delegated Proof of Stake (DPoS)?," *Bitcoin - Open source P2P money*. https://learn.bybit.com/blockchain/delegated-proof-of-stake-dpos/ (accessed Jan. 29, 2023).

[33] Y. Xiao, N. Zhang, W. Lou, and Y. T. Hou, "A Survey of Distributed Consensus Protocols for Blockchain Networks," *IEEE Commun. Surv. Tutor.*, vol. 22, no. 2, pp. 1432–1465, 2020, doi: 10.1109/COMST.2020.2969706.

[34] G. Zhang *et al.*, "Reaching Consensus in the Byzantine Empire: A Comprehensive Review of BFT Consensus Algorithms." arXiv, Aug. 27, 2022. doi: 10.48550/arXiv.2204.03181.

[35] "Introduction to smart contracts," *ethereum.org*. https://ethereum.org/en/developers/docs/smart-contracts/ (accessed Jan. 29, 2023).

[36] N. Kappert, E. Karger, and M. Kureljusic, "Quantum Computing - The Impending End for the Blockchain?" Rochester, NY, Jun. 24, 2021. Accessed: Jan. 29, 2023. [Online]. Available: https://papers.ssrn.com/abstract=4075591

[37] A. Krishnakumar, *Quantum Computing and Blockchain in Business: Exploring the applications, challenges, and collision of quantum computing and blockchain*. Packt Publishing, 2020.

[38] A. Einstein, B. Podolsky, and N. Rosen, "Can Quantum-Mechanical Description of Physical Reality Be Considered Complete?," *Phys. Rev.*, vol. 47, no. 10, pp. 777–780, May 1935, doi: 10.1103/PhysRev.47.777.

[39] "The Nobel Prize in Physics 2022," *NobelPrize.org*. https://www.nobelprize.org/prizes/physics/2022/press-release/ (accessed Jan. 29, 2023).

[40] P. A. M. Dirac and R. H. Fowler, "The quantum theory of the electron," *Proc. R. Soc. Lond. Ser. Contain. Pap. Math. Phys. Character*, vol. 117, no. 778, pp. 610–624, Jan. 1997, doi: 10.1098/rspa.1928.0023.

[41] J. R. Johansson, P. D. Nation, and F. Nori, "QuTiP 2: A Python framework for the dynamics of open quantum systems," *Comput. Phys. Commun.*, vol. 184, no. 4, pp. 1234–1240, Apr. 2013, doi: 10.1016/j.cpc.2012.11.019.

[42] Z. Yang, M. Zolanvari, and R. Jain, "A Survey of Important Issues in Quantum Computing and Communications," *IEEE Commun. Surv. Tutor.*, vol. 25, no. 2, pp. 1059–1094, 2023, doi: 10.1109/COMST.2023.3254481.

[43] M. A. Nielsen and I. L. Chuang, *Quantum Computation and Quantum Information: 10th Anniversary Edition*, Anniversary edition. Cambridge ; New York: Cambridge University Press, 2011.

[44] "Ethereum Foundation," *ethereum.org*. https://ethereum.org (accessed Jan. 29, 2023).

[45] J. J. Kearney and C. A. Perez-Delgado, "Vulnerability of blockchain technologies to quantum attacks," *Array*, vol. 10, p. 100065, Jul. 2021, doi: 10.1016/j.array.2021.100065.

[46] "QRL: The Quantum Resistant Ledger," *QRL*. https://www.theqrl.org/ (accessed Jan. 29, 2023).

[47] N. Anhao, "Bitcoin post-quantum," *bitcoinpq. org, Tech. Rep*, 2018. https://bitcoinpq.org/download/bitcoinpq-whitepaper-english.pdf (accessed Jan. 29, 2023).

[48] "JPMorgan unveils research on quantum resistant blockchain network," *Cointelegraph*, Feb. 18, 2022. https://cointelegraph.com/news/jpmorgan-unveils-research-on-quantum-resistant-blockchain-network (accessed Jan. 29, 2023).

[49] D. Rajan and M. Visser, "Quantum Blockchain Using Entanglement in Time," *Quantum Rep.*, vol. 1, no. 1, Art. no. 1, Sep. 2019, doi: 10.3390/quantum1010002.

[50] R. C. Merkle, "A Digital Signature Based on a Conventional Encryption Function," in *Advances in Cryptology — CRYPTO '87*, C. Pomerance,

[50] Ed., in Lecture Notes in Computer Science. Berlin, Heidelberg: Springer, 1988, pp. 369–378. doi: 10.1007/3-540-48184-2_32.
[51] M. Ajtai, "Generating hard instances of lattice problems," in *Proceedings of the twenty-eighth annual ACM symposium on Theory of computing*, 1996, pp. 99–108.
[52] M. S. Şahin and S. Akleylek, "A survey of quantum secure group signature schemes: Lattice-based approach," *J. Inf. Secur. Appl.*, vol. 73, p. 103432, Mar. 2023, doi: 10.1016/j.jisa.2023.103432.
[53] R. Overbeck and N. Sendrier, "Code-based cryptography," in *Post-Quantum Cryptography*, D. J. Bernstein, J. Buchmann, and E. Dahmen, Eds., Berlin, Heidelberg: Springer, 2009, pp. 95–145. doi: 10.1007/978-3-540-88702-7_4.
[54] E. Berlekamp, R. McEliece, and H. van Tilborg, "On the inherent intractability of certain coding problems (Corresp.)," *IEEE Trans. Inf. Theory*, vol. 24, no. 3, pp. 384–386, May 1978, doi: 10.1109/TIT.1978.1055873.
[55] T. Richardson and R. Urbanke, *Modern Coding Theory*. Cambridge University Press, 2008.
[56] J. Ding and B.-Y. Yang, "Multivariate Public Key Cryptography," in *Post-Quantum Cryptography*, D. J. Bernstein, J. Buchmann, and E. Dahmen, Eds., Berlin, Heidelberg: Springer, 2009, pp. 193–241. doi: 10.1007/978-3-540-88702-7_6.
[57] D. Jao and L. De Feo, "Towards Quantum-Resistant Cryptosystems from Supersingular Elliptic Curve Isogenies," in *Post-Quantum Cryptography*, B.-Y. Yang, Ed., in Lecture Notes in Computer Science. Berlin, Heidelberg: Springer, 2011, pp. 19–34. doi: 10.1007/978-3-642-25405-5_2.
[58] W. Castryck, T. Lange, C. Martindale, L. Panny, and J. Renes, "CSIDH: An Efficient Post-Quantum Commutative Group Action," in *Advances in Cryptology – ASIACRYPT 2018*, T. Peyrin and S. Galbraith, Eds., in Lecture Notes in Computer Science. Cham: Springer International Publishing, 2018, pp. 395–427. doi: 10.1007/978-3-030-03332-3_15.
[59] O. Regev, "On lattices, learning with errors, random linear codes, and cryptography," *J. ACM*, vol. 56, no. 6, p. 34:1-34:40, Sep. 2009, doi: 10.1145/1568318.1568324.
[60] I. T. L. Computer Security Division, "Post-Quantum Cryptography | CSRC | CSRC," *CSRC | NIST*, Jan. 03, 2017. https://csrc.nist.gov/Projects/post-quantum-cryptography (accessed Jan. 29, 2023).
[61] C. Boutin, "NIST Announces First Four Quantum-Resistant Cryptographic Algorithms," *NIST*, Jul. 05, 2022. https://www.nist.gov/news-events/news/2022/07/nist-announces-first-four-quantum-resistant-cryptographic-algorithms (accessed Jan. 29, 2023).
[62] R. Avanzi *et al.*, "CRYSTALS-Kyber algorithm specifications and supporting documentation," *NIST PQC Round*, vol. 2, no. 4, pp. 1–43, 2017.
[63] L. Ducas *et al.*, "CRYSTALS-Dilithium: A Lattice-Based Digital Signature Scheme," *IACR Trans. Cryptogr. Hardw. Embed. Syst.*, pp. 238–268, Feb. 2018, doi: 10.13154/tches.v2018.i1.238-268.
[64] P.-A. Fouque *et al.*, "Falcon: Fast-Fourier lattice-based compact signatures over NTRU," *Submiss. NIST's Post-Quantum Cryptogr. Stand. Process*, vol. 36, no. 5, 2018.
[65] P. Schwabe, "SPHINCS+," *SPHINCS+*. https://sphincs.org/ (accessed Jan. 29, 2023).
[66] J. Buchmann, E. Dahmen, and A. Hülsing, "XMSS - A Practical Forward Secure Signature Scheme Based on Minimal Security Assumptions," in *Post-Quantum Cryptography*, B.-Y. Yang, Ed., in Lecture Notes in Computer Science. Berlin, Heidelberg: Springer, 2011, pp. 117–129. doi: 10.1007/978-3-642-25405-5_8.
[67] L. Lamport, "Constructing Digital Signatures from a One Way Function," CSL-98, Oct. 1979. [Online]. Available: https://www.microsoft.com/en-us/research/publication/constructing-digital-signatures-one-way-function/
[68] F. Shahid and A. Khan, "Smart Digital Signatures (SDS): A post-quantum digital signature scheme for distributed ledgers," *Future Gener. Comput. Syst.*, vol. 111, pp. 241–253, Oct. 2020, doi: 10.1016/j.future.2020.04.042.
[69] A. Hülsing, "W-OTS+ – Shorter Signatures for Hash-Based Signature Schemes," in *Progress in Cryptology – AFRICACRYPT 2013*, A. Youssef, A. Nitaj, and A. E. Hassanien, Eds., in Lecture Notes in Computer Science. Berlin, Heidelberg: Springer, 2013, pp. 173–188. doi: 10.1007/978-3-642-38553-7_10.
[70] D. J. Bernstein *et al.*, "SPHINCS: Practical Stateless Hash-Based Signatures," in *Advances in Cryptology -- EUROCRYPT 2015*, E. Oswald and M. Fischlin, Eds., in Lecture Notes in Computer Science. Berlin, Heidelberg: Springer, 2015, pp. 368–397. doi: 10.1007/978-3-662-46800-5_15.
[71] J. Hoffstein, J. Pipher, and J. H. Silverman, "NTRU: A ring-based public key cryptosystem," in *Algorithmic Number Theory*, J. P. Buhler, Ed., in Lecture Notes in Computer Science. Berlin, Heidelberg: Springer, 1998, pp. 267–288. doi: 10.1007/BFb0054868.
[72] C. Peikert, "Public-key cryptosystems from the worst-case shortest vector problem: extended abstract," in *Proceedings of the forty-first annual ACM symposium on Theory of computing*, in STOC '09. New York, NY, USA: Association for Computing Machinery, May 2009, pp. 333–342. doi: 10.1145/1536414.1536461.
[73] J. H. Cheon, D. Kim, J. Lee, and Y. Song, "Lizard: Cut Off the Tail! A Practical Post-quantum Public-Key Encryption from LWE and LWR," in *Security and Cryptography for Networks*, D. Catalano and R. De Prisco, Eds., in Lecture Notes in Computer Science. Cham: Springer International Publishing, 2018, pp. 160–177. doi: 10.1007/978-3-319-98113-0_9.
[74] Z. Brakerski and V. Vaikuntanathan, "Efficient Fully Homomorphic Encryption from (Standard) LWE," *SIAM J. Comput.*, vol. 43, no. 2, pp. 831–871, Jan. 2014, doi: 10.1137/120868669.
[75] Z. Brakerski, C. Gentry, and S. Halevi, "Packed Ciphertexts in LWE-Based Homomorphic Encryption," in *Public-Key Cryptography – PKC 2013*, K. Kurosawa and G. Hanaoka, Eds., in Lecture Notes in Computer Science. Berlin, Heidelberg: Springer, 2013, pp. 1–13. doi: 10.1007/978-3-642-36362-7_1.
[76] J. Ding, X. Xie, and X. Lin, "A Simple Provably Secure Key Exchange Scheme Based on the Learning with Errors Problem." 2012. Accessed: Jan. 29, 2023. [Online]. Available: https://eprint.iacr.org/2012/688
[77] S. Fluhrer, "Cryptanalysis of ring-LWE based key exchange with key share reuse." 2016. Accessed: Jan. 29, 2023. [Online]. Available: https://eprint.iacr.org/2016/085
[78] V. Lyubashevsky, C. Peikert, and O. Regev, "On Ideal Lattices and Learning with Errors over Rings," *J. ACM*, vol. 60, no. 6, p. 43:1-43:35, Nov. 2013, doi: 10.1145/2535925.
[79] T. Güneysu, V. Lyubashevsky, and T. Pöppelmann, "Practical Lattice-Based Cryptography: A Signature Scheme for Embedded Systems," in *Cryptographic Hardware and Embedded Systems – CHES 2012*, E. Prouff and P. Schaumont, Eds., in Lecture Notes in Computer Science. Berlin, Heidelberg: Springer, 2012, pp. 530–547. doi: 10.1007/978-3-642-33027-8_31.
[80] V. Lyubashevsky, "Fiat-Shamir with Aborts: Applications to Lattice and Factoring-Based Signatures | SpringerLink," presented at the Advances in Cryptology–ASIACRYPT 2009: 15th International Conference on the Theory and Application of Cryptology and Information Security, Tokyo, Japan: Springer, pp. 598–616. Accessed: Jan. 29, 2023. [Online]. Available: https://link.springer.com/chapter/10.1007/978-3-642-10366-7_35
[81] R. J. McEliece, "A public-key cryptosystem based on algebraic," *Deep Space Netw. Prog. Rep.*, vol. 42–44, pp. 114–116, 1978.
[82] H. Niederreiter, "Knapsack-type cryptosystems and algebraic coding theory," *SciSpace*, vol. 15, no. 2, pp. 157–166, Jan. 1986.
[83] M. Naor and M. Yung, "Universal one-way hash functions and their cryptographic applications," in *Proceedings of the twenty-first annual ACM symposium on Theory of computing*, in STOC '89. New York, NY, USA: Association for Computing Machinery, Feb. 1989, pp. 33–43. doi: 10.1145/73007.73011.
[84] J.-C. Faugère and A. Joux, "Algebraic Cryptanalysis of Hidden Field Equation (HFE) Cryptosystems Using Gröbner Bases," in *Advances in Cryptology - CRYPTO 2003*, D. Boneh, Ed., in Lecture Notes in Computer Science. Berlin, Heidelberg: Springer, 2003, pp. 44–60. doi: 10.1007/978-3-540-45146-4_3.
[85] L. Bettale, J.-C. Faugère, and L. Perret, "Cryptanalysis of HFE, multi-HFE and variants for odd and even characteristic," *Des. Codes Cryptogr.*, vol. 69, no. 1, pp. 1–52, Oct. 2013, doi: 10.1007/s10623-012-9617-2.
[86] N. T. Courtois, "The security of hidden field equations (HFE)," in *Topics in Cryptology—CT-RSA 2001: The Cryptographers' Track at RSA Conference 2001 San Francisco, CA, USA, April 8–12, 2001 Proceedings*, Springer, 2001, pp. 266–281.
[87] J. Patarin, N. Courtois, and L. Goubin, "FLASH, a Fast Multivariate Signature Algorithm," in *Topics in Cryptology — CT-RSA 2001*, D. Naccache, Ed., in Lecture Notes in Computer Science. Berlin, Heidelberg: Springer, 2001, pp. 298–307. doi: 10.1007/3-540-45353-9_22.

[88] J. Patarin, "Hidden fields equations (HFE) and isomorphisms of polynomials (IP): Two new families of asymmetric algorithms," in *Advances in Cryptology—EUROCRYPT'96: International Conference on the Theory and Application of Cryptographic Techniques Saragossa, Spain, May 12–16, 1996 Proceedings 15*, Springer, 1996, pp. 33–48.

[89] A. Kipnis and A. Shamir, "Cryptanalysis of the oil and vinegar signature scheme," in *Advances in Cryptology — CRYPTO '98*, H. Krawczyk, Ed., in Lecture Notes in Computer Science. Berlin, Heidelberg: Springer, 1998, pp. 257–266. doi: 10.1007/BFb0055733.

[90] A. Kipnis, J. Patarin, and L. Goubin, "Unbalanced oil and vinegar signature schemes," in *Eurocrypt*, Springer, 1999, pp. 206–222.

[91] W. Diffie and M. E. Hellman, "New Directions in Cryptography," in *Democratizing Cryptography: The Work of Whitfield Diffie and Martin Hellman*, 1st ed.New York, NY, USA: Association for Computing Machinery, 2022, pp. 365–390. Accessed: Jan. 29, 2023. [Online]. Available: https://doi.org/10.1145/3549993.3550007

[92] S. Alghamdi and S. Almuhammadi, "The Future of Cryptocurrency Blockchains in the Quantum Era," in *2021 IEEE International Conference on Blockchain (Blockchain)*, Dec. 2021, pp. 544–551. doi: 10.1109/Blockchain53845.2021.00082.

[93] K. Ikeda, "Chapter Seven - Security and Privacy of Blockchain and Quantum Computation," in *Advances in Computers*, P. Raj and G. C. Deka, Eds., in Blockchain Technology: Platforms, Tools and Use Cases, vol. 111. Elsevier, 2018, pp. 199–228. doi: 10.1016/bs.adcom.2018.03.003.

[94] IBM, "IBM builds the technology you need for a quantum-safe future," *IBM Quantum Computing | Quantum Safe*, Oct. 01, 2015. https://www.ibm.com/quantum/quantum-safe (accessed Jan. 29, 2023).

[95] I. Corporation, "Crypto-Agile and Quantum-safe Security Solutions," *ISARA Corporation*. https://www.isara.com// (accessed Jan. 29, 2023).

[96] "Hello future," *Hedera*. https://hedera.com/ (accessed Jan. 29, 2023).

[97] "IOTA Tangle," *IOTA*. https://www.iota.org (accessed Jan. 29, 2023).

[98] S. Popov, "The tangle," *IOTA White paper*, 2018. http://cryptoverze.s3.us-east-2.amazonaws.com/wp-content/uploads/2018/11/10012054/IOTA-MIOTA-Whitepaper.pdf (accessed Jan. 29, 2023).

[99] "IOTA still wants to build a better blockchain and get it right this time," *ZDNET*. https://www.zdnet.com/finance/blockchain/iota-still-wants-to-build-a-better-blockchain-and-get-it-right-this-time/ (accessed Jan. 29, 2023).

[100] J. Buchmann, E. Dahmen, S. Ereth, A. Hülsing, and M. Rückert, "On the Security of the Winternitz One-Time Signature Scheme.," *Africacrypt*, vol. 11, pp. 363–378, 2011.

[101] E. Heilman *et al.*, "Cryptanalysis of Curl-P and Other Attacks on the IOTA Cryptocurrency." 2019. Accessed: Jan. 29, 2023. [Online]. Available: https://eprint.iacr.org/2019/344

[102] Abelian, "A Quantum-Resistant Cryptocurrency Balancing Privacy and Accountability," *Abelian White paper*. https://www.abelian.info/whitepaper.pdf (accessed Jan. 29, 2023).

[103] A. Langlois and D. Stehlé, "Worst-case to average-case reductions for module lattices," *Des. Codes Cryptogr.*, vol. 75, no. 3, pp. 565–599, Jun. 2015, doi: 10.1007/s10623-014-9938-4.

[104] A. K. Lenstra and E. R. Verheul, "Selecting cryptographic key sizes," in *Public Key Cryptography: Third International Workshop on Practice and Theory in Public Key Cryptosystems, PKC 2000, Melbourne, Victoria, Australia, January 18-20, 2000. Proceedings 3*, Springer, 2000, pp. 446–465.

[105] T. Renduchintala, H. Alfauri, Z. Yang, R. D. Pietro, and R. Jain, "A Survey of Blockchain Applications in the FinTech Sector," *J. Open Innov. Technol. Mark. Complex.*, vol. 8, no. 4, Art. no. 4, Dec. 2022, doi: 10.3390/joitmc8040185.

[106] "Corda | Leading DLT Platform for Regulated Industries," *Corda*. https://corda.net/ (accessed Jan. 29, 2023).

[107] K. Chalkias, J. Brown, M. Hearn, T. Lillehagen, I. Nitto, and T. Schroeter, "Blockchained Post-Quantum Signatures," in *2018 IEEE International Conference on Internet of Things (iThings) and IEEE Green Computing and Communications (GreenCom) and IEEE Cyber, Physical and Social Computing (CPSCom) and IEEE Smart Data (SmartData)*, Jul. 2018, pp. 1196–1203. doi: 10.1109/Cybermatics_2018.2018.00213.

[108] S. Josefsson and I. Liusvaara, "Edwards-Curve Digital Signature Algorithm (EdDSA)," Internet Engineering Task Force, Request for Comments RFC 8032, Jan. 2017. doi: 10.17487/RFC8032.

[109] J. Polge, J. Robert, and Y. Le Traon, "Permissioned blockchain frameworks in the industry: A comparison," *ICT Express*, vol. 7, no. 2, pp. 229–233, Jun. 2021, doi: 10.1016/j.icte.2020.09.002.

[110] C.-Y. Li, X.-B. Chen, Y.-L. Chen, Y.-Y. Hou, and J. Li, "A New Lattice-Based Signature Scheme in Post-Quantum Blockchain Network," *IEEE Access*, vol. 7, pp. 2026–2033, 2019, doi: 10.1109/ACCESS.2018.2886554.

[111] D. Cash, D. Hofheinz, E. Kiltz, and C. Peikert, "Bonsai Trees, or How to Delegate a Lattice Basis," *J. Cryptol.*, vol. 25, no. 4, pp. 601–639, Oct. 2012, doi: 10.1007/s00145-011-9105-2.

[112] Y.-L. Gao, X.-B. Chen, Y.-L. Chen, Y. Sun, X.-X. Niu, and Y.-X. Yang, "A Secure Cryptocurrency Scheme Based on Post-Quantum Blockchain," *IEEE Access*, vol. 6, pp. 27205–27213, 2018, doi: 10.1109/ACCESS.2018.2827203.

[113] M. C. Semmouni, A. Nitaj, and M. Belkasmi, "Bitcoin Security with Post Quantum Cryptography," in *Networked Systems*, M. F. Atig and A. A. Schwarzmann, Eds., in Lecture Notes in Computer Science. Cham: Springer International Publishing, 2019, pp. 281–288. doi: 10.1007/978-3-030-31277-0_19.

[114] P. S. L. M. Barreto, P. Longa, M. Naehrig, J. E. Ricardini, and G. Zanon, "Sharper Ring-LWE Signatures." 2016. Accessed: Jan. 29, 2023. [Online]. Available: https://eprint.iacr.org/2016/1026

[115] J.-P. Aumasson, S. Neves, Z. Wilcox-O'Hearn, and C. Winnerlein, "BLAKE2: Simpler, Smaller, Fast as MD5," in *Applied Cryptography and Network Security*, M. Jacobson, M. Locasto, P. Mohassel, and R. Safavi-Naini, Eds., in Lecture Notes in Computer Science. Berlin, Heidelberg: Springer, 2013, pp. 119–135. doi: 10.1007/978-3-642-38980-1_8.

[116] M. J. Dworkin, "SHA-3 Standard: Permutation-Based Hash and Extendable-Output Functions," National Institute of Standards and Technology, NIST FIPS PUB 202, Aug. 2015. Accessed: Jan. 29, 2023. [Online]. Available: https://www.nist.gov/publications/sha-3-standard-permutation-based-hash-and-extendable-output-functions

[117] F. Shahid, A. Khan, S. U. R. Malik, and K.-K. R. Choo, "WOTS-S: A Quantum Secure Compact Signature Scheme for Distributed Ledger," *Inf. Sci.*, vol. 539, pp. 229–249, Oct. 2020, doi: 10.1016/j.ins.2020.05.024.

[118] R. Saha *et al.*, "A Blockchain Framework in Post-Quantum Decentralization," *IEEE Trans. Serv. Comput.*, vol. 16, no. 1, pp. 1–12, Jan. 2023, doi: 10.1109/TSC.2021.3116896.

[119] R. Shen, H. Xiang, X. Zhang, B. Cai, and T. Xiang, "Application and Implementation of Multivariate Public Key Cryptosystem in Blockchain (Short Paper)," in *Collaborative Computing: Networking, Applications and Worksharing*, X. Wang, H. Gao, M. Iqbal, and G. Min, Eds., in Lecture Notes of the Institute for Computer Sciences, Social Informatics and Telecommunications Engineering. Cham: Springer International Publishing, 2019, pp. 419–428. doi: 10.1007/978-3-030-30146-0_29.

[120] "Zero-Knowledge Proofs: STARKs vs SNARKs," *ConsenSys*. https://consensys.net/blog/blockchain-explained/zero-knowledge-proofs-starks-vs-snarks/ (accessed Jan. 29, 2023).

[121] "Zero-Knowledge rollups," *ethereum.org*. https://ethereum.org/en/developers/docs/scaling/zk-rollups/ (accessed Jan. 29, 2023).

[122] C. P. Schnorr, "Efficient Identification and Signatures for Smart Cards," in *Advances in Cryptology — CRYPTO' 89 Proceedings*, G. Brassard, Ed., in Lecture Notes in Computer Science. New York, NY: Springer, 1990, pp. 239–252. doi: 10.1007/0-387-34805-0_22.

[123] A. Fiat and A. Shamir, "How to Prove Yourself: Practical Solutions to Identification and Signature Problems.," in *Crypto*, Springer, 1986, pp. 186–194.

[124] E. Ben-Sasson, I. Bentov, Y. Horesh, and M. Riabzev, "Scalable, transparent, and post-quantum secure computational integrity." 2018. Accessed: Jan. 29, 2023. [Online]. Available: https://eprint.iacr.org/2018/046

[125] N. Bitansky, R. Canetti, A. Chiesa, and E. Tromer, "From extractable collision resistance to succinct non-interactive arguments of knowledge, and back again," in *Proceedings of the 3rd Innovations in Theoretical Computer Science Conference*, in ITCS '12. New York, NY, USA: Association for Computing Machinery, Jan. 2012, pp. 326–349. doi: 10.1145/2090236.2090263.

[126] "STARKWARE," *Starkware*. https://starkware.co/stark/ (accessed Jan. 29, 2023).

[127] E. Ben Sasson *et al.*, "Zerocash: Decentralized Anonymous Payments from Bitcoin," in *2014 IEEE Symposium on Security and Privacy*, May 2014, pp. 459–474. doi: 10.1109/SP.2014.36.

[128] E. Ben-Sasson, A. Chiesa, A. Gabizon, M. Riabzev, and N. Spooner, "Interactive Oracle Proofs with Constant Rate and Query Complexity." 2016. Accessed: Jan. 29, 2023. [Online]. Available: https://eprint.iacr.org/2016/324

[129] Protocol Labs, "Filecoin: A Decentralized Storage Network," *Filecoin White paper*, 2017. https://filecoin.io/filecoin.pdf (accessed Jan. 29, 2023).

[130] X. Sun, F. R. Yu, P. Zhang, Z. Sun, W. Xie, and X. Peng, "A Survey on Zero-Knowledge Proof in Blockchain," *IEEE Netw.*, vol. 35, no. 4, pp. 198–205, Jul. 2021, doi: 10.1109/MNET.011.2000473.

[131] A.-E. Panait and R. F. Olimid, "On Using zk-SNARKs and zk-STARKs in Blockchain-Based Identity Management," in *Innovative Security Solutions for Information Technology and Communications*, D. Maimut, A.-G. Oprina, and D. Sauveron, Eds., in Lecture Notes in Computer Science. Cham: Springer International Publishing, 2021, pp. 130–145. doi: 10.1007/978-3-030-69255-1_9.

[132] A. M. Pinto, "An Introduction to the Use of zk-SNARKs in Blockchains," in *Mathematical Research for Blockchain Economy*, P. Pardalos, I. Kotsireas, Y. Guo, and W. Knottenbelt, Eds., in Springer Proceedings in Business and Economics. Cham: Springer International Publishing, 2020, pp. 233–249. doi: 10.1007/978-3-030-37110-4_16.

[133] M. F. Esgin, R. K. Zhao, R. Steinfeld, J. K. Liu, and D. Liu, "MatRiCT: Efficient, Scalable and Post-Quantum Blockchain Confidential Transactions Protocol," in *Proceedings of the 2019 ACM SIGSAC Conference on Computer and Communications Security*, in CCS '19. New York, NY, USA: Association for Computing Machinery, Nov. 2019, pp. 567–584. doi: 10.1145/3319535.3354200.

[134] M. F. Esgin, R. Steinfeld, and R. K. Zhao, "MatRiCT+: More Efficient Post-Quantum Private Blockchain Payments," in *2022 IEEE Symposium on Security and Privacy (SP)*, May 2022, pp. 1281–1298. doi: 10.1109/SP46214.2022.9833655.

[135] S. Noether, A. Mackenzie, and the M. R. Lab, "Ring Confidential Transactions," *Ledger*, vol. 1, pp. 1–18, Dec. 2016, doi: 10.5195/ledger.2016.34.

[136] S. Ling, K. Nguyen, D. Stehlé, and H. Wang, "Improved Zero-Knowledge Proofs of Knowledge for the ISIS Problem, and Applications," in *Public-Key Cryptography – PKC 2013*, K. Kurosawa and G. Hanaoka, Eds., in Lecture Notes in Computer Science. Berlin, Heidelberg: Springer, 2013, pp. 107–124. doi: 10.1007/978-3-642-36362-7_8.

[137] S. Gao, D. Zheng, R. Guo, C. Jing, and C. Hu, "An Anti-Quantum E-Voting Protocol in Blockchain With Audit Function," *IEEE Access*, vol. 7, pp. 115304–115316, 2019, doi: 10.1109/ACCESS.2019.2935895.

[138] H. Yu and W. Hui, "Certificateless ring signature from NTRU lattice for electronic voting," *J. Inf. Secur. Appl.*, vol. 75, p. 103496, Jun. 2023, doi: 10.1016/j.jisa.2023.103496.

[139] J. D. Preece and J. M. Easton, "Towards Encrypting Industrial Data on Public Distributed Networks," in *2018 IEEE International Conference on Big Data (Big Data)*, Dec. 2018, pp. 4540–4544. doi: 10.1109/BigData.2018.8622246.

[140] W. Yin, Q. Wen, W. Li, H. Zhang, and Z. Jin, "An Anti-Quantum Transaction Authentication Approach in Blockchain," *IEEE Access*, vol. 6, pp. 5393–5401, 2018, doi: 10.1109/ACCESS.2017.2788411.

[141] H. An and K. Kim, "QChain: Quantum-resistant and decentralized PKI using blockchain," in *2018 Symposium on Cryptography and Information Security (SCIS 2018)*, IEICE Technical Committee on Information Security, 2018.

[142] B. Bhushan, C. Sahoo, P. Sinha, and A. Khamparia, "Unification of Blockchain and Internet of Things (BIoT): requirements, working model, challenges and future directions," *Wirel. Netw.*, vol. 27, no. 1, pp. 55–90, Jan. 2021, doi: 10.1007/s11276-020-02445-6.

[143] X. Hao, W. Ren, Y. Fei, T. Zhu, and K.-K. R. Choo, "A Blockchain-Based Cross-Domain and Autonomous Access Control Scheme for Internet of Things," *IEEE Trans. Serv. Comput.*, vol. 16, no. 2, pp. 773–786, Mar. 2023, doi: 10.1109/TSC.2022.3179727.

[144] H. Yi, "Secure Social Internet of Things Based on Post-Quantum Blockchain," *IEEE Trans. Netw. Sci. Eng.*, vol. 9, no. 3, pp. 950–957, May 2022, doi: 10.1109/TNSE.2021.3095192.

[145] A. Al-Fuqaha, M. Guizani, M. Mohammadi, M. Aledhari, and M. Ayyash, "Internet of Things: A Survey on Enabling Technologies, Protocols, and Applications," *IEEE Commun. Surv. Tutor.*, vol. 17, no. 4, pp. 2347–2376, 2015, doi: 10.1109/COMST.2015.2444095.

[146] Z. Yang and T. Nakajima, "Connecting Smart Objects in IoT Architectures by Screen Remote Monitoring and Control," *Computers*, vol. 7, no. 4, Art. no. 4, Dec. 2018, doi: 10.3390/computers7040047.

[147] K. Seyhan, T. N. Nguyen, S. Akleylek, K. Cengiz, and S. K. H. Islam, "Bi-GISIS KE: Modified key exchange protocol with reusable keys for IoT security," *J. Inf. Secur. Appl.*, vol. 58, p. 102788, May 2021, doi: 10.1016/j.jisa.2021.102788.

[148] J. Chen, W. Gan, M. Hu, and C.-M. Chen, "On the construction of a post-quantum blockchain for smart city," *J. Inf. Secur. Appl.*, vol. 58, p. 102780, May 2021, doi: 10.1016/j.jisa.2021.102780.

[149] A. E. Azzaoui and J. H. Park, "Post-Quantum Blockchain for a Scalable Smart City," *J. Internet Technol.*, vol. 21, no. 4, Art. no. 4, Jul. 2020.

[150] J.-C. Faugére, "A new efficient algorithm for computing Gröbner bases (F4)," *J. Pure Appl. Algebra*, vol. 139, no. 1, pp. 61–88, Jun. 1999, doi: 10.1016/S0022-4049(99)00005-5.

[151] S. Krendelev and P. Sazonova, "Parametric Hash Function Resistant to Attack by Quantum Computer," in *2018 Federated Conference on Computer Science and Information Systems (FedCSIS)*, Sep. 2018, pp. 387–390.

[152] S. Balogh, O. Gallo, R. Ploszek, P. Špaček, and P. Zajac, "IoT Security Challenges: Cloud and Blockchain, Postquantum Cryptography, and Evolutionary Techniques," *Electronics*, vol. 10, no. 21, Art. no. 21, Jan. 2021, doi: 10.3390/electronics10212647.

[153] A. H. Karbasi and S. Shahpasand, "A post-quantum end-to-end encryption over smart contract-based blockchain for defeating man-in-the-middle and interception attacks," *Peer--Peer Netw. Appl.*, vol. 13, no. 5, pp. 1423–1441, Sep. 2020, doi: 10.1007/s12083-020-00901-w.

[154] V. Chamola, A. Jolfaei, V. Chanana, P. Parashari, and V. Hassija, "Information security in the post quantum era for 5G and beyond networks: Threats to existing cryptography, and post-quantum cryptography," *Comput. Commun.*, vol. 176, pp. 99–118, Aug. 2021, doi: 10.1016/j.comcom.2021.05.019.

[155] M. Nassar, K. Salah, M. H. ur Rehman, and D. Svetinovic, "Blockchain for explainable and trustworthy artificial intelligence," *WIREs Data Min. Knowl. Discov.*, vol. 10, no. 1, p. e1340, 2020, doi: 10.1002/widm.1340.

[156] S. Letzgus, P. Wagner, J. Lederer, W. Samek, K.-R. Müller, and G. Montavon, "Toward Explainable Artificial Intelligence for Regression Models: A methodological perspective," *IEEE Signal Process. Mag.*, vol. 39, no. 4, pp. 40–58, Jul. 2022, doi: 10.1109/MSP.2022.3153277.

[157] M. Zolanvari, Z. Yang, K. Khan, R. Jain, and N. Meskin, "TRUST XAI: Model-Agnostic Explanations for AI With a Case Study on IIoT Security," *IEEE Internet Things J.*, vol. 10, no. 4, pp. 2967–2978, Feb. 2023, doi: 10.1109/JIOT.2021.3122019.

[158] S. Dolev and Z. Wang, "SodsBC: Stream of Distributed Secrets for Quantum-safe Blockchain," in *2020 IEEE International Conference on Blockchain (Blockchain)*, Nov. 2020, pp. 247–256. doi: 10.1109/Blockchain50366.2020.00038.

[159] S. Dolev, B. Guo, J. Niu, and Z. Wang, "SodsBC: A Post-Quantum by Design Asynchronous Blockchain Framework," *IEEE Trans. Dependable Secure Comput.*, pp. 1–16, 2023, doi: 10.1109/TDSC.2023.3243588.

[160] "Dev Status Update — November, 2020," *IOTA Foundation Blog*, Nov. 12, 2020. http://blog.iota.org/dev-status-update-november-2020-80e28a27f7bb (accessed Jan. 29, 2023).

[161] K. (Konstantinos) Chalkias, "R3 publishes a new post-quantum signature algorithm tailored to blockchains," *Corda*, Aug. 10, 2018. https://medium.com/corda/r3-publishes-a-new-post-quantum-signature-algorithm-tailored-to-blockchains-51719c64fd4c (accessed Jan. 29, 2023).

[162] C. H. Bennett and G. Brassard, "Quantum cryptography: Public key distribution and coin tossing," *Theor. Comput. Sci.*, vol. 560, pp. 7–11, Dec. 2014, doi: 10.1016/j.tcs.2014.05.025.

[163] A. K. Ekert, "Quantum cryptography based on Bell's theorem," *Phys. Rev. Lett.*, vol. 67, no. 6, pp. 661–663, Aug. 1991, doi: 10.1103/PhysRevLett.67.661.

[164] C. H. Bennett, "Quantum cryptography using any two nonorthogonal states," *Phys. Rev. Lett.*, vol. 68, no. 21, pp. 3121–3124, May 1992, doi: 10.1103/PhysRevLett.68.3121.

[165] V. Scarani, A. Acín, G. Ribordy, and N. Gisin, "Quantum Cryptography Protocols Robust against Photon Number Splitting Attacks for Weak Laser Pulse Implementations," *Phys. Rev. Lett.*, vol. 92, no. 5, p. 057901, Feb. 2004, doi: 10.1103/PhysRevLett.92.057901.

[166] K. Inoue, E. Waks, and Y. Yamamoto, "Differential-phase-shift quantum key distribution using coherent light," *Phys. Rev. A*, vol. 68, no. 2, p. 022317, Aug. 2003, doi: 10.1103/PhysRevA.68.022317.

[167] A. I. Nurhadi and N. R. Syambas, "Quantum Key Distribution (QKD) Protocols: A Survey," in *2018 4th International Conference on Wireless*

[167] *and Telematics (ICWT)*, Jul. 2018, pp. 1–5. doi: 10.1109/ICWT.2018.8527822.
[168] V. Padamvathi, B. V. Vardhan, and A. V. N. Krishna, "Quantum Cryptography and Quantum Key Distribution Protocols: A Survey," in *2016 IEEE 6th International Conference on Advanced Computing (IACC)*, Feb. 2016, pp. 556–562. doi: 10.1109/IACC.2016.109.
[169] H.-K. Lo, M. Curty, and K. Tamaki, "Secure quantum key distribution," *Nat. Photonics*, vol. 8, no. 8, Art. no. 8, Aug. 2014, doi: 10.1038/nphoton.2014.149.
[170] J. G. Rarity, P. R. Tapster, P. M. Gorman, and P. Knight, "Ground to satellite secure key exchange using quantum cryptography," *New J. Phys.*, vol. 4, no. 1, p. 82, Oct. 2002, doi: 10.1088/1367-2630/4/1/382.
[171] E. Villaseñor, M. He, Z. Wang, R. Malaney, and M. Z. Win, "Enhanced Uplink Quantum Communication With Satellites via Downlink Channels," *IEEE Trans. Quantum Eng.*, vol. 2, pp. 1–18, 2021, doi: 10.1109/TQE.2021.3091709.
[172] J. S. Sidhu *et al.*, "Advances in space quantum communications," *IET Quantum Commun.*, vol. 2, no. 4, pp. 182–217, 2021, doi: 10.1049/qtc2.12015.
[173] Z. Yang, A. Ghubaish, D. Unal, and R. Jain, "Factors Affecting the Performance of Sub-1 GHz IoT Wireless Networks," *Wirel. Commun. Mob. Comput.*, vol. 2021, p. e8870222, Jun. 2021, doi: 10.1155/2021/8870222.
[174] D. Gottesman and I. Chuang, "Quantum Digital Signatures." arXiv, Nov. 14, 2001. doi: 10.48550/arXiv.quant-ph/0105032.
[175] A. S. Holevo, "Bounds for the quantity of information transmitted by a quantum communication channel," *Probl. Peredachi Informatsii*, vol. 9, no. 3, pp. 3–11, 1973.
[176] S. Singh, N. K. Rajput, V. K. Rathi, H. M. Pandey, A. K. Jaiswal, and P. Tiwari, "Securing Blockchain Transactions Using Quantum Teleportation and Quantum Digital Signature," *Neural Process. Lett.*, Jun. 2020, doi: 10.1007/s11063-020-10272-1.
[177] C. H. Bennett, G. Brassard, C. Crépeau, R. Jozsa, A. Peres, and W. K. Wootters, "Teleporting an unknown quantum state via dual classical and Einstein-Podolsky-Rosen channels," *Phys. Rev. Lett.*, vol. 70, no. 13, pp. 1895–1899, Mar. 1993, doi: 10.1103/PhysRevLett.70.1895.
[178] A. Miranowicz and K. Tamaki, "An Introduction to Quantum Teleportation." arXiv, Feb. 14, 2003. doi: 10.48550/arXiv.quant-ph/0302114.
[179] M. Naor, "Bit commitment using pseudorandomness," *J. Cryptol.*, vol. 4, no. 2, pp. 151–158, Jan. 1991, doi: 10.1007/BF00196774.
[180] L. Hardy and A. Kent, "Cheat Sensitive Quantum Bit Commitment," *Phys. Rev. Lett.*, vol. 92, no. 15, p. 157901, Apr. 2004, doi: 10.1103/PhysRevLett.92.157901.
[181] D. Mayers, "Unconditionally Secure Quantum Bit Commitment is Impossible," *Phys. Rev. Lett.*, vol. 78, no. 17, pp. 3414–3417, Apr. 1997, doi: 10.1103/PhysRevLett.78.3414.
[182] X. Sun, Q. Wang, P. Kulicki, and M. Sopek, "A Simple Voting Protocol on Quantum Blockchain," *Int. J. Theor. Phys.*, vol. 58, no. 1, pp. 275–281, Jan. 2019, doi: 10.1007/s10773-018-3929-6.
[183] H. Krawczyk, "New Hash Functions for Message Authentication," in *Advances in Cryptology — EUROCRYPT '95*, L. C. Guillou and J.-J. Quisquater, Eds., in Lecture Notes in Computer Science. Berlin, Heidelberg: Springer, 1995, pp. 301–310. doi: 10.1007/3-540-49264-X_24.
[184] N. D. Truong, J. Y. Haw, S. M. Assad, P. K. Lam, and O. Kavehei, "Machine Learning Cryptanalysis of a Quantum Random Number Generator," *IEEE Trans. Inf. Forensics Secur.*, vol. 14, no. 2, pp. 403–414, Feb. 2019, doi: 10.1109/TIFS.2018.2850770.
[185] W. McCutcheon *et al.*, "Experimental verification of multipartite entanglement in quantum networks," *Nat. Commun.*, vol. 7, no. 1, Art. no. 1, Nov. 2016, doi: 10.1038/ncomms13251.
[186] Q. Wang, X. Yu, Q. Zhu, Y. Zhao, and J. Zhang, "Quantum key pool construction and key distribution scheme in multi-domain QKD optical networks (QKD-ON)," in *4th Optics Young Scientist Summit (OYSS 2020)*, SPIE, Feb. 2021, pp. 509–512. doi: 10.1117/12.2591316.
[187] K. Li, R. Shi, M. Wu, Y. Li, and X. Zhang, "A novel privacy-preserving multi-level aggregate signcryption and query scheme for Smart Grid via mobile fog computing," *J. Inf. Secur. Appl.*, vol. 67, p. 103214, Jun. 2022, doi: 10.1016/j.jisa.2022.103214.
[188] X. Sun, Q. Wang, P. Kulicki, and X. Zhao, "Quantum-enhanced Logic-based Blockchain I: Quantum Honest-success Byzantine Agreement and Qulogicoin." arXiv, Jul. 16, 2018. doi: 10.48550/arXiv.1805.06768.
[189] X. Sun, M. Sopek, Q. Wang, and P. Kulicki, "Towards Quantum-Secured Permissioned Blockchain: Signature, Consensus, and Logic," *Entropy*, vol. 21, no. 9, Art. no. 9, Sep. 2019, doi: 10.3390/e21090887.
[190] X. Sun, P. Kulicki, and M. Sopek, "Lottery and Auction on Quantum Blockchain," *Entropy*, vol. 22, no. 12, Art. no. 12, Dec. 2020, doi: 10.3390/e22121377.
[191] W. Wang, Y. Yu, and L. Du, "Quantum blockchain based on asymmetric quantum encryption and a stake vote consensus algorithm," *Sci. Rep.*, vol. 12, no. 1, Art. no. 1, May 2022, doi: 10.1038/s41598-022-12412-0.
[192] C. Tan and L. Xiong, "DPoSB: Delegated Proof of Stake with node's behavior and Borda Count," in *2020 IEEE 5th Information Technology and Mechatronics Engineering Conference (ITOEC)*, Jun. 2020, pp. 1429–1434. doi: 10.1109/ITOEC49072.2020.9141744.
[193] X. Xin, Q. Yang, and F. Li, "Quantum public-key signature scheme based on asymmetric quantum encryption with trapdoor information," *Quantum Inf. Process.*, vol. 19, no. 8, p. 233, Jul. 2020, doi: 10.1007/s11128-020-02736-z.
[194] Z. Yang, T. Salman, R. Jain, and R. D. Pietro, "Decentralization Using Quantum Blockchain: A Theoretical Analysis," *IEEE Trans. Quantum Eng.*, vol. 3, pp. 1–16, 2022, doi: 10.1109/TQE.2022.3207111.
[195] K. Ikeda, "qBitcoin: A Peer-to-Peer Quantum Cash System," in *Intelligent Computing*, K. Arai, S. Kapoor, and R. Bhatia, Eds., in Advances in Intelligent Systems and Computing. Cham: Springer International Publishing, 2019, pp. 763–771. doi: 10.1007/978-3-030-01174-1_58.
[196] A. A. Abd El-Latif, B. Abd-El-Atty, I. Mehmood, K. Muhammad, S. E. Venegas-Andraca, and J. Peng, "Quantum-Inspired Blockchain-Based Cybersecurity: Securing Smart Edge Utilities in IoT-Based Smart Cities," *Inf. Process. Manag.*, vol. 58, no. 4, p. 102549, Jul. 2021, doi: 10.1016/j.ipm.2021.102549.
[197] S. Banerjee, A. Mukherjee, and P. K. Panigrahi, "Quantum blockchain using weighted hypergraph states," *Phys. Rev. Res.*, vol. 2, no. 1, p. 013322, Mar. 2020, doi: 10.1103/PhysRevResearch.2.013322.
[198] A. J. Bennet and S. Daryanoosh, "Energy efficient mining on a quantum-enabled blockchain using light," *Ledger*, vol. 4, Jul. 2019, doi: 10.5195/ledger.2019.143.
[199] X.-J. Wen, Y.-Z. Chen, X.-C. Fan, W. Zhang, Z.-Z. Yi, and J.-B. Fang, "Blockchain consensus mechanism based on quantum zero-knowledge proof," *Opt. Laser Technol.*, vol. 147, p. 107693, Mar. 2022, doi: 10.1016/j.optlastec.2021.107693.
[200] S. Gupta, A. Gupta, I. Y. Pandya, A. Bhatt, and K. Mehta, "End to end secure e-voting using blockchain & quantum key distribution," *Mater. Today Proc.*, Aug. 2021, doi: 10.1016/j.matpr.2021.07.254.
[201] H. Zhu, X. Wang, C.-M. Chen, and S. Kumari, "Two novel semi-quantum-reflection protocols applied in connected vehicle systems with blockchain," *Comput. Electr. Eng.*, vol. 86, p. 106714, Sep. 2020, doi: 10.1016/j.compeleceng.2020.106714.
[202] N. Gisin, G. Ribordy, W. Tittel, and H. Zbinden, "Quantum cryptography," *Rev. Mod. Phys.*, vol. 74, no. 1, pp. 145–195, Mar. 2002, doi: 10.1103/RevModPhys.74.145.
[203] Z. Cai, J. Qu, P. Liu, and J. Yu, "A Blockchain Smart Contract Based on Light-Weighted Quantum Blind Signature," *IEEE Access*, vol. 7, pp. 138657–138668, 2019, doi: 10.1109/ACCESS.2019.2941153.
[204] L. Lin and Y. Guo, "Constructions of Quantum Blind Signature Based on Two-Particle-Entangled System," in *2009 Second International Symposium on Information Science and Engineering*, Dec. 2009, pp. 355–358. doi: 10.1109/ISISE.2009.39.
[205] X.-Y. Li, Y. Chang, S.-B. Zhang, J.-Q. Dai, and T. Zheng, "Quantum Blind Signature Scheme Based on Quantum Walk," *Int. J. Theor. Phys.*, vol. 59, no. 7, pp. 2059–2073, Jul. 2020, doi: 10.1007/s10773-020-04478-1.
[206] D. Chaum, "Blind Signatures for Untraceable Payments," in *Advances in Cryptology*, D. Chaum, R. L. Rivest, and A. T. Sherman, Eds., Boston, MA: Springer US, 1983, pp. 199–203. doi: 10.1007/978-1-4757-0602-4_18.
[207] M. Bhavin, S. Tanwar, N. Sharma, S. Tyagi, and N. Kumar, "Blockchain and quantum blind signature-based hybrid scheme for healthcare 5.0 applications," *J. Inf. Secur. Appl.*, vol. 56, p. 102673, Feb. 2021, doi: 10.1016/j.jisa.2020.102673.
[208] A. Coladangelo and O. Sattath, "A Quantum Money Solution to the Blockchain Scalability Problem," *Quantum*, vol. 4, p. 297, Jul. 2020, doi: 10.22331/q-2020-07-16-297.

[209] M. Zhandry, "Quantum Lightning Never Strikes the Same State Twice. Or: Quantum Money from Cryptographic Assumptions," *J. Cryptol.*, vol. 34, no. 1, p. 6, Jan. 2021, doi: 10.1007/s00145-020-09372-x.
[210] J. Jogenfors, "Quantum Bitcoin: An Anonymous, Distributed, and Secure Currency Secured by the No-Cloning Theorem of Quantum Mechanics," in *2019 IEEE International Conference on Blockchain and Cryptocurrency (ICBC)*, May 2019, pp. 245–252. doi: 10.1109/BLOC.2019.8751473.
[211] S. Wiesner, "Conjugate coding," *ACM Sigact News*, vol. 15, no. 1, pp. 78–88, 1983.
[212] Q. Li, J. Wu, J. Quan, J. Shi, and S. Zhang, "Efficient Quantum Blockchain With a Consensus Mechanism QDPoS," *IEEE Trans. Inf. Forensics Secur.*, vol. 17, pp. 3264–3276, 2022, doi: 10.1109/TIFS.2022.3203316.
[213] J. Wang, Y. Ding, N. N. Xiong, W.-C. Yeh, and J. Wang, "GSCS: General Secure Consensus Scheme for Decentralized Blockchain Systems," *IEEE Access*, vol. 8, pp. 125826–125848, 2020, doi: 10.1109/ACCESS.2020.3007938.
[214] A. A. Clerk, M. H. Devoret, S. M. Girvin, F. Marquardt, and R. J. Schoelkopf, "Introduction to quantum noise, measurement, and amplification," *Rev. Mod. Phys.*, vol. 82, no. 2, pp. 1155–1208, Apr. 2010, doi: 10.1103/RevModPhys.82.1155.
[215] D. Deutsch and R. Penrose, "Quantum theory, the Church–Turing principle and the universal quantum computer," *Proc. R. Soc. Lond. Math. Phys. Sci.*, vol. 400, no. 1818, pp. 97–117, Jan. 1997, doi: 10.1098/rspa.1985.0070.
[216] "IBM Unveils 400 Qubit-Plus Quantum Processor and Next-Generation IBM Quantum System Two," *IBM Newsroom*. https://newsroom.ibm.com/2022-11-09-IBM-Unveils-400-Qubit-Plus-Quantum-Processor-and-Next-Generation-IBM-Quantum-System-Two (accessed Jan. 29, 2023).
[217] S. Bäuml and K. Azuma, "Fundamental limitation on quantum broadcast networks," *Quantum Sci. Technol.*, vol. 2, no. 2, p. 024004, May 2017, doi: 10.1088/2058-9565/aa6d3c.
[218] J. F. Fitzsimons, "Private quantum computation: an introduction to blind quantum computing and related protocols," *Npj Quantum Inf.*, vol. 3, no. 1, Art. no. 1, Jun. 2017, doi: 10.1038/s41534-017-0025-3.
[219] A. Hayashi, M. Horibe, and T. Hashimoto, "Quantum pure-state identification," *Phys. Rev. A*, vol. 72, no. 5, p. 052306, Nov. 2005, doi: 10.1103/PhysRevA.72.052306.
[220] D. Aggarwal, G. K. Brennen, T. Lee, M. Santha, and M. Tomamichel, "Quantum attacks on Bitcoin, and how to protect against them," *Ledger*, vol. 3, Oct. 2018, doi: 10.5195/ledger.2018.127.
[221] Q. Zhu, S. W. Loke, R. Trujillo-Rasua, F. Jiang, and Y. Xiang, "Applications of Distributed Ledger Technologies to the Internet of Things: A Survey," *ACM Comput. Surv.*, vol. 52, no. 6, p. 120:1-120:34, Nov. 2019, doi: 10.1145/3359982.
[222] C. Dods, N. P. Smart, and M. Stam, "Hash Based Digital Signature Schemes," in *Cryptography and Coding*, N. P. Smart, Ed., in Lecture Notes in Computer Science. Berlin, Heidelberg: Springer, 2005, pp. 96–115. doi: 10.1007/11586821_8.
[223] T. H. Troyer Thomas Hner, Matthias, "Disentangling Hype from Practicality: On Realistically Achieving Quantum Advantage," *Communications of the ACM*. https://cacm.acm.org/magazines/2023/5/272276-disentangling-hype-from-practicality-on-realistically-achieving-quantum-advantage/fulltext (accessed Jan. 29, 2023).

**Zebo Yang** (Graduate Student Member, IEEE) received a B.E. degree in computer engineering from the Guangdong University of Foreign Studies, Guangzhou, China, in 2012 and an M.E. degree in computer engineering from the Waseda University, Tokyo, Japan, in 2019. He is working toward a Ph.D. in computer science at Washington University in St. Louis, St. Louis, MO, USA.
From 2011 to 2017, he worked as a Software Engineer with Tencent, Inc., Baidu, Inc., and DJI, Inc. Since 2019, he has worked as a Graduate Research Assistant with Washington University in St. Louis. His research interests include quantum computing, blockchains, computer networks, network and system security, machine learning, and the Internet of Things.

**Haneen Alfauri** received a B.S. degree in Electrical Engineering (minor in Telecommunications) from Princes Sumaya University, Amman, Jordan, in 2014 and the M.S. degree (Hons.) in Electrical and Computer Engineering from Saint Louis University, Saint Louis, MO, USA, in 2021. She is pursuing a Ph.D. in computer engineering at Washington University, Saint Louis, MO, USA. Her research interests include network and system security, blockchains, quantum computing, distributed systems, AI and machine learning.

**Behrooz Farkiani** received B.Sc. and M.Sc. degrees in computer engineering from the Amirkabir University of Technology in 2012 and 2014, respectively. He is currently pursuing a Ph.D. degree in Computer Science at Washington University in St. Louis. His research interests include networked systems and the management of softwarized networks.

**Raj Jain** (Life Fellow, IEEE) received a B.E. degree in electrical engineering from APS University, Rewa, India, an M.E. degree in Automation from Indian Institute of Science, Bangalore, India, and a Ph. D. degree in Applied Maths (computer science) from Harvard University, Cambridge Massachusetts, USA, in 1972, 1974 and 1978 respectively. He is the Barbara J. and Jerome R. Cox, Jr., Professor with the Department of Computer Science and Engineering, Washington University in St Louis, St Louis, MO, USA. He co-founded Nayna Networks, Inc., San Jose, CA, USA, a next-generation telecommunications systems company in San Jose. He was a Senior Consulting Engineer with Digital Equipment Corporation, Littleton, MA, USA, and then a Computer and Information Sciences Professor at Ohio State University, Columbus, OH, USA.
Prof. Jain is a recipient of the 2018 James B. Eads Award from the St. Louis Academy of Science, the 2017 ACM SIGCOMM Life-Time Achievement Award, and the 2015 A. A. Michelson Award from Computer Measurement Group. He ranks among the Most Cited Authors in Computer Science. He has authored the Art of Computer Systems Performance Analysis, which won the 1991 "Best-Advanced How-to Book, Systems" award from the Computer Press Association. He is a Fellow of the IEEE, ACM, and AAAS.

**Roberto Di Pietro** (Fellow, IEEE; Distinguished Scientist, ACM) is a Full Professor in Cybersecurity with the KAUST-CEMSE, KSA. Previously, he was a Professor in Cybersecurity and founder of the CRI-Lab at HBKU-CES, Doha-Qatar. He also served as Global Head of Security Research for Nokia Bell Labs. He has been working in the cybersecurity field for more than 25 years, leading technology-oriented and research-focused teams in the private sector, government, and academia (MoD, United Nations HQ, EUROJUST, IAEA, and WIPO). Besides being involved in M&A of startups—and having founded one (exited)—he is a board member of a few research institutions. He has produced 270+ scientific papers, three books, and 16 patents in the cybersecurity field.

In 2011–2012, he received a Chair of Excellence from the University Carlos III, Madrid. In 2020, he received the Jean-Claude Laprie Award for significantly influencing the theory and practice of Dependable Computing. In 2022, he was awarded the Individual Innovation Award from HBKU. His education accounts for an M.S. in Computer Science (1994), an M.S. in informatics (2003), a Specialization Diploma in Operations Research and Strategic Decisions (2003), and a Ph.D. degree in Computer Science (2004).

**Aiman Erbad** (Senior Member, IEEE) is an Associate Professor and ICT Division Head in the College of Science and Engineering, Hamad Bin Khalifa University (HBKU). Dr. Erbad obtained a Ph.D. in Computer Science from the University of British Columbia (Canada) in 2012, a Master of Computer Science in embedded systems and robotics from the University of Essex (UK) in 2005, and a BSc in Computer Engineering from the University of Washington, Seattle in 2004. He received the 2020 Best Research Paper Award from Computer Communications (COMCOM), the IWCMC 2019 Best Paper Award, and the IEEE CCWC 2017 Best Paper Award. He published more than 160 papers in reputable international conferences and journals. He is the general chair for ISNCC 2023. He also served as a Program Chair of IWCMC 2022 and IWCMC 2019, as a Publicity chair of ACM MoVid 2015, as a Local Arrangement Chair of NOSSDAV 2011, and as a Technical Program Committee member in various IEEE and ACM international conferences (GlobeCom, ICC, NOSSDAV, MMSys, ACMMM, IC2E, and ICNC). His research interests span cloud computing, quantum networks, edge intelligence, Internet of Things (IoT), and private and secure networks. He is a senior member of IEEE and ACM.